%% file: texla_ar.tex
\documentstyle[amssymb,secnumtab]{fbsart11}

\begin{document}

\def\bsig{\mbox{\boldmath{$\sigma$}}}
\def\bl{{\bf l}}
\def\haf{\textstyle{1\over2}}

\input prepictex
\input pictex
\input postpictex

\title{New Three-Nucleon-Force Terms in the Three-Nucleon System
\footnote{Dedicated to the 60th birthday of Walter Gl\"ockle.}}
\author{D. H\"uber\instnr{1,2}, J.L. Friar\instnr{1},
A. Nogga\instnr{3}, H. Wita\l a\instnr{4}, U. van Kolck\instnr{5}}

\instlist{
Theoretical Division, Los Alamos National Laboratory,
M.S. B283, Los Alamos, NM 87545, USA
\and
living systems AG, Roggenbach Str. 1, D-78050 VS-Villingen, Germany; email:
dirk.hueber@living-systems.de
\and
Institut f\"ur theoretische Physik II, Ruhr-Universit\"at Bochum,
D-44780 Bochum, Germany
\and
Institute of Physics, Jagellonian University, PL-30059 Cracow,
Poland
\and
Kellogg Radiation Laboratory, 106-38, California Institute of
Technology, Pasadena, CA 91125, USA
}

\runningauthor{D. H\"uber et al.}
\runningtitle{New Three-Nucleon-Force Terms}

\sloppy
\maketitle

\begin{abstract}
We include two new three-nucleon-force terms of pion-range -- short-range form
in our momentum-space calculations for the three-nucleon continuum. These two
terms are expected by chiral perturbation theory to be non-negligible. We study
the effects of these terms in elastic neutron-deuteron scattering and pay
special attention to the neutron vector analyzing power $A_y$.
\end{abstract}
\newpage

\section{Introduction}
\label{intro}

Recently it became possible to explain differences between data and
predictions
of modern nucleon-nucleon (NN) potentials for the total neutron-deuteron
(nd)
cross section and the minimum of the differential nd cross section at
higher
energies by incorporating the $2\pi$-exchange Tucson-Melbourne (TM)
three-nucleon force (3NF) \cite{total} \cite{diff}. However, the puzzling
discrepancy between theory and data for the low-energy analyzing power
\cite{report} cannot be explained by that 3NF. Since it was shown in
\cite{Ay}
that this low-energy analyzing-power puzzle cannot be solved by reasonable
changes in the NN potentials, one obviously needs new 3NF mechanisms in
order
to resolve this long-standing mystery.

The number of possible operators that can be used in constructing a 3NF is
much larger than in the NN-force case, and it is not practicable to
examine them all in order to see which are important for the low-energy
analyzing powers and which are not.
Rather, we need a systematic scheme that tells us which terms are likely
candidates to be of importance
and which are not. One such approach is chiral perturbation theory
($\chi$PT), which provides a power-counting scheme for the strength of the
various 3NF terms.

As we will explain below, $\chi$PT predicts in lowest non-vanishing order
(beside the usual $2\pi$-exchange terms) two terms of pion-range --
short-range
nature and three terms of short-range -- short-range nature. Short range
means,
for example, that a meson heavier than a pion is exchanged between two of
the
three nucleons. Since the naive expectation is that pion-range --
short-range
3NF terms are more important than the ones of short-range -- short-range
nature,
in this paper we will deal only with the two pion-range -- short-range 3NF
terms, which we will include in our momentum-space calculations for the
3N continuum. Our interest will be concentrated on the low-energy
vector analyzing powers.

In Section \ref{2pi3nf} we review the history of the $2\pi$-exchange
3NF and its effect in the 3N continuum. In Section \ref{heavy} we discuss
those
3NF terms that include the exchange of mesons heavier than pions and that
have been tested in the 3N continuum thus far.

The two 3NF terms of pion-range -- zero-range nature that are predicted
in lowest non-vanishing order of $\chi$PT are introduced in Section
\ref{newterms}. In this section we also explain how we adapt these terms,
making them of finite range
in order to be consistent with the traditional potentials that we use
 as our NN force, and also with the TM force that we use as our
$2\pi$-exchange
3NF in this paper. At the end of this section we discuss what conventional
nuclear field-theory models say about these two 3NF terms.

Our approach to solving the Faddeev equations for the 3N continuum is
reviewed briefly in Section \ref{3nc}.

The results for the nd elastic-scattering observables that incorporate the
new
3NF terms are presented and discussed in Section \ref{results}.

Finally we sum up and conclude in Section \ref{sum}.

In \ref{PWD} we present the partial-wave decomposition (PWD) for the two
new 3NF terms.

\section{Short Review of $2\pi$-Exchange Three-Nucleon Forces}
\label{2pi3nf}

\begin{figure}
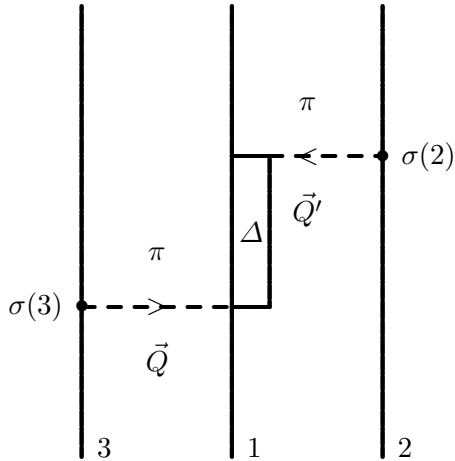

\begin{center}
\hskip0truecm\hbox{\input logoFM.tex }
\end{center}
\caption{\label{figFM} The Fujita-Miyazawa 3NF.}
\end{figure}

Three-nucleon forces have been part of nuclear physics for more than 40
years.
Realistic models began with the Fujita-Miyazawa (FM) force \cite{FM},
which describes
the exchange of two pions with an intermediate $\Delta$-isobar as depicted
in Fig.~\ref{figFM}. Figure~\ref{figFM} shows only that part of the 3NF for
which nucleon 1 is the middle nucleon (i.e., the nucleon at which the
virtual
pion scattering takes place). We will call this configuration $V_4^{(1)}$
\footnote{The use of the index ``4'' for a 3NF is common usage for
three-nucleon calculations, where the indices ``1''-``3'' denote the
three NN-pair forces.}. The full 3NF is then given by
\begin{equation}
\label{eq1}
V_4=V_4^{(1)}+V_4^{(2)}+V_4^{(3)}
\end{equation}

\begin{figure}
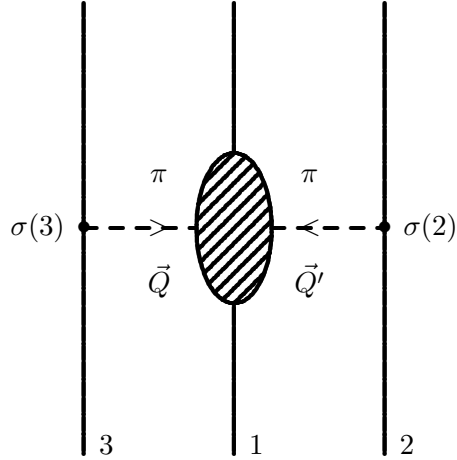

\begin{center}
\hskip0truecm\hbox{\input logo3NF.tex }
\end{center}
\caption{\label{fig3NF} The $2\pi$-exchange 3NF.}
\end{figure}

Scattering the pion in Fig.~\ref{figFM} from nucleon 1 via a virtual
$\Delta$-isobar is not the most general process that leads to a
$2\pi$-exchange 3NF. The many ways to accomplish this are indicated by the
blob in Fig.~\ref{fig3NF}. Several Ans\"atze have
been used to derive the $2\pi$-exchange 3NF up to now. A short overview is
given in Table~\ref{tab3NF}. We will not comment here on the different
techniques and the underlying physical ideas that lead to these 3NF models.
For that we refer the interested reader to our recent paper \cite{cs3nf}.
Rather, we will concentrate here on the differences between these models in
terms of operator form and the effects of these different models in 3N
calculations.
\begin{table}
\begin{tabular}{|l|l|l|c|c|c|c|}
\hline
year&3NF&characteristic&$a'$&$b$&$c$&$d$\\
\hline
\hline
1957&Fujita-Miyazawa\protect\cite{FM}&isobars&0&-1.15&0&-0.29\\
\hline
1979&Tucson-Melbourne\protect\cite{TM}&current
algebra&-1.03&-2.62&1.03&-0.60\\
\hline
1983&Brazil\protect\cite{Br}&chiral Lagrangian
&-1.05&-2.29&(1.05)&-0.77\\
&&+ (current algebra)&&&&\\
\hline
1983&Urbana\protect\cite{Ur}&isobars with additional&0&-1.20&0&-0.30\\
&&phenomenological&&&&\\
&&medium-range term&&&&\\
\hline
1993&Texas\protect\cite{Texas}&chiral perturbation
theory&-1.87&-3.82&0&-1.12\\
\hline
1996&RuhrPot\protect\cite{Rp}&non-chiral Lagrangian&-0.51&-1.82&0&-0.48\\
\hline
\end{tabular}
\caption{\label{tab3NF} Various $2\pi$-exchange 3NF models in use today.
}
\end{table}

Let us begin with the types of operators contained in various 3NF models in
use
today. We will use the familiar language of the Tucson-Melbourne
force \cite{TM}, which labels four $2\pi$-exchange operators by its
parameters
$a$, $b$, $c$, and $d$.
The operator form of these terms (neglecting all overall factors and form
factors) is given by
\begin{eqnarray}
\label{eq2}
V_4^{(1)}|_a&=&{1\over {\vec Q}^2+m_\pi^2}\ {1\over {\vec {Q'}}^2+m_\pi^2}\
              \vec \tau_2 \cdot \vec \tau_3\\
\label{eq3}
V_4^{(1)}|_b&=&{1\over {\vec Q}^2+m_\pi^2}\ {1\over {\vec {Q'}}^2+m_\pi^2}\
              \vec Q \cdot \vec {Q'}\ \vec \tau_2 \cdot \vec \tau_3\\
\label{eq4}
V_4^{(1)}|_c&=&{1\over {\vec Q}^2+m_\pi^2}\ {1\over {\vec {Q'}}^2+m_\pi^2}\
              ({\vec Q}^2 + {\vec {Q'}}^2)\ \vec \tau_2 \cdot \vec \tau_3\\
\label{eq5}
V_4^{(1)}|_d&=&{1\over {\vec Q}^2+m_\pi^2}\ {1\over {\vec {Q'}}^2+m_\pi^2}\
              \vec \sigma_1 \cdot \vec Q \times \vec {Q'}\
              \vec \tau_1 \cdot \vec \tau_2 \times \vec \tau_3
\end{eqnarray}
where we have used the notation implicit in Figs.~\ref{figFM} and
\ref{fig3NF}.

The main difference between all these models (in terms of operator
structure)
is the $c$-term in the TM force. Let's have a closer look at the $c$-term
and
rewrite it (neglecting the isospin dependence in Eq.~(\ref{eq4})) as
\begin{eqnarray}
\label{eq6}
V_4^{(1)}|_c&\propto&
{\vec Q^2\over \vec Q^2+m_\pi^2}\
{1\over \vec {Q'}^2+m_\pi^2}+(Q\leftrightarrow Q')\nonumber\\
&=&{\vec Q^2+m_\pi^2-m_\pi^2\over \vec Q^2+m_\pi^2}\
{1\over \vec {Q'}^2+m_\pi^2}+(Q\leftrightarrow Q')\nonumber\\
&=&\left( \underbrace{1}_{SR}-
\underbrace{m_\pi^2\over \vec Q^2+m_\pi^2}_{\pi{\rm -range}}\right) \
{1\over \vec {Q'}^2+m_\pi^2}+(Q\leftrightarrow Q')
\end{eqnarray}
Thus the $c$-term can be decomposed into a $2\pi$-exchange term with the
same
operator
structure as the $a$-term plus a short-range -- $\pi$-range term (marked
``$SR$'' in Eq.~(\ref{eq6})). The
inclusion
of the $2\pi$-exchange part of the $c$-term leads to a redefinition of $a$
as
\begin{equation}
\label{eq7}
a'=a-2\ m_\pi^2 \ c
\end{equation}
which essentially means a change of sign for $a$: $a'\approx -a$. Therefore
the difference between the TM model and the other  $2\pi$-exchange 3NFs
is the former's short-range -- $\pi$-range part of the $c$-term. Arguments
developed in \cite{cs3nf} using chiral symmetry show that
this short-range -- $\pi$-range part of the $c$-term should be
dropped. Doing this (and accordingly replacing $a$ by $a^{\prime}$), one
gets a
``corrected'' TM force that we will call TM$^{\prime}$ in what follows.

Since the FM force has included only the $2\pi$-$\Delta$ mechanism, it has
only the $b$- and $d$-terms. Also, for the same reason, the values of $b$
and
$d$ are roughly half the size of the corresponding TM-model values.

The Brazil force \cite{Br} is very closely related to the TM model;
however,
it doesn't have the $c$-term. (That is, in it's original form, although a
$c$-term was later included in order to agree with the TM model; this is
indicated by the brackets in Table~\ref{tab3NF}.)
The values for the parameters $a'$, $b$, and $d$
are very close to the TM values.

The Urbana model \cite{Ur}
has only the $b$- and $d$-terms with values for $b$ and $d$ like those of
the FM model, since it also incorporates only the $\Delta$-mechanism.
In addition it has a phenomenological medium-range term, whose inclusion
was motivated by nuclear-matter calculations.
It turns out that this term plays no important role in the 3N system.

Like the Brazil force the Texas force \cite{Texas} has the $a$-, $b$-, and
$d$-terms, but with values for $a'$, $b$, and $d$ obtained from a fit to
$\pi-N$
scattering and substantially larger than
those
used in the Brazil or TM models.

Finally, the RuhrPot model \cite{Rp} also employs the $a'$-, $b$-, and
$d$-terms,
but with significantly smaller values for the parameters than those used in
the
TM force.

Consequently one does not expect major differences in the effects of these
$2\pi$-exchange 3NFs in the 3N system  and, indeed, no major differences
have been found; (except for the Texas force) all of the above-mentioned
$2\pi$-exchange 3NF models have been tested in 3N bound-state and continuum
calculations, either by the Bochum-Cracow group or the Pisa group or by
both.
Even the original TM force with the
$c$-term gives results similar to all the other models at low energies,
for reasons explained below. The only differences to be found are related
to the
different strengths of these models due to different values for the
parameters,
as well as due to different choices for the $\pi$NN form factors.

Let us now study the effects of the different terms individually. For the
triton this has already been done in Ref.~\cite{triton}. There it was found
that the largest contribution to the binding energy (60\%-70\%) comes from
the $b$-term, whereas the $a^{\prime}$-term can be neglected. The rest
comes
from the $c$- and $d$-terms, where the relative importance of these two
terms
depends
strongly on the chosen NN interaction (Ref.~\cite{triton} used the RSC and
AV14
potentials).

In order to get the experimental value of the triton binding energy in
conjunction with various NN force models, one can adjust the cut-off
parameter
$\Lambda$ in the $\pi$NN form factors of the TM and Brazil
forces \cite{trifit}. The $\pi$NN form factors of these models have the
form
\begin{equation}
\label{eq8}
F(Q^2)={\Lambda^2-m_\pi^2\over\Lambda^2+Q^2}
\end{equation}
Not only does the denominator in Eq.~(\ref{eq8}) suppress momenta large
compared to the cut-off parameter $\Lambda$,
but because $F$ is normalized to one at $Q^2=-m_\pi^2$, the numerator (and
thus $F$ at low momenta) also varies with $\Lambda$.
This form factor therefore acts to a certain degree
like a strength factor of the 3NF, which allows one to adjust the
value of the triton binding energy.

The different versions of the Urbana 3NF are fitted to give the correct
value
for the triton binding energy when used with one of the Argonne NN
potentials.

Most remarkable is the RuhrPot model, because the RuhrPot $2\pi$-exchange
3NF
(together with the RuhrPot NN force) gives essentially the experimental
value
for the triton binding energy without having any adjustable parameters
\cite{ich}.

\begin{figure}
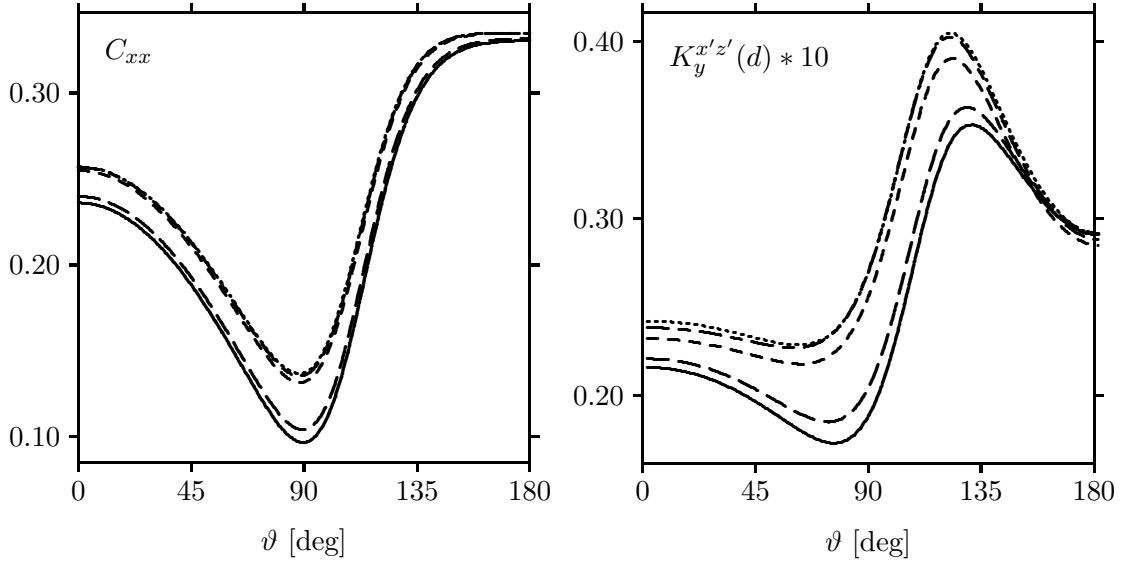

\begin{center}
\hskip0truecm\hbox{\input figabcd.tex }
\end{center}
\caption{\label{figabcd} Effects of the various terms of the TM
$2\pi$-exchange 3NF on the vector spin-correlation coefficient
$C_{xx}$ and the nucleon-to-deuteron tensor spin-transfer
coefficient $K^{x'z'}_{y}$ at $E_{lab}=3$ MeV. Predictions are: AV18 (solid
line), AV18+b (short-dashed line), AV18+d (long-dashed line), AV18+abd
(dotted line) and AV18+abcd (long-short dashed line). The 3NFs are switched
on
only in the $J^\Pi=1/2^\pm$ channels with $j_{max}=2$.}
\end{figure}

For the 3N continuum we have studied the effects of the individual 3NF
terms
($a^{\prime},b,d$), in conjunction with the AV18 np potential \cite{AV18},
on elastic-scattering observables at $E_{lab}=3$ MeV.
The result is that the most important role in the $2\pi$-exchange 3NF is
played
by the $b$-term. The other terms may have significant contributions to some
(small) observables, but they are always smaller than the $b$-term
contribution.
As typical examples we depict the vector spin-correlation coefficient
$C_{xx}$
and the nucleon-to-deuteron tensor spin-transfer coefficient
$K^{x'z'}_{y}$ in Fig.~\ref{figabcd}.

This dominance of the $b$-term explains why all $2\pi$-exchange 3NFs,
including the TM force with the $c$-term, give essentially the same
results for the 3N continuum after being fitted to the triton binding
energy, even for those
observables that do not scale with the triton binding energy. Of course,
this
picture might change if we go to higher energies.

\begin{figure}
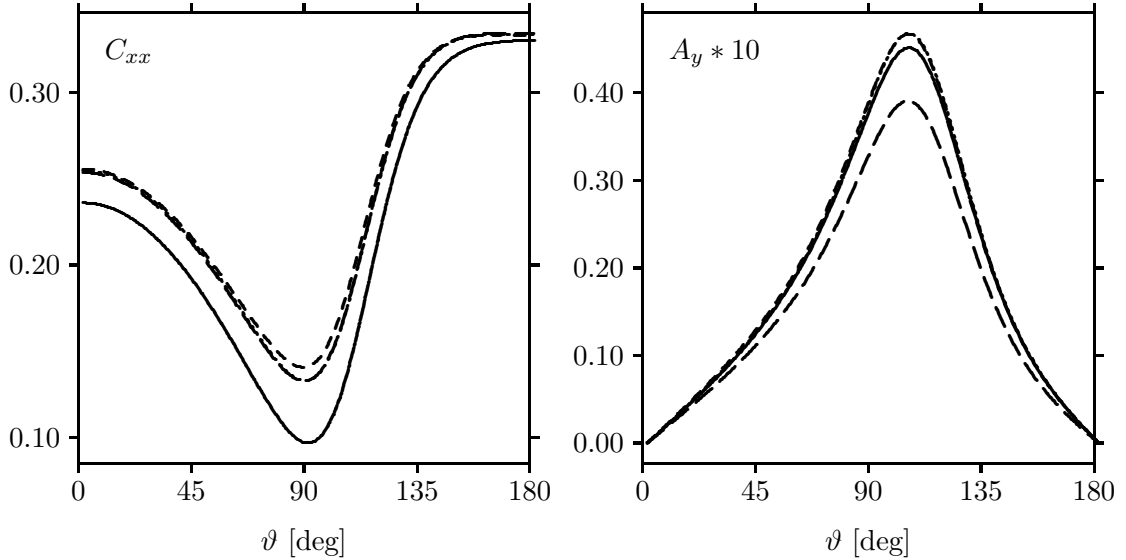

\begin{center}
\hskip0truecm\hbox{\input figTM+.tex }
\end{center}
\caption{\label{figTM+} Contributions of the TM 3NF in various channels for
the scaling observable $C_{xx}$ and the non-scaling observable $A_y$. The
prediction without 3NF is the solid line (AV18 only). The TM 3NF is
switched on
for $J^\Pi=1/2^+$ (short-dashed line), $J^\Pi=1/2^\pm$ (long-dashed line),
and $J^\Pi\le 5/2^\pm$ (dotted line), respectively.}
\end{figure}

Interesting within this context is that many low-energy observables,
especially
in elastic scattering, show the just-mentioned scaling behavior with the
triton
binding energy \cite{report} (i.e., all predictions for these observables
using
Hamiltonians with different NN potentials and a 3NF fitted to the triton
binding energy agree with each other). We had a closer look at this scaling
phenomenon and found (not unexpectedly) that the scaling observables are
those that show a nonnegligible 3NF effect {\bf only} in $J^\Pi=1/2^+$
($J$ being the total 3-body angular momentum and $\Pi$ the parity), whereas
non-scaling observables show 3NF effects also (or only) for other values of
$J^\Pi$. As typical examples we show $C_{xx}$ and $A_y$ in
Fig.~\ref{figTM+}.

\section{Three-Nucleon Forces Including Heavier Mesons}
\label{heavy}

The TM model was extended to incorporate
the exchange of $\rho$-mesons in \cite{rhos},
which leads to a $\pi$-$\rho$ and a $\rho$-$\rho$ 3NF \cite{TMrhos}. These
forces were studied together with the TM $\pi$-$\pi$ 3NF in the 3N elastic
scattering and breakup process \cite{3Nrhos}.

The result of this study was that the $\rho$-$\rho$ 3NF has no visible
effects, whereas the $\pi$-$\rho$ 3NF always has an effect opposite in sign
to the effect of the $\pi$-$\pi$ 3NF, but smaller. Moreover, it appears
that the effect of the $\pi$-$\rho$ 3NF is more or less proportional to
the effect of the $\pi$-$\pi$ 3NF.

That result is somewhat surprising. It means that replacing a $\pi$ with
a $\rho$ in the 3NF produces the same effects, though smaller and in
the opposite direction. In other words, the effective physics in the
$\pi$-$\rho$ 3NF is roughly the same as in the $\pi$-$\pi$ 3NF.
Given that the $\rho$ and the $\pi$ interact very differently, this is a
surprising conclusion.

In order to understand this let's recall that the most important term in a
$\pi$-$\pi$ 3NF is the $b$-term. If we examine the operator structure of
the
$\pi$-$\rho$ 3NF (see, for example, the last paper of Ref.~\cite{TMrhos}),
we
find that this force also contains a term with the same operator structure
as
the $b$-term of the $\pi$-$\pi$-exchange TM 3NF, but with different
parameters
and form factors and with the $\pi$-mass in one of the meson propagators
replaced by the $\rho$-mass. Moreover, we find that the $\pi$-$\rho$
$b$-term
has the opposite sign to the $\pi$-$\pi$ $b$-term. So it seems plausible
that
the TM $\pi$-$\rho$ 3NF could be dominated by its $b$-term, also; at least,
that
would explain the pattern found in the 3N scattering observables. (As a
side
remark, we note that the $\pi$-$\rho$ 3NF also has a $d$-term, which has
the
same sign as the $\pi$-$\pi$ $d$-term.)

Whether our supposition about the $\pi$-$\rho$ $b$-term is true or not, the
finding in \cite{3Nrhos} about the effect of the $\pi$-$\rho$ TM 3NF on 3N
scattering observables shows that the $\pi$-$\rho$ 3NF of \cite{TMrhos}
cannot
be expected to contribute to the explanation of any discrepancies between
experiment and theory like the $A_y$-puzzle. This $\pi$-$\rho$ 3NF just
weakens
the $\pi$-$\pi$ 3NF, but does not lead to any new effects. In order to
explain
current puzzles we need new 3NF terms coming from other physics.

\section{The Texas Force and the Texas -- Los Alamos Three-Nucleon Force}
\label{newterms}

In addition to the conventional $2\pi$-exchange 3NF terms (the
$a^{\prime}$-,
$b$-, and $d$-terms in the TM language), $\chi$PT in first non-vanishing
order
predicts
two terms of $\pi$-range -- zero-range nature (called $d_1$- and
$d_2$-terms
in the language of the Texas force) and three terms of zero-range --
zero-range
nature (called $e_1$-, $e_2$-, and $e_3$-terms) \cite{Texas}. Since the
zero-range -- zero-range terms may be less important than the
$\pi$-range -- zero-range terms (as the $\rho$-$\rho$ terms in the TM force
are
much less important than the $\pi$-$\rho$ terms), we discuss here only the
$\pi$-range -- zero-range terms, and defer until later the purely
short-range
terms. The form of the former terms in momentum space is given by
\begin{eqnarray}
\label{eq9}
V_4^{(1)}&=&{d_1\over (2\pi)^6}\
{g_A\over 2f_\pi^2}\ \vec\sigma_1\cdot\vec{Q'}\
\vec\sigma_2\cdot\vec{Q'}\ {1\over {Q'}^2+m_\pi^2}\
\vec\tau_1\cdot\vec\tau_2\nonumber\\
&+&{d_1\over (2\pi)^6}\
{g_A\over 2f_\pi^2}\ \vec\sigma_1\cdot\vec{Q}\
\vec\sigma_3\cdot\vec{Q}\ {1\over {Q}^2+m_\pi^2}\
\vec\tau_1\cdot\vec\tau_3
\end{eqnarray}
and
\begin{eqnarray}
\label{eq10}
V_4^{(1)}&=&-{d_2\over (2\pi)^6}\ {g_A\over 4f_\pi^2}\
\vec\sigma_1\times\vec\sigma_3\cdot\vec{Q'}\ \vec\sigma_2\cdot\vec{Q'}\
{1\over {Q'}^2+m_\pi^2}\
\vec\tau_1\cdot\vec\tau_2\times\vec\tau_3\nonumber\\
&\phantom=&-{d_2\over (2\pi)^6}\ {g_A\over 4f_\pi^2}\
\vec\sigma_1\times\vec\sigma_2\cdot\vec{Q}\ \vec\sigma_3\cdot\vec{Q}\
{1\over {Q}^2+m_\pi^2}\
\vec\tau_1\cdot\vec\tau_3\times\vec\tau_2
\end{eqnarray}

\begin{figure}
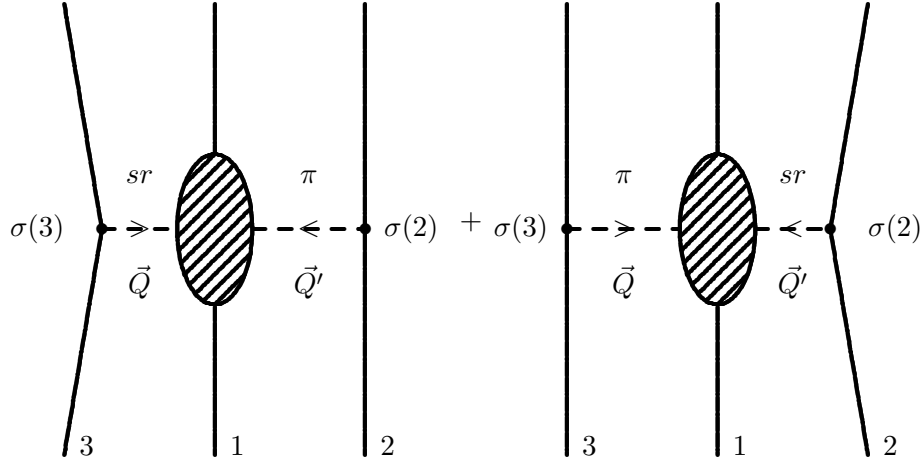

\begin{center}
\hskip0truecm\hbox{\input logoTLA.tex $+$\input logoTLA2.tex }
\end{center}
\caption{\label{figTLA} $\pi$-range -- short-range 3NF terms.}
\end{figure}

In a traditional Hamiltonian, of course, there are no zero-range forces.
These
are an artifact of $\chi$PT. A realistic force would contain short-range
components from the exchange of heavy mesons. Indeed, one can construct a
$d_1$-term from the exchange of $\omega$ or $\sigma$ mesons,
whereas the $d_2$-term gets contributions from $\rho$ and $A_1$ exchanges.
We can envision a zero-range $d_1$-term as caused by the exchange of a very
heavy isoscalar meson, and the corresponding $d_2$ term by a very heavy
isovector meson.
Consequently we extend the zero-range forces to finite range by filling in
meson
propagators and adding form factors. Equation~(\ref{eq9})
is then modified to
\begin{eqnarray}
\label{eq11}
V_4^{(1)}&=&{d_1\over (2\pi)^6}\
{g_A\over 2f_\pi^2}\ \vec\sigma_1\cdot\vec{Q'}\
\vec\sigma_2\cdot\vec{Q'}\ {F({Q'}^2)\over {Q'}^2+m_\pi^2}\
{\cal O}_{SR}(Q^2)\ \vec\tau_1\cdot\vec\tau_2\nonumber\\
&+&{d_1\over (2\pi)^6}\
{g_A\over 2f_\pi^2}\ {\cal O}_{SR}({Q'}^2)\ \vec\sigma_1\cdot\vec{Q}\
\vec\sigma_3\cdot\vec{Q}\ {F({Q}^2)\over {Q}^2+m_\pi^2}\
\vec\tau_1\cdot\vec\tau_3
\end{eqnarray}
while Eq.~(\ref{eq10}) becomes
\begin{eqnarray}
\label{eq12}
V_4^{(1)}&=&-{d_2\over (2\pi)^6}\ {g_A\over 4f_\pi^2}\
\vec\sigma_1\times\vec\sigma_3\cdot\vec{Q'}\ \vec\sigma_2\cdot\vec{Q'}\
{F({Q'}^2)\over {Q'}^2+m_\pi^2}\ {\cal O}_{SR}(Q^2)\
\vec\tau_1\cdot\vec\tau_2\times\vec\tau_3\nonumber\\
&\phantom=&-{d_2\over (2\pi)^6}\ {g_A\over 4f_\pi^2}\ {\cal
O}_{SR}({Q'}^2)\
\vec\sigma_1\times\vec\sigma_2\cdot\vec{Q}\ \vec\sigma_3\cdot\vec{Q}\
{F({Q}^2)\over {Q}^2+m_\pi^2}\
\vec\tau_1\cdot\vec\tau_3\times\vec\tau_2
\end{eqnarray}
where for our purposes ${\cal O}_{SR}(Q^2)$ can be taken to be one of the
following choices:
\begin{eqnarray}
\label{eq13}
{\cal O}_{SR}(Q^2)&=&{m_{sr}^2\over m_{sr}^2+Q^2}\\
\label{eq14}
{\cal O}_{SR}(Q^2)&=&{m_{sr}^2\over m_{sr}^2+Q^2}\
\left( {\Lambda_{sr}^2\over \Lambda_{sr}^2+Q^2}\right)^2\\
\label{eq15}
{\cal O}_{SR}(Q^2)&=&{m_{sr}^2\over m_{sr}^2+Q^2}\
\left( {\Lambda_{sr}^2\over \Lambda_{sr}^2+Q^2}\right)^4
\end{eqnarray}
Equation~(\ref{eq13}) is simply a heavy-meson propagator that is
normalized to 1 at $Q^2 = 0$; in
Eq.~(\ref{eq14}) that propagator is multiplied by the product of two
monopole
form factors; and in Eq.~(\ref{eq15}) it is multiplied by the product of
two
dipole form factors. These terms are depicted in Fig.~\ref{figTLA}.

The form factors in Eqs.~(\ref{eq14}) and (\ref{eq15}) are chosen in such a
way that the resulting 3NF matrix elements at low momenta
will not depend strongly on the
value of the cut-off parameter $\Lambda_{sr}$. In order to stay consistent
with the TM$^{\prime}$ $2\pi$-exchange 3NF, which we will use, we keep the
same form for the $\pi NN$ form factors used in the TM 3NF:
\begin{equation}
\label{eq16}
F({Q}^2)=\left( {\Lambda^2-m_\pi^2\over \Lambda^2+Q^2}\right)^2
\end{equation}
which is normalized to 1 at the pion pole ($Q^2=-m^2_{\pi}$). Due to the
$-m_\pi^2$ factor in the numerator of Eq.~(\ref{eq16}), the size of a 3NF
matrix
element containing this form factor depends even for low momenta
on the value of
the cut-off parameter $\Lambda$, just as the TM 3NF does.

Since the $d_1$- and $d_2$-terms can be associated with the exchange of
many
different heavy mesons, we interpret them as effective forces subsuming
the
effects of all such mesons contributing to the respective term. Thus the
exact value for the mass $m_{sr}$ in the propagator is not important; it
just
has to be roughly the right size. We choose $m_{sr}$ to be the
$\omega$-meson
mass.

Another quantity that is unknown for the $d_1$- and $d_2$-terms is their
strength, since $\chi$PT cannot predict it. The only thing $\chi$PT can say
about the strength of these terms is that they should be ``natural''.

In order to see what naturalness means we rewrite the dimensionful coupling
constants $d_1$ and $d_2$ in terms of dimensionless ones \cite{Texas}
\begin{eqnarray}
\label{eq16ab}
d_1&=&{c_1\over f_\pi^3\ \Lambda}\\
d_2&=&{c_2\over f_\pi^3\ \Lambda}
\end{eqnarray}
where $\Lambda=1$ GeV, $f_{\pi}=92.4$ MeV, and $c_1$ and $c_2$ are
dimensionless. For $c_1$ and $c_2$ to be natural means that their value
should
be on the order of 1. In practical terms the absolute values of $c_1$ or
$c_2$
should typically (and roughly) be numbers comparable to 1,2,... (or the
inverse); their signs, however, are not known.

So in order to use the $d_1$- and $d_2$-terms predicted by a $\chi$PT-based
(i.e., the Texas) force, we have to make adaptations such as those
described
above. In order to differentiate between that zero-range force and the
finite-range terms written down in Eqs.~(\ref{eq11}) and (\ref{eq12}), we
refer to the latter as the Texas - Los Alamos force.

At this point we should mention that the $d_1$- and $d_2$-terms are
predicted
not only by $\chi$PT, but by conventional nuclear field-theory models as
well.
For example, the $d_1$-term is contained in \cite{d1meson} as $\pi$ and
$\sigma$ exchanges or $\pi$ and $\omega$ exchanges with an intermediate
N(1440)
resonance (Eqs.~(3.2a) and (3.2b) in \cite{d1meson}), while the $d_2$-term
is included in \cite{rhos} as a $\pi$-$\rho$ Kroll-Ruderman term
(Eq.~(2.13f) together with the last term, the ``4'', of Eq.~(2.15a) in
\cite{rhos}).

The $d_2$-term is also included in the meson-theoretical RuhrPot
$\pi$-$\rho$
3NF \cite{Rp} as Eq.~(A7), among many other terms. Here the
mechanism for the $d_2$-term is a $NN\pi\rho$ vertex at one of the three
nucleons.

However, these papers include many other 3NF terms, as well, that are of
higher
order in power counting (i.e., they are smaller) than the $d_1$ and $d_2$
terms.
Reference \cite{rhos} includes 126 terms, for example, and
the $d_2$-term is part of one of them. After having written down these
terms
the authors conclude that the $d_2$-part is the dominant one. This exhibits
the
advantage of a power-counting scheme like the one that is part of $\chi$PT:
one knows from the very beginning which terms should be important and which
should not. With conventional nuclear field-theory models there is no easy
way
to know this beforehand.

The $d_1$- and $d_2$-terms are therefore not new in the sense that they
have
never been written down before. But they are new in the sense that for the
first time they have been identified as the leading-order terms of the
pion-range -- short-range 3NF and can be used (see below) in a realistic
calculation of the 3N continuum.

\section{Calculating the Three-Nucleon Continuum}
\label{3nc}

It became possible for the first time in \cite{first} to include a
realistic
3NF in a rigorous calculation of the 3N continuum for energies above the
deuteron breakup threshold. The algorithm used in \cite{first} was later
replaced by the more effective one developed in \cite{algorithm}.

The Faddeev equation with a 3NF included reads
\begin{eqnarray}
\label{eq17}
T&=&tP\phi +(1+tG_0)V_4^{(1)}(1+P)\phi\nonumber\\
&+&tPG_0T+(1+tG_0)V_4^{(1)}(1+P)G_0T
\end{eqnarray}
where $T$ is the Faddeev amplitude for which Eq.~(\ref{eq17}) has to be
solved, $t$ is the 2-body t-matrix, $G_0$ is the free 3-nucleon propagator,
$P$
is the sum of a cyclic and an anti-cyclic permutation (operator) for the
three
nucleons \cite{book}, $V_4^{(1)}$ has already been defined in Section
\ref{2pi3nf}, and $\phi$ is the incoming state composed of a free nucleon
and
a free deuteron. Once Eq.~(\ref{eq17}) is solved one gets the elastic
transition amplitude via \begin{equation}
\label{eq18}
U=PG_0^{-1}+PT+V_4^{(1)}(1+P)\phi +V_4^{(1)}(1+P)G_0T
\end{equation}
and the transition amplitude for the breakup process via
\begin{equation}
\label{eq19}
U_0=(1+P)T
\end{equation}
The 3NF is built into Eq.~(\ref{eq17}) in a perturbative way. Solving
Eq.~(\ref{eq17}) by iteration not only gives the different orders in $T$,
but
the different orders in $V_4^{(1)}$ as well. This will become important
later
on. Of course we always iterate Eq.~(\ref{eq17}) until we reach full
convergence.

In addition to using the new algorithm for the Faddeev equations, we have
also
replaced the PWD for the 3NF used in \cite{first} by one developed later
in \cite{pwd}. The reason is that the original PWD is numerically unstable
for
higher partial waves than those required in the early work. The PWD for the
$d_1$- and $d_2$-terms is performed in \ref{PWD} For details on the
numerics
see Ref.~\cite{report}.

\section{Results}
\label{results}

The first thing one realizes when dealing with the $d_1$- and $d_2$-terms
is
that these terms appear to be big. By big we mean that one cannot get
convergence for the iteration of Eq.~(\ref{eq17}) when $c_1$ and $c_2$ are
significantly larger than 1 and 0.5, respectively. Indeed, the convergence
criterion (Eq.~(5.15) of Ref.~\cite{PhD} - the difference between the last and
before-last result of the iteration of Eq.~(\ref{eq17}), averaged in a 
specific way, has to become less than an $\epsilon$) cannot usually be 
fulfilled for
$\epsilon=10^{-4}$ (our usual value for $\epsilon$), but only for
$\epsilon=10^{-3}$ or larger.

\begin{figure}
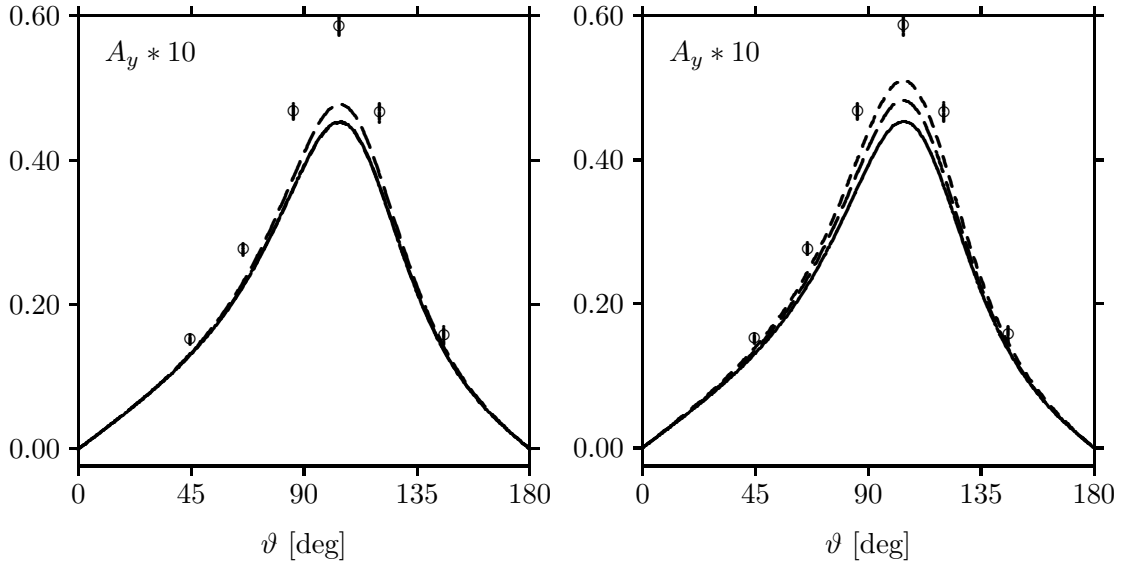

\begin{center}
\hskip0truecm\hbox{\input figay3.tex }
\end{center}
\caption{\label{figay3}
The elastic neutron analyzing power $A_y$ at $E_{lab}=3$ MeV. The solid
line
in both figures represents the prediction for the NN potential (AV18)
alone.
In the left-hand figure the long-dashed line is the prediction for AV18
plus
the $d_1$-term 3NF with $c_1=-1$. The short-dashed line is the prediction
of
AV18 plus the $d_2$-term 3NF with $c_2=0.5$. (Note that the short-dashed
line
overlaps almost completely with the solid line, and is nearly invisible.)
In
the right-hand figure the
long-dashed line is the prediction of AV18 plus the TM$^{\prime}$ 3NF with
the
cut-off parameter $\Lambda =5.215\,m_\pi$. For the short-dashed line on top
of
TM$^{\prime}$
the $d_1$- and $d_2$-terms have been added with parameters $c_1=-1$ and
$c_2=0.5$, respectively. Form factors and other parameters have been chosen
as described in Section \protect{\ref{newterms}}.
The circles are nd data from Ref.~\protect{\cite{ay3data}}.
}
\end{figure}
This slower (or lack of) convergence can easily be related to the size of
these
3NF terms. For example, if one slowly increases the value of $c_1$
from below 1 or $c_2$ from below 0.5, one can examine how the
convergence becomes worse around (or above) 1 or 0.5, respectively.

One finds that the smallest $\epsilon$ for which the convergence
criterion can be fulfilled increases with increasing $c_1$ and $c_2$.
If one increases the values for $c_1$ and $c_2$ further one
finally reaches the point where convergence is totally lost.

The above-described loss of convergence happens first for $J^\Pi=1/2^+$,
whose convergence is always slowest due to the presence of
the bound state. However,
even for larger
values of $c_1$ and $c_2$ where there is no longer convergence
in the $1/2^+$ channels, the 3NF contribution to the Faddeev amplitude $T$
is still significantly smaller than the NN-force contribution.
Thus naively one would expect no difficulty with the convergence.
However,
iterating Eq.~(\ref{eq17}) mixes the perturbation series for the
3NF with the iteration series for the solution of $T$ and therefore we lose
convergence even for relatively small 3NFs.

It might be that this behavior is an artifact of our way of parameterizing
the new 3NF terms (e.g., of our choice for the form factors).
Unfortunately, we didn't have the resources necessary to study this.
Since we intend to study only the cardinal effects of these new 3NF terms,
this
problem is not particularly important. We simply confine ourselves to
smaller
values for $c_1$ and $c_2$ for which we can present fully converged
results.

Even if the convergence problem turns out not to arise from our choice
of parameterization, it is still not a problem of principle. We would just
have to use a different algorithm for the incorporation of a 3NF into the
Faddeev equations, either the one we used in the past \cite{first} or, due
to
the advance of super-computers, just treat the 3NF in a straight-forward
way
(i.e., solve the Lippmann-Schwinger equation driven by $V_4^{(1)}$ and then
use the corresponding t-matrix in the Faddeev equations).

All calculations presented in the following have been performed with a
somewhat reduced accuracy. That means that the maximally allowed angular
momentum for the two-body subsystem has been reduced to $j_{max}=2$
everywhere except for the inner states during the calculation of the 3NF
matrix elements (see Ref.~\cite{pwd} for details), which have been reduced
to
$j_{max}=3$. With these restrictions we are still within 2-3\% of the
results
for a fully converged calculation at 3 and 10 MeV and not much worse at 50
MeV.
This is sufficient for our purpose.

In the following we concentrate on the neutron analyzing power $A_y$. Of
course, if we want to find a solution for the $A_y$-puzzle we also need to
look
at the deuteron analyzing power $iT_{11}$. We don't do this here for
several
reasons. First, we do not intend to present a solution, but rather to see
if a solution is possible. Second, there are no nd data for $iT_{11}$, but
only
pd data. Since we cannot include the Coulomb force in our calculations it
makes more sense to concentrate on $A_y$ here. Third, the effects on
$iT_{11}$
are always very similar to the effects on $A_y$; there are no significant
differences. This suggests that if we are able to describe $A_y$ we
probably
will also describe $iT_{11}$.

\begin{figure}
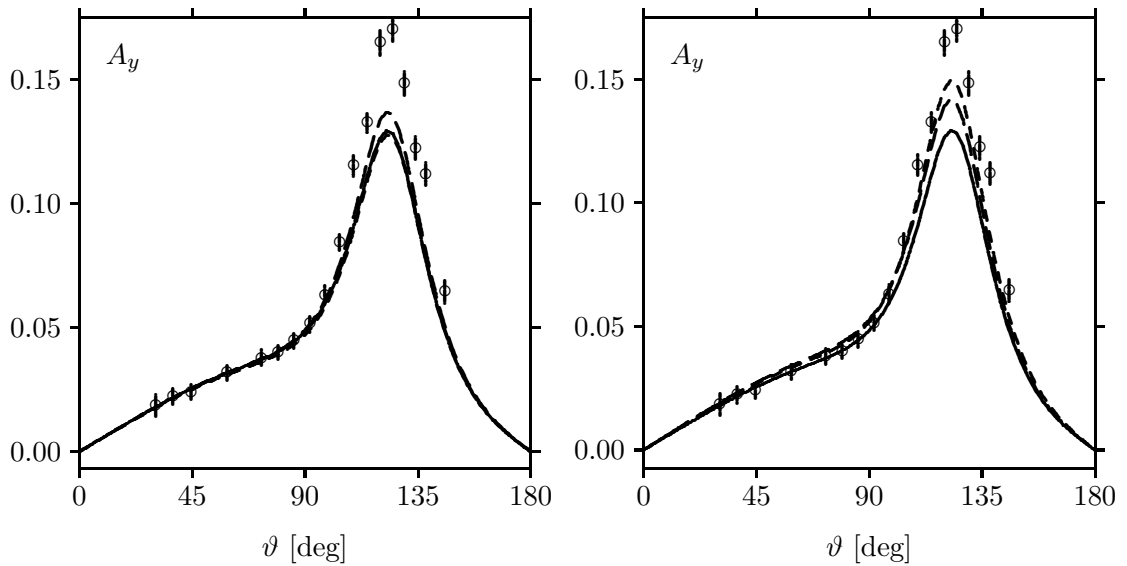

\begin{center}
\hskip0truecm\hbox{\input figay10.tex }
\end{center}
\caption{\label{figay10}
Same as Figure \protect{\ref{figay3}} but for $E_{lab}=10$ MeV.
The circles are nd data from Ref.~\protect{\cite{ay10data}}.
}
\end{figure}
\begin{figure}
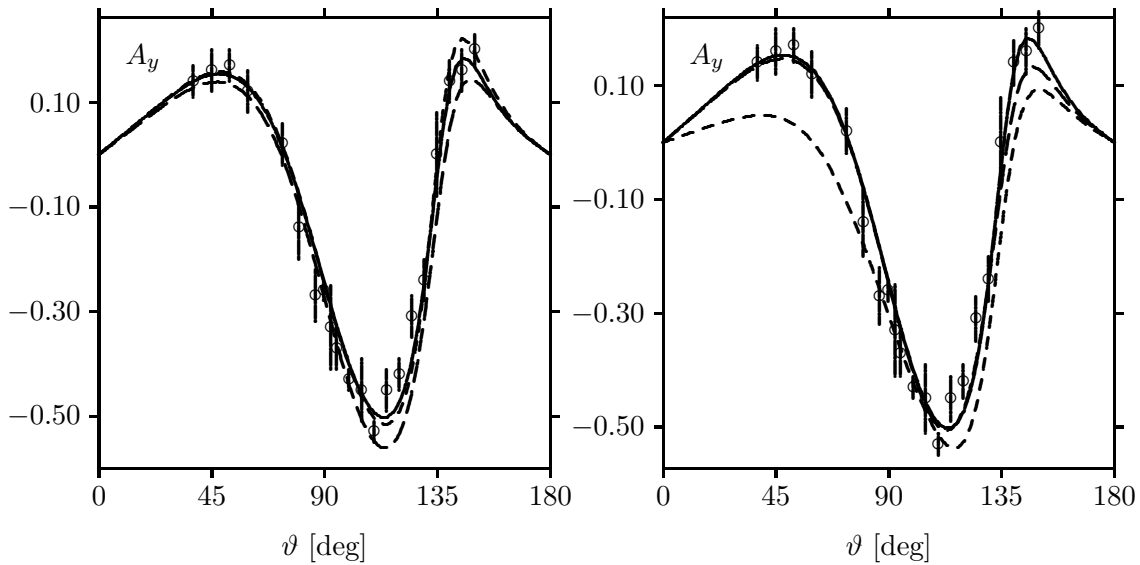

\begin{center}
\hskip0truecm\hbox{\input figay50.tex }
\end{center}
\caption{\label{figay50}
Same as Figure \protect{\ref{figay3}} but for $E_{lab}=50$ MeV.
The circles are nd data from Ref.~\protect{\cite{ay50data}}.
}
\end{figure}

In Figs.~\ref{figay3}-\ref{figay50} we show the nucleon analyzing power
$A_y$
for elastic
neutron-deuteron scattering. The negative sign for $c_1$ was chosen in
order
to get an enhancement in the maximum of $A_y$ instead of a decrease.
Although the enhancement by $d_1$ of the maximum of $A_y$ at 3 MeV
(depicted
in Fig.~\ref{figay3}) is not terribly large and $A_y$ calculated with
the $d_1$-term is still far from the data, this is nevertheless a very
promising result, since one can easily reach the experimental points with a
larger value for $c_1$ (i.e., more negative than $-1$) that is still
natural.
We will return to this point.

As can be seen from Fig.~\ref{figay3} the $d_2$-term has practically no
effect
on $A_y$ at 3 MeV. This is ideal since one can fix the value of $c_1$ at 3
MeV
without worrying about effects from $d_2$.

One should note that the $2\pi$-exchange 3NF (TM$^{\prime}$) already causes
a
visible increase in the maximum of $A_y$. However, plausible choices for
the
parameters in the $2\pi$-exchange 3NF will not be able to explain the
$A_y$-puzzle, since these parameters are much more restricted than they are
for the new 3NF terms. Our choice for the cut-off parameter of the $\pi NN$
form factors in the TM$^{\prime}$ 3NF ($\Lambda =5.215\,m_\pi$) probably
makes
our $2\pi$-exchange 3NF somewhat too small. But even with the recommended
choice
of the Tucson-Melbourne group, one can only expect that the
effect of the $2\pi$-exchange 3NF would be larger by about 20\%. This means
that with a $2\pi$-exchange 3NF one can come nowhere close to the
experimental
values.

At 10 MeV the situation is similar to 3 MeV.  One can see,
however, that the $d_2$-term has a small effect on $A_y$ at this higher
energy.

At 50 MeV the situation has totally changed. Firstly, there is no
$A_y$-puzzle
at this energy (the predictions of all modern NN forces agree reasonably
well with the data). However, it should be emphasized that the experimental
situation is not as clear as at the lower energies, since the error bars
are
significantly larger. This means that there is still some room for small
3NF
effects at this energy.

Secondly, the form of $A_y$ has changed. The maximum that we have seen at
the
two lower energies has become a minimum here, and there are two
less-pronounced
maxima, one at each side of the minimum. This is due to the fact
that at this energy the
mechanisms that build up the analyzing power have changed. We know
\cite{report} that at 50 MeV the P-waves have become less important and
that
D-waves play a role.

If we now examine Fig.~\ref{figay50} we see that both the $d_1$- and the
$d_2$-terms have large effects on $A_y$ at this energy. This does not come
as a surprise. We know that a 3NF always is of shorter range than the
one-pion-exchange part of the NN force. Therefore one naively expects that
3NFs
become more and more important with increasing energy. That this is indeed
true for the $2\pi$-exchange 3NF has been shown in Ref.~\cite{report} and
more recently in Refs.~\cite{total} and \cite{diff} for the total and
differential elastic cross sections, respectively. Since the new 3NF terms
$d_1$ and $d_2$ are of even shorter range than the $2\pi$-exchange 3NF, one
expects that their effects grow even faster with increasing energy. Our
calculations suggest that this is true.

Of course, in this case we will face serious problems in our attempt to
find a solution for the $A_y$-puzzle via 3NFs of shorter range than the
$2\pi$-exchange 3NF. We will discuss this below.

There is one other interesting aspect of Fig.~\ref{figay50}. If we look at
the left maximum in the left-hand figure of Fig.~\ref{figay50} we see that
the effects both of the $d_1$- and $d_2$-terms in that maximum are
moderate.
However, if we add these two terms on top of the $2\pi$-exchange 3NF this
combined 3NF has very large effects on that maximum, as can be seen in the
right-hand figure of Fig.~\ref{figay50}, whereas the effect of the
$2\pi$-exchange 3NF alone is almost zero. This means that in this special
case
a very strong constructive interference develops between the various 3NF
terms.
It tells us that in dealing with 3NFs one has
to be very careful if one excludes certain 3NF terms, since even if their
individual effects are small, their interference with other 3NF terms might
lead to surprisingly large effects.

\begin{table}
\begin{center}
\begin{tabular}{lllll}
&3 MeV&10 MeV&50 MeV&50 MeV\\
&$A_y|_{max}$($105^o$)&$A_y|_{max}$($122.5^o$)
&$A_y|_{max}$($47.5^o$)&$A_y|_{min}$($112.5^o$)\\
\hline
AV18*&0.04549&0.1275&0.1599&-0.5066\\
AV18&0.04536&0.1294&0.1539&-0.5016\\
\hline
+TM*&0.04857(+6.8\%)&0.1328(+4.2\%)&0.1543(-3.6\%)&-0.4953(-2.3\%)\\
+TM$^{\prime}$&0.04825(+6.4\%)&0.1420(+9.7\%)&0.1472(-4.5\%)&-0.5066(+1.0\%
)\\
\hline
$+d_1$&0.04343(-4.4\%)&0.1231(-5.1\%)&0.1589(+3.2\%)&-0.4306(-16.5\%)\\
$-d_1$&0.04778(+5.3\%)&0.1370(+5.9\%)&0.1388(-10.9\%)&-0.5606(+11.8\%)\\
$-2d_1$&0.05094(+12.3\%)\\
\hline
$+0.5d_2$&0.04521(-0.3\%)&0.1277(-1.3\%)&0.1578(+2.5\%)&-0.5157(+2.8\%)\\
$-0.5d_2$&0.04561(+0.6\%)&0.1469(+13.5\%)&0.1503(-2.4\%)&-0.4832(-3.8\%)\\
\hline
TM$^{\prime}-d_1$&0.05106(+12.6\%)&0.1497(+15.5\%)&0.04804&-0.5367(+7.0\%)
\\
$+0.5d_2$\\
\end{tabular}
\end{center}
\caption{\label{tabay}
Theoretical predictions
for $A_y$ at the extrema for various energies. The c.m. angles of the
extrema are given in brackets in the second header line of the table. For
calculations where a 3NF is added to AV18, the deviation of that result
from the calculation with AV18 alone is given in per cent
in brackets as well. All calculations were performed with $j_{max}=2$
except
the ones in the two lines marked with *, which were performed with
$j_{max}=3$.
For simplicity we denoted different values for the dimensionless constants
$c_1$ and $c_2$ as multiplicative factors in front of $d_1$ and $d_2$ in
the
first column of the table (e.g., -2$d_1$ means $c_1 = -2$).
}
\end{table}
Next we want to quantify our findings. For this purpose we list in
Table \ref{tabay} the effects of the various 3NFs alone and together in the
extrema of $A_y$. This table includes more calculations than we have shown
in the figures. We included one calculation with $c_1=-2$ at 3 MeV. For
this
calculation we could only reach a convergence with $\epsilon=10^{-3}$.
Therefore we did not repeat this calculation for the higher energies.

The first thing we see in Table \ref{tabay} is that the effect of the
elimination of the short-range part of the c-term in the original TM 3NF
is negligible at 3 MeV, but already significant at 10 MeV. This means that
older calculations with the original TM 3NF should be repeated using
TM$^{\prime}$ for energies above the deuteron breakup threshold.

Next we see that the effect of the $d_1$-term at 3 and 10 MeV is more or
less
linear with the value of $c_1$, but not so at 50 MeV. On the other hand the
effect of the $d_2$-term is not linear with the value of $c_2$ at 3 MeV and
is dramatically less so at 10 MeV, but becomes more or less linear at 50
MeV.

A look at the last line of Table \ref{tabay} reveals that at all energies
there are some interference effects between the various 3NF terms, though
they
are strongest in the maximum at 50 MeV. Interestingly, we have a
significant
destructive interference in the minimum of $A_y$ at 50 MeV.

Next we want to see if it is possible to find a combination of $d_1$ and
$d_2$
with which it is possible to come close to the experimental data for $A_y$.
Of course, due to the nature of our calculations, any such combination that
we might find can only give a rough estimate for the values of $c_1$ and
$c_2$, but that is all we want. If we can find such a combination we would
have
shown
that a solution of the $A_y$-puzzle using these two 3NF terms of pion-range
--
short-range nature is possible.

So let us start with 3 MeV. Here the situation is relatively simple, since
we
can neglect the effect of the $d_2$-term on $A_y$. Taking into account that
we probably underestimate the effect of the $2\pi$-exchange 3NF somewhat
(as
we stated above), we can extrapolate from Table \ref{tabay} that a value
for
$c_1$ of about $-3$ would probably be able to close the gap ($\sim$ 30\%)
between the data and the predictions of the NN potentials.

If we move on to 10 MeV we see that (not taking into account $d_2$ for the
moment) a combination of TM$^{\prime}$$-3d_1$ (this notation means that we
are
using TM$^{\prime}$ and a $d_1$ force with $c_1=-3$) would probably
overestimate the
data a little bit, since the gap is also about 30\% at this energy, but the
$2\pi$-exchange 3NF has a somewhat larger effect at 10 MeV than at 3 MeV,
whereas the effect of $d_1$ is more or less the same at both energies.
However, at 10 MeV the $d_2$-term has a small but visible effect on $A_y$.
So we can use the $d_2$-term to counterbalance the increased effect of the
$2\pi$-exchange 3NF. This leads us to a value for $c_2$ that should be
positive and small, perhaps 0.5 or 1. Thus a combination like TM$^{\prime}$
$-3d_1+0.5d_2$ or TM$^{\prime}$$-3d_1+d_2$ would be able to bring the
theoretical prediction for $A_y$ close to the experimental data at 3 and 10
MeV.

It might be, however, that the values we just gave for $c_1$ and $c_2$
are even smaller in reality. The reason is that, as mentioned above, we
used
the $\pi NN$ form factor Eq.~(\ref{eq16}) for the $d_1$- and $d_2$-terms
with
the same small value for $\Lambda$ as for the $2\pi$-exchange part of the
3NF. This might cause us to underestimate the strength of $d_1$ and $d_2$
somewhat.

Now let's look at 50 MeV. We see immediately from Table \ref{tabay} that
the
combination of 3NF terms mentioned above does not work at this energy. The
effects would be far too large. If that is so, is there any way out of this
dilemma?

One might argue that 3NF terms of even shorter range than $d_1$ and
$d_2$ might become important at this higher energy. If that is the case
these additional 3NF terms will provide additional parameters with which
one might be able to describe the data at 50 MeV. As we mentioned
in Section \ref{newterms} such terms exist: the so-called $e_1$-, $e_2$-,
and $e_3$-terms, which are of short-range -- short-range nature.
However, we believe that the inclusion of these terms will lead to serious
problems for two (connected) reasons.

We face the conceptual and practical problem that the short-range
$e_1$-, $e_2$-, and $e_3$-terms are indistinguishable from very short-range
(i.e., $\delta$-function) parts of the $2\pi$-exchange 3NF. Disentangling
these
terms is usually made more serious by the lack of consistency between the
NN-force and the 3NF models we use. In addition, accumulating too many
parameters (the values of $e_1$-, $e_2$-, and $e_3$ are unknown) in a
three-nucleon problem is self-defeating.

This leads us immediately to a deeper, basic question: how high in  energy
do models for the nuclear force based on meson exchange make sense?
Meson-exchange models have been extremely successful, even at
much higher energies than we are considering here. Could it be that the
meson-exchange picture - or rather the way we usually implement it -
breaks down (i.e., becomes excessively complicated) at an energy as low as
50
MeV as soon as one starts to look into (admittedly small) details of the
nuclear force? To answer this question one
would have to investigate whether the short-range --
short-range 3NF terms play any role in $A_y$ (and other observables) at 50
MeV.
Since this is such a basic question it might be worthwhile to do so.

However, there is one other possibility. We have not yet mentioned the
so-called 3NF Born terms \cite{cf,cs3nf}. These terms are also predicted by
$\chi$PT to be the same order in power counting as all the other 3NF terms
mentioned here. These terms are model dependent in the sense that they
depend on how the ``off-shellness'' of the NN potential has been defined,
and
they thus depend on the details of the NN potential that one uses
\cite{cf,cs3nf}. Usually these
Born terms are neglected, on the one hand due to the complications they
bring (such as nonlocality), and on the other hand because a subset of
them are known to be small \cite{TM}. One of the 16 Born terms is of
spin-orbit nature \cite{cf},
however. This particular Born term might be important for $A_y$, since that
observable is very sensitive to the spin-orbit force \cite{Ay}. The next
step
in exploring the $A_y$-puzzle should be to take into account (at least)
this
particular Born term.

We have found that the effects of the $d_1$- and $d_2$-terms on other
observables in elastic nd scattering are usually smaller than their effects
on $A_y$ and $iT_{11}$. Only for some tensor-polarization observables are
the
effects comparable to those on $A_y$ and $iT_{11}$.

\section{Summary and Conclusions}
\label{sum}

In Section \ref{2pi3nf} we have reviewed the current state of affairs for
the
$2\pi$-exchange 3NF. Due to the elimination of the pion-range --
short-range
part of the $c$-term of the TM 3NF, the question of the operator form of
the
$2\pi$-exchange 3NF is now settled. We showed that in low-energy elastic
nd scattering the effect of the $2\pi$-exchange 3NF is dominated by the
$b$-term, as in the bound state. In addition we found that those
observables
that scale with the triton binding energy show 3NF effects only for the
channels for which the 3N system is bound (viz., those with $J^\Pi=1/2^+$).

Next we commented in Section
\ref{heavy} on the results found so far using 3NFs that include the
exchange of one or more mesons that are heavier than the pion. Due to the
nature of these effects we suspect that the $\pi$-$\rho$ TM 3NF is also
dominated by its $b$-term and therefore cannot explain the $A_y$-puzzle.

In Section \ref{newterms} we introduced the Texas force, which is based
on $\chi$PT. We explained which of the 3NF terms of the Texas force are
of
interest to us, and we extended the zero-range parts of these terms to
finite
range in order to connect with the traditional models we use for the NN
force
and the $2\pi$-exchange 3NF. Thereafter we gave a
meson-exchange interpretation to the new terms. Finally we commented on the
appearance of those terms in calculations that are not based on $\chi$PT.

A brief review of our approach to solving the Faddeev equations for the 3N
continuum has been given in Section \ref{3nc}. For the PWD of the new 3NF
terms we refer to \ref{PWD}

Finally we presented our results in Section \ref{results}. We studied the
effects of the new 3NF terms on $A_y$ at 3, 10, and 50 MeV. For the two
lower energies we could find a combination of the $d_1$- and $d_2$-terms
that (together with the NN force and the $2\pi$-exchange 3NF) would be able
to describe the $A_y$ data. Although this is only a
qualitative finding, it is of considerable importance because we have
developed for
the first time a microscopic model of the nuclear force that has the
potential
to describe the
low-energy vector-analyzing-power data. To make this model quantitative
will involve considerable effort. We typically obtain the $b$ and $d$
parameters from fits to $\pi-N$ scattering, and in principle the $d_1$ and
$d_2$
parameters can be obtained from pion production in NN scattering, although
the
latter remains to be seen. If that proves impractical we would have to fit
$d_1$
and $d_2$ (and the cut-off parameter $\Lambda$ in the $\pi NN$ form factor)
to the triton binding energy, to $A_y$, and to other nd
observables for which we have nd data. Given the complexity of
three-nucleon
calculations, this three-parameter search would be very time consuming.

However, as discussed in Section \ref{results}, such a 3NF would not be
able
to describe the $A_y$ data (and possibly data for other observables) at 50
MeV and higher. This might be due to the fact that other 3NF terms than
those
included in this paper become important at that energy. One such 3NF term
could be one of the Born terms that is of spin-orbit type, and therefore is
a
candidate to be of importance for $A_y$. Up to now the Born terms have been
largely neglected, but it seems obvious that at least the one just
mentioned
has to be included in 3N continuum calculations as a next step towards a
solution of the $A_y$-puzzle.

Other candidates for 3NF terms that might become important at 50 MeV
are the ones of short-range -- short-range type, which are predicted by
$\chi$PT to be the same
order as the $d_1$- and $d_2$-terms and the $2\pi$-exchange 3NF terms
that have been included in this study. If it turns out that these terms
do become important at 50 MeV we face serious problems with meson-exchange
models.

Unfortunately, another problem would also arise: we would have to fit
many more terms, which on the one hand is very difficult for a 3N problem,
and on the other hand we might loose any predictive power by accumulating
too
many parameters with increasing energy. Since these are important
questions, it
might be worthwhile to check first whether or not the short-range --
short-range 3NF terms predicted by $\chi$PT in lowest non-vanishing
order
have any effect at 50 MeV.

In this paper we believe that we have made an important step forward toward
the
solution of the long-standing $A_y$-puzzle by identifying new 3NF terms
that
have a significant effect on $A_y$ even at low energies. We found these
terms
using a systematic approach for the classification of the 3NF terms (i.e.,
the power-counting scheme of $\chi$PT). In doing so new questions
arose that require further testing of our models of nuclear forces.
However, we have to leave the answers to these questions to future work.

\begin{acknowledge}
The work of D.H. and J.L.F.was performed under the auspices of the U.S.
Department of Energy. The work of D.H. was supported in part by the
Deutsche Forschungsgemeinschaft under Project No. Hu 746/1-3.
The work of A.N. was supported by the
Deutsche Forschungsgemeinschaft under Project No. Gl 87/27-1. The work of
U.v.K.
was supported in part under NSF grant PHY 94-20470.
The numerical calculations have been performed on the Cray T90 and Cray T3E
of the H\"ochstleistungsrechenzentrum in J\"ulich,
Germany.
\end{acknowledge}

\appendix
\section{Partial-Wave Decomposition}
\label{PWD}
\newfont{\Yfont}{cmti10 scaled 2074}
\newcommand{\Y}{\hbox{{\Yfont y}\phantom.}}

For the partial-wave decomposition (PWD) of the new 3NF terms
we will closely follow
Ref.~\cite{pwd}. We will not repeat here the principles and ideas behind
the PWD as developed in Ref.~\cite{pwd} but rather apply them to the new
$d_1$ and $d_2$ terms. We will often refer to equations from
Ref.~\cite{pwd} in the form ($xx$\cite{pwd}), where $xx$ is the
equation number in Ref.~\cite{pwd}. In addition, we refer
to Ref.~\cite{pwd} for the notation used here.

\subsection{The $d_1$-Term}

Let us begin with the $d_1$-term. It is given in Eq.~(\ref{eq11}) as
\begin{eqnarray}
\label{A1}
V_4^{(1)}&=&{d_1\over (2\pi)^6}\
{g_A\over 2f_\pi^2}\ \vec\sigma_1\cdot\vec{Q'}\
\vec\sigma_2\cdot\vec{Q'}\ {F({Q'}^2)\over {Q'}^2+m_\pi^2}\
{\cal O}_{SR}(Q^2)\ \vec\tau_1\cdot\vec\tau_2\nonumber\\
&+&{d_1\over (2\pi)^6}\
{g_A\over 2f_\pi^2}\ {\cal O}_{SR}({Q'}^2)\ \vec\sigma_1\cdot\vec{Q}\
\vec\sigma_3\cdot\vec{Q}\ {F({Q}^2)\over {Q}^2+m_\pi^2}\
\vec\tau_1\cdot\vec\tau_3
\end{eqnarray}
For the notation see Fig.~\ref{figTLA}.
As in Ref.~\cite{pwd} we will deal with the momentum - spin matrix elements
and the isospin matrix elements separately (see Eq.~(35\cite{pwd})).

We start with the momentum - spin matrix elements.
In analogy to Eqs.~($28$\cite{pwd}), ($31$\cite{pwd}) and ($35$\cite{pwd})
we
split the momentum - spin dependent part of the
$d_1$-term into two quasi-two-body operators. This leads to matrix
elements
\begin{eqnarray}
\label{A2}
M_{d_1}^{J,2}&=&_2\left< pq\alpha_J\right|\vec\sigma_1\cdot\vec Q\
\vec\sigma_3\cdot\vec Q\ {F(Q^2)\over Q^2+m_\pi^2}
\left|p_1q_1\alpha_{1J}\right>_2\nonumber\\
&=&{\delta(q-q_1)\over {q}^2}\ \delta_{\lambda\lambda_1}\ \delta_{II_1}\
\sum_{mm_1} C(j m\ I M-m,J M)\
C(j_1 m_1\ I_1 M_1-m_1,J_1 M_1)\nonumber\\
&\times&\underbrace{_2\left< pjm\right|\vec\sigma_1\cdot\vec Q\
\vec\sigma_3\cdot\vec Q\ {F(Q^2)\over Q^2+m_\pi^2}
\left|p_1j_1m_1\right>_2}_{\equiv M_{d_1}^{j,2}}\\
\label{A3}
\tilde M_{d_1}^{J,3}&=&_3\left<p_2q_2\alpha_{2J}\right|{\cal O}({Q'}^2)
\left|p'q'\alpha'_J\right>_3\nonumber\\
&=&{\delta(q_2-q')\over {q'}^2}\ \delta_{\lambda'\lambda_2}\
\delta_{I'I_2}\
\sum_{m_2m'} C(j_2 m_2\ I_2 M_2-m_2,J_2 M_2)\
C(j' m'\ I' M'-m',J' M')\nonumber\\
&\times&\underbrace{_3\left<p_2j_2m_2\right|{\cal O}({Q'}^2)
\left|p'j'm'\right>_3}_{\equiv \tilde M_{d_2}^{j,3}}
\end{eqnarray}
for the second term in Eq.~(\ref{A1}). The momentum transfers $\vec Q$ and
$\vec {Q'}$ are given by
\begin{eqnarray}
\label{A4}
\vec Q&=&\vec p-\vec p_1\\
\label{A5}
\vec{Q'}&=&\vec{p'}-\vec p_2
\end{eqnarray}

For $M_{d_1}^{j,2}$ we have to decompose the operator
\begin{eqnarray}
\label{A6}
&&\vec\sigma_1\cdot\vec Q\ \vec\sigma_3\cdot\vec Q\nonumber\\
&=&4\pi\ |\vec p-\vec p_1|^2\
\left\{\left\{\sigma_1,Y_1(\widehat{\vec p-\vec p_1}\right\}^0,
\left\{\sigma_1,Y_1(\widehat{\vec p-\vec
p_1}\right\}^0\right\}^0\nonumber\\
&=&\sqrt{4\pi}\ |\vec p-\vec p_1|^2\
\sum_i C(10\ 10,i0)\ \sum_{a+b=i} {p^a(-p_1)^b\over |\vec p-\vec p_1|^i}\
\sqrt{4\pi\ \hat i!\over \hat a!\ \hat b!}\
\left\{\left\{\sigma_1,\sigma_3\right\}^i,
\Y^i_{ab}(\hat p,\hat{p_1})\right\}^0\nonumber\\
&=&4\pi\ \left[ -{1\over\sqrt3}\ |\vec p-\vec p_1|^2\
\left\{\left\{\sigma_1,\sigma_3\right\}^0,
\Y^0_{00}(\hat p,\hat{p_1})\right\}^0\right.\nonumber\\
&\phantom=&\phantom{4\pi\ }\left.+4\sqrt5\
\sum_{a+b=2} {p^a(-p_1)^b\over \sqrt{\hat a!\ \hat b!}}\
\left\{\left\{\sigma_1,\sigma_3\right\}^2,
\Y^2_{ab}(\hat p,\hat{p_1})\right\}^0\right]
\end{eqnarray}

For the expansion of the angular dependence in the propagator and form
factor
we use Eqs.~(43-47\cite{pwd}) and (59\cite{pwd}). Putting this together
with
the result (\ref{A6}) we get
\begin{eqnarray}
\label{A7}
M_{d_1}^{j,2}&=& {(4\pi)^2\over 2}\
\left[ \sum_{\bar l} (-)^{\bar l}\ \sqrt{\hat {\bar l}}\
\tilde H_{\bar l}(pp_1)\
_2\left< pjm\right|\left\{\left\{\sigma_1,\sigma_3\right\}^0,
\Y^0_{00}(\hat p,\hat{p_1})\right\}^0\
\Y^{00}_{\bar l\bar l}(\hat p,\hat{p_1})\left|p_1j_1m_1\right>_2
\right.\nonumber\\
&\phantom=&\phantom{{(4\pi)^2\over 2}\ }+\sqrt{2\over 3}\ \sqrt{5!}\
\sum_{a+b=2} {p^a(-p_1)^b\over \sqrt{\hat a!\ \hat b!}}\
\sum_{\bar l} (-)^{\bar l}\ \sqrt{\hat {\bar l}}\ H_{\bar l}(pp_1)\
\nonumber\\
&\phantom=&\phantom{{(4\pi)^2\over 2}\ }\left.\times\
_2\left< pjm\right|\left\{\left\{\sigma_1,\sigma_3\right\}^2,
\Y^2_{ab}(\hat p,\hat{p_1})\right\}^0\
\Y^{00}_{\bar l\bar l}(\hat p,\hat{p_1})\left|p_1j_1m_1\right>_2\right]
\end{eqnarray}

Next we need to recouple the spherical harmonics that appear in the two
matrix
elements of Eq.~(\ref{A7}):
\begin{eqnarray}
\label{A8}
&&\left\{\left\{\sigma_1,\sigma_3\right\}^0,
\Y^0_{00}(\hat p,\hat{p_1})\right\}^0\
\Y^{00}_{\bar l\bar l}(\hat p,\hat{p_1})
={1\over 4\pi}\ \left\{\sigma_1,\sigma_3\right\}^0\
\Y^{00}_{\bar l\bar l}(\hat p,\hat{p_1})\\
\label{A9}
&&\left\{\left\{\sigma_1,\sigma_3\right\}^2,
\Y^2_{ab}(\hat p,\hat{p_1})\right\}^0\
\Y^{00}_{\bar l\bar l}(\hat p,\hat{p_1})
={1\over 4\pi}\ \sum_{i_1i_2} (-)^{\bar l+a+i_2}\
\sqrt{\hat a\hat b\hat{\bar l}}\
\left\{ \matrix{i_1&i_2&2\cr b&a&\bar l}\right\}\nonumber\\
&&\times\
C(a0\ \bar l0,i_10)\ C(b0\ \bar l0,i_20)\
\left\{\left\{\sigma_1,\sigma_3\right\}^2,
\Y^2_{i_1i_2}(\hat p,\hat{p_1})\right\}^0
\end{eqnarray}
and therefore $M_{d_1}^{j,2}$ becomes
\begin{eqnarray}
\label{A10}
M_{d_1}^{j,2}&=&-{4\pi\over 2}\ {1\over \sqrt3}\
\sum_{\bar l} (-)^{\bar l}\ \sqrt{\hat {\bar l}}\
\tilde H_{\bar l}(pp_1)\
_2\left< pjm\right|\left\{\sigma_1,\sigma_3\right\}^0\
\Y^{00}_{\bar l\bar l}(\hat p,\hat{p_1})\left|p_1j_1m_1\right>_2
\nonumber\\
&+&8\pi\ \sqrt5\
\sum_{\bar l} (-)^{\bar l}\ \sqrt{\hat {\bar l}}\ H_{\bar l}(pp_1)\
\sum_{a+b=2} {p^a(-p_1)^b\over \sqrt{\hat a!\ \hat b!}}\
\sum_{i_1i_2} (-)^{\bar l+a+i_2}\
\sqrt{\hat a\hat b\hat{\bar l}}\
\left\{ \matrix{i_1&i_2&2\cr b&a&\bar l}\right\}\nonumber\\
&\times&
C(a0\ \bar l0,i_10)\ C(b0\ \bar l0,i_20)\
_2\left< pjm\right|\left\{\left\{\sigma_1,\sigma_3\right\}^2,
\Y^2_{i_1i_2}(\hat p,\hat{p_1})\right\}^0\left|p_1j_1m_1\right>_2
\end{eqnarray}

Now we have to evaluate the two matrix elements in Eq.~(\ref{A10}):
\begin{eqnarray}
\label{A11}
&&_2\left< pjm\right|\left\{\sigma_1,\sigma_3\right\}^0\
\Y^{00}_{\bar l\bar l}(\hat p,\hat{p_1})\left|p_1j_1m_1\right>_2
\nonumber\\
&&=\delta_{jj_1}\ \delta_{mm_1}\ \delta_{ll_1}\ \delta_{ss_1}\
\delta_{l\bar l}\ 2\sqrt3\ (-)^{l+s}\ {1\over\sqrt{\hat l}}\
\left\{ \matrix{1/2&1/2&s\cr 1/2&1/2&1}\right\}\\
\label{A12}
&&_2\left< pjm\right|\left\{\left\{\sigma_1,\sigma_3\right\}^2,
\Y^2_{i_1i_2}(\hat p,\hat{p_1})\right\}^0\left|p_1j_1m_1\right>_2
\nonumber\\
&&=\delta_{jj_1}\ \delta_{mm_1}\ \delta_{li_1}\ \delta_{l_1i_2}\
6\sqrt5\ (-)^{j+s_1}\ \sqrt{\hat s\hat s_1}\
\left\{ \matrix{l_1&s_1&j\cr s&l&2}\right\}\
\left\{ \matrix{1&1&2\cr 1/2&1/2&s\cr 1/2&1/2&s_1}\right\}
\end{eqnarray}
Inserting these results into Eq.~(\ref{A10}) yields
\begin{eqnarray}
\label{A13}
M_{d_1}^{j,2}&=&\delta_{jj_1}\ \delta_{mm_1}\ \delta_{ll_1}\ \delta_{ss_1}\
4\pi\ (-)^{s+1}\ \tilde H_l(pp_1)\
\left\{ \matrix{1/2&1/2&s\cr 1/2&1/2&1}\right\}\nonumber\\
&+&\delta_{jj_1}\ \delta_{mm_1}\ 240\pi\
(-)^{j_1+l_1+s_1}\ \sqrt{\hat s\hat s_1}\
\left\{ \matrix{l_1&s_1&j\cr s&l&2}\right\}\
\left\{ \matrix{1&1&2\cr 1/2&1/2&s\cr 1/2&1/2&s_1}\right\}\nonumber\\
&\times&\sum_{\bar l} \hat{\bar l}\ H_{\bar l}(pp_1)\
\sum_{a+b=2} {p^a(-p_1)^b\over \sqrt{\hat a!\ \hat b!}}\
\left\{ \matrix{l&l_1&2\cr b&a&\bar l}\right\}\
C(a0\ \bar l0,l0)\ C(b0\ \bar l0,l_10)
\end{eqnarray}
Thus the final result for $M_{d_1}^{J,2}$, after evaluating the sums over
$m$ and $m_1$ in Eq.~(\ref{A2}), becomes
\begin{eqnarray}
\label{A14}
M_{d_1}^{J,2}&=&{\delta (q_1-q)\over q^2}\ \delta_{JJ_1}\ \delta_{MM_1}\
\delta_{jj_1}\ \delta_{\lambda\lambda_1}\ \delta_{II_1}\nonumber\\
&\times&\left[ \delta_{ll_1}\ \delta_{ss_1}\ 4\pi\ (-)^{s+1}\ \tilde
H_l(pp_1)\
\left\{ \matrix{1/2&1/2&s\cr 1/2&1/2&1}\right\}\right.\nonumber\\
&+&240\pi\ (-)^{j_1+l_1+s_1}\ \sqrt{\hat s\hat s_1}\
\left\{ \matrix{l_1&s_1&j\cr s&l&2}\right\}\
\left\{ \matrix{1&1&2\cr 1/2&1/2&s\cr 1/2&1/2&s_1}\right\}\nonumber\\
&\times&\left.\sum_{\bar l} \hat{\bar l}\ H_{\bar l}(pp_1)\
\sum_{a+b=2} {p^a(-p_1)^b\over \sqrt{\hat a!\ \hat b!}}\
\left\{ \matrix{l&l_1&2\cr b&a&\bar l}\right\}\
C(a0\ \bar l0,l0)\ C(b0\ \bar l0,l_10)\right]
\end{eqnarray}

A useful test for the correctness of the result (\ref{A14}) is a comparison
with the result for $M_{d}^{J,2}$ in \cite{pwd}. The difference between
$M_{d_1}^{J,2}$ and $M_{d}^{J,2}$ is that the first matrix element has an
operator
$\vec\sigma_1\cdot\vec Q=4\pi\ \left\{\sigma_1,Y^1(\hat Q)\right\}^{00}$,
whereas the latter matrix element has the operator
$-i\ \sqrt2\ \sqrt{4\pi\over 3}\ \left\{\sigma_1,Y^1(\hat
Q)\right\}^{1\mu}$.
Thus the difference between $M_{d_1}^{J,2}$ and $M_{d}^{J,2}$, besides a
different factor in front, is that in one case we have a rank-0 operator
and in the other case the same operator, but with rank-1 this time. Of
course
the replacement of the rank-1 operator in the result for the $d$-term with
the rank-0 operator in order to get the result for the $d_1$-term is not
as straightforward as it might appear and has to be done with great care.
Nevertheless, one finds that the result (\ref{A14}) for  $M_{d_1}^{J,2}$

is consistent with the result (90\cite{pwd}) for $M_{d}^{J,2}$.

Besides $M_{d_1}^{J,2}$ we will also need the matrix element
\begin{equation}
\label{A15}
M_{d_1}^{J,3}=_3\left<p_2q_2\alpha_{2J}\right|\vec\sigma_1\cdot\vec {Q'}\
\vec\sigma_2\cdot\vec {Q'}\ {F(Q'^2)\over Q'^2+m_\pi^2}
\left|p'q'\alpha'_J\right>_3
\end{equation}
for the first term in Eq.~(\ref{A1}).
$M_{d_1}^{J,3}$ can easily be obtained from $M_{d_1}^{J,2}$ via the
symmetry relation
\begin{equation}
\label{A16}
M_{d_1}^{J,2}(pq\alpha,p_1q_1\alpha_1)
=M_{d_1}^{J,3}(p_1q_1\alpha_1,pq\alpha)
\end{equation}

The second momentum - spin matrix element occurring in the $d_1$-term is
$\tilde M_{d_1}^{J,3}$, Eq.~(\ref{A3}). For the moment we do not need to
specify
which choice of Eqs.~(\ref{eq13}-\ref{eq15}) we want to make for the
short-range operator ${\cal O}({Q'}^2)$. That means we just use
Eq.~(46\cite{pwd}) for the expansion of the angular dependence and do not
yet
specify how to determine $H$. With this we get
\begin{equation}
\label{A17}
\tilde M_{d_1}^{j,3}=2\pi\ \sum_{\bar l} \sqrt{\hat{\bar l}}\
\bar H_{\bar l}(p'p'_2)\
_3\left<p_2q_2\alpha_{2J}\right|\Y^{00}_{\bar l\bar l}(\hat{p'}\hat{p_2})
\left|p'q'\alpha'_J\right>_3
\end{equation}
The matrix element in Eq.~(\ref{A16}) can easily be evaluated as
\begin{equation}
\label{A18}
_3\left<p_2q_2\alpha_{2J}\right|\Y^{00}_{\bar l\bar l}(\hat{p'}\hat{p_2})
\left|p'q'\alpha'_J\right>_3
=\delta_{j'j_2}\ \delta_{l'l_2}\ \delta_{s's_2}\ \delta_{m'm_2}\
{(-)^{l'}\over\sqrt{\hat{l'}}}
\end{equation}
and with this we get
\begin{equation}
\label{A19}
\tilde M_{d_1}^{J,3}={\delta (q_2-q')\over {q'}^2}\ \delta_{J'J_2}\
\delta_{M'M_2}\ \delta_{j'j_2}\ \delta_{l'l_2}\ \delta_{s's_2}\
\delta_{\lambda'\lambda_2}\ \delta_{I'I_2}\ 2\pi\ \bar H_{l'}(p'p_2)
\end{equation}
The symmetry relation to obtain $\tilde M_{d_1}^{J,3}$ is given by
\begin{equation}
\label{A20}
\tilde M_{d_1}^{J,2}(pq\alpha,p_1q_1\alpha_1)=
\tilde M_{d_1}^{J,3}(p_1q_1\alpha_1,pq\alpha)
\end{equation}

Let us now determine $\bar H$ for the different choices
Eqs.~(\ref{eq13}-\ref{eq15}) for ${\cal O}(Q^2)$. Choosing ${\cal O}(Q^2)$
according to Eq.~(\ref{eq13}) as a propagator only
we get instead of Eqs.~(43-44\cite{pwd})
\begin{equation}
\label{A21}
f(x)={m_{sr}^2\over (\vec p-\vec{p_1})^2+m_{sr}^2}
={m_{sr}^2\over 2pp_1}\ {1\over B_{m_{sr}}-x}
\end{equation}
and therefore $\bar H$ becomes for this case
\begin{equation}
\label{A22}
\bar H_l(pp_1)={m_{sr}^2\over pp_1}\ Q_l(B_{m_{sr}})
\end{equation}

The second case, Eq.~(\ref{eq14}), where the propagator is multiplied by
monopole form factors, has already been calculated in
Eqs.~(43-47\cite{pwd}). Thus for this case one has
\begin{equation}
\label{A23}
\bar H_l(pp_1)=m_{sr}^2\left\{
{1\over pp_1}\ [Q_l(B_{m_{sr}})-Q_l(B_{\Lambda_{sr}})]
+{\Lambda_{sr}^2-m_{sr}^2\over 2(pp_1)^2}\ Q'_l(B_{\Lambda_{sr}})
\right\}
\end{equation}

Finally we have the case where the propagator is multiplied by dipole
form factors, Eq.~(\ref{eq15}):
\begin{eqnarray}
\label{A24}
f(x)&=&{m_{sr}^2\over (\vec p-\vec{p_1})^2+m_{sr}^2}\
\left({\Lambda_{sr}^2\over\Lambda_{sr}^2+Q^2}\right)^4\nonumber\\
&=&{m_{sr}^2\ \Lambda_{sr}^8\over(2pp_1)^5}\
\left\{ \left({2pp_1\over \Lambda_{sr}^2-m_{sr}^2}\right)^4\
\left[ {1\over B_{m_{sr}}-x}-{1\over B_{\Lambda_{sr}}-x}\right]
- \left({2pp_1\over \Lambda_{sr}^2-m_{sr}^2}\right)^3\
{1\over (B_{\Lambda_{sr}}-x)^2} \right.\nonumber\\
&-& \left. \left({2pp_1\over \Lambda_{sr}^2-m_{sr}^2}\right)^2\
{1\over (B_{\Lambda_{sr}}-x)^3}
- \left({2pp_1\over \Lambda_{sr}^2-m_{sr}^2}\right)\
{1\over (B_{\Lambda_{sr}}-x)^4} \right\}
\end{eqnarray}
This leads to
\begin{eqnarray}
\label{A25}
\bar H_l(pp_1)&=&{m_{sr}^2\over (\Lambda_{sr}^2-m_{sr}^2)^4}\
\left\{ {1\over pp_1}\ [Q_l(B_{m_{sr}})-Q_l(B_{\Lambda_{sr}})]
+{\Lambda_{sr}^2-m_{sr}^2\over 2(pp_1)^2}\ Q'_l(B_{\Lambda_{sr}})
\right.\nonumber\\
&-&\left.
{(\Lambda_{sr}^2-m_{sr}^2)^2\over (2pp_1)^2}\ Q''_l(B_{\Lambda_{sr}})
+{(\Lambda_{sr}^2-m_{sr}^2)^3\over 3(2pp_1)^4}\ Q'''_l(B_{\Lambda_{sr}})
\right\}
\end{eqnarray}

The only pieces still missing for the PWD of the $d_1$-term are the isospin
matrix elements. They are given by
\begin{eqnarray}
\label{A26}
I_{d_1}&\equiv&_2\left<(t1/2)TM_T\right|\vec\tau_1\cdot\vec\tau_2
\left|(t'1/2)T'M_T'\right>_3\nonumber\\
&=&\delta_{TT'}\ \delta_{M_T M'_T}\ (-6)\ \sqrt{\hat t\hat{t'}}\
\left\{ \matrix{1/2&1/2&t\cr 1/2&T&t'}\right\}\
\left\{ \matrix{1/2&1/2&t'\cr 1/2&1/2&1}\right\}\\
\label{A27}
\tilde I_{d_1}&\equiv&_2\left<(t1/2)TM_T\right|\vec\tau_3\cdot\vec\tau_1
\left|(t'1/2)T'M_T'\right>_3\nonumber\\
&=&\delta_{TT'}\ \delta_{M_T M'_T}\ (-)^{t+t'+1}\ 6\ \sqrt{\hat t\hat{t'}}\
\left\{ \matrix{1/2&1/2&t\cr 1/2&T&t'}\right\}\
\left\{ \matrix{1/2&1/2&t\cr 1/2&1/2&1}\right\}\nonumber\\
&=&\cases{\phantom-I_{d_1}& for $t=t'$\cr -I_{d_1}& for $t\ne t'$\cr}
\end{eqnarray}

Now we can put all parts together in order to obtain the $d_1$ matrix
element.
In order to do this we will use an obvious symbolic notation:
\begin{equation}
\label{A28}
_1\left< V_4^{(1)}|_{d_1}\right>_1=P_{1\leftrightarrow2}\
\{ [ \tilde M_{d_1}^{J,2}\ P_{2\leftrightarrow3}\ M_{d_1}^{J,3}]\  I_{d_1}
+ [ M_{d_1}^{J,2}\ P_{2\leftrightarrow3}\ \tilde M_{d_1}^{J,3}]\
\tilde I_{d_1}
\}\ P_{3\leftrightarrow1}
\end{equation}

\subsection{The $d_2$-Term}

The $d_2$-term is given in Eq.~(\ref{eq12}) as
\begin{eqnarray}
\label{A29}
V_4^{(1)}&=&-{d_2\over (2\pi)^6}\ {g_A\over 4f_\pi^2}\
\vec\sigma_1\times\vec\sigma_3\cdot\vec{Q'}\ \vec\sigma_2\cdot\vec{Q'}\
{F({Q'}^2)\over {Q'}^2+m_\pi^2}\ {\cal O}_{SR}(Q^2)\
\vec\tau_1\cdot\vec\tau_2\times\vec\tau_3\nonumber\\
&\phantom=&-{d_2\over (2\pi)^6}\ {g_A\over 4f_\pi^2}\ {\cal
O}_{SR}({Q'}^2)\
\vec\sigma_1\times\vec\sigma_2\cdot\vec{Q}\ \vec\sigma_3\cdot\vec{Q}\
{F({Q}^2)\over {Q}^2+m_\pi^2}\
\vec\tau_1\cdot\vec\tau_3\times\vec\tau_2
\end{eqnarray}

The isospin dependence is the same as in the $d$-term. The isospin matrix
element for the $d_2$-term is therefore already given by
Eq.~(34\cite{pwd}):
$I_{d_2}=I_d$ and $\tilde I_{d_2}=-I_d$.

The $d_2$-term does not fall into two parts as naturally as the $d_1$-term
does. Nonetheless we can still split it into two quasi-two-body operators
by rewriting (for the second term in Eq.~(\ref{A29}))
\begin{eqnarray}
\label{A30}
\vec\sigma_1\times\vec\sigma_2\cdot\vec Q&=&
i\sqrt6\ \sqrt{4\pi\over 3}\ Q\
\left\{\left\{\sigma_1,\sigma_2\right\}^1,
Y^1(\widehat{\vec p-\vec{p_1}})\right\}^{00}\nonumber\\
&=&-i\sqrt6\ \sqrt{4\pi\over 3}\ Q\
\left\{\sigma_2,\left\{\sigma_1,Y^1{\hat
Q}\right\}^1\right\}^{00}\nonumber\\
&=&i\sqrt2\ \sqrt{4\pi\over 3}\ Q\
\sum_\mu (-)^\mu\ \sigma_2^\mu\ \left\{\sigma_1,Y^1(\hat
Q)\right\}^{1,-\mu}
\end{eqnarray}
in analogy to Eq.~(77\cite{pwd}). Therefore we have to calculate the
following
two matrix elements for the second term of Eq.~(\ref{A29}):
\begin{eqnarray}
\label{A31}
M_{d_2}^{J,2}&=&_2\left<pq\alpha\right| i\sqrt{2}\ \sqrt{4\pi\over3}\ Q\
\left\{\sigma_1,\Y_{ab}^1(\hat p\hat{p_1})\right\}^{1,-\mu}\
{\vec\sigma_3\cdot\vec Q\over Q^2+m_\pi^2}\ F(Q^2)
\left|p_1q_1\alpha_1\right>_2\nonumber\\
&=&{\delta(q-q_1)\over {q}^2}\ \delta_{\lambda\lambda_1}\ \delta_{II_1}\
\sum_{mm_1} C(j m\ I M-m,J M)\
C(j_1 m_1\ I_1 M_1-m_1,J_1 M_1)\nonumber\\
&\times&\underbrace{
_2\left<pjm\right| i\sqrt{2}\ \sqrt{4\pi\over3}\ Q\
\left\{\sigma_1,\Y_{ab}^1(\hat p\hat{p_1})\right\}^{1,-\mu}\
{\vec\sigma_3\cdot\vec Q\over Q^2+m_\pi^2}\ F(Q^2)
\left|p_1j_1m_1\right>_2}_{\equiv M_{d_2}^{j,2}}\\
\label{A32}
\tilde M_{d_2}^{J,3}&=&_3\left<p_2q_2\alpha_2\right|\sigma_2^\mu\
{\cal O}(Q')\left|p'q'\alpha'\right>_3\nonumber\\
&=&{\delta(q_2-q')\over {q'}^2}\ \delta_{\lambda'\lambda_2}\
\delta_{I'I_2}\
\sum_{m'm_2} C(j_2 m_2\ I_2 M_2-m_2,J_2 M_2)\
C(j' m'\ I' M'-m',J' M')\nonumber\\
&\times&\underbrace{
_3\left<p_2j_2m_2\right|\sigma_2^\mu\
{\cal O}(Q')\left|p'j'm'\right>_3}_{\equiv \tilde M_{d_2}^{j,3}}
\end{eqnarray}
The sum over $\mu$ from Eq.~(\ref{A30}) will be performed later on after we
will have put together all the pieces.

The first matrix element, $M_{d_2}^{J,2}$ Eq.~(\ref{A31}),
occurred already in the $d$-term (with opposite sign)
and is given by Eq.~(90\cite{pwd}):
\begin{eqnarray}
\label{A33}
M_{d_2}^{J,2}&=&{\delta(q-q_1)\over q^2}\
\delta_{\lambda\lambda_1}\ \delta_{II_1}\
(-)^{I+J}\ \sqrt{\hat j\hat j_1\hat s\hat s_1\hat J_1}\
C(1-\mu\ J_1M_1,JM)\
\left\{\matrix{1&j_1&j\cr I&J&J_1}\right\} \nonumber\\
&\times&\left[ \delta_{ll_1}\ i4\pi\ \sqrt6\ (-)^{l+s+1}\ \tilde H_l(pp_1)\
\left\{\matrix{l&s&j\cr 1&j_1&s_1}\right\}\
\left\{\matrix{1&1&1\cr 1/2&1/2&s_1\cr 1/2&1/2&s}\right\}
\right.\nonumber\\
&+& i240\pi\ \sqrt6\ (-)^{j_1}\ \sum_{\bar l} \hat{\bar l}\ H_{\bar
l}(pp_1)\
\sum_{a+b=2} {p^a\ p_1^b\over \sqrt{(2a)!\ (2b)!}} \nonumber\\
&\times& \left\{\matrix{a&b&2\cr l_1&l&\bar l}\right\}\
C(a0\ \bar l0,l0)\ C(b0\ \bar l0,l_10) \nonumber\\
&\times& \left. \sum_{i_1} \hat i_1\
\left\{\matrix{2&i_1&1\cr 1&1&1}\right\}\
\left\{\matrix{2&i_1&1\cr l_1&s_1&j_1\cr l&s&j}\right\}\
\left\{\matrix{1&1&i_1\cr 1/2&1/2&s_1\cr 1/2&1/2&s}\right\} \right]
\nonumber\\
&=&-M_d^{J,2}
\end{eqnarray}
(Note that there are two misprints in Eq.~(90\cite{pwd}):
The phase in the second line of Eq.~(90\cite{pwd}) must read $(-)^{l+s+1}$,
and the phase $(-)^i$ in the third line has to be eliminated.).

For the first term in Eq.~(\ref{A29}) we need
\begin{eqnarray}
\label{A34}
\vec\sigma_1\times\vec\sigma_3\cdot\vec {Q'}&=&
i\sqrt6\ \sqrt{4\pi\over 3}\ Q'\
\left\{\left\{\sigma_1,\sigma_3\right\}^1,
Y^1(\widehat{\vec {p'}-\vec{p_2}})\right\}^{00}\nonumber\\
&=&-i\sqrt6\ \sqrt{4\pi\over 3}\ Q'\
\left\{\sigma_3,\left\{\sigma_1,Y^1{\hat
{Q'}}\right\}^1\right\}^{00}\nonumber\\
&=&i\sqrt2\ \sqrt{4\pi\over 3}\ Q'\
\sum_\mu (-)^\mu\ \sigma_3^\mu\ \left\{\sigma_1,Y^1(\hat
{Q'})\right\}^{1,-\mu}
\end{eqnarray}
which leads us to calculate the matrix element
\begin{eqnarray}
\label{A35}
M_{d_2}^{J,3}&=&_3\left<p_2q_2\alpha_2\right| i\sqrt{2}\ \sqrt{4\pi\over3}\
Q'\
\left\{\sigma_1,\Y_{ab}^1(\hat p'\hat{p_2})\right\}^{1,-\mu}\
{\vec\sigma_2\cdot\vec {Q'}\over Q'^2+m_\pi^2}\ F(Q'^2)
\left|p'q'\alpha'\right>_3\nonumber\\
\end{eqnarray}
and $M_{d_2}^{J,3}=-M_d^{J,3}$.
Eq.~(91\cite{pwd}) gives $M_{d_2}^{J,3}$
and Eqs.~(92-93\cite{pwd}) the relation between $M_{d_2}^{J,2}$ and
$M_{d_2}^{J,3}$. Note that
Eq.~(93\cite{pwd}) has a misprint in the phase; the correct phase must be
$(-)^{s+s_1}$.

For the second matrix element, $\tilde M_{d_2}^{J,3}$ in Eq.~(\ref{A32}),
we need
\begin{eqnarray}
\label{A36}
\tilde M_{d_2}^{j,3}&=&{\delta(p_2-p')\over {p'}^2}\ \delta_{l_2l'}\
C(1\mu\ j'm',j_2m_2)\ \sqrt{6}\ 2\pi\ \bar H_{l'}(p_2p')\nonumber\\
&\times&(-)^{j'+l'+s'+s_2}\ \sqrt{\hat{j'}\hat{s'}\hat s_2}\
\left\{\matrix{s'&j'&l'\cr j_2&s_2&1}\right\}\
\left\{\matrix{1/2&1/2&s'\cr 1&s_2&1/2}\right\}
\end{eqnarray}
Again, what we have to insert for $\bar H$ depends on our choice for
${\cal O}$.

With Eq.~(\ref{A36}) we get after performing the sums over $m_2$ and $m'$
in Eq.~(\ref{A32})
\begin{eqnarray}
\label{A37}
\tilde M_{d_2}^{J,3}&=&{\delta(p_2-p')\over {p'}^2}\
{\delta(q_2-q')\over {q'}^2}\
\delta_{l_2l'}\ \delta_{\lambda_2\lambda'}\ \delta_{I_2I'}\
\sqrt{6}\ 2\pi\ \bar H_{l'}(p_2p')\nonumber\\
&\times&(-)^{1+l'+s'+s_2+I'+J_2}\
\sqrt{\hat{j'}\hat j_2\hat{s'}\hat s_2\hat J_2}\
\left\{\matrix{s'&j'&l'\cr j_2&s_2&1}\right\}\
\left\{\matrix{1/2&1/2&s'\cr 1&s_2&1/2}\right\}\nonumber\\
&\times&\left\{\matrix{1&j'&j_2\cr I'&J_2&J'}\right\}\
C(1\mu\ J'M',J_2M_2)
\end{eqnarray}

Similarly we get
\begin{eqnarray}
\label{A38}
\tilde M_{d_2}^{J,2}&=&{\delta(p-p_1)\over {p}^2}\
{\delta(q-q_1)\over {q}^2}\
\delta_{ll_1}\ \delta_{\lambda\lambda_1}\ \delta_{II_1}\
\sqrt{6}\ 2\pi\ \bar H_{l}(pp_1)\nonumber\\
&\times&(-)^{1+l+I+J}\
\sqrt{\hat{j}\hat j_1\hat{s}\hat s_1\hat J_1}\
\left\{\matrix{s&j&l\cr j_1&s_1&1}\right\}\
\left\{\matrix{1/2&1/2&s\cr 1&s_1&1/2}\right\}\nonumber\\
&\times&\left\{\matrix{1&j&j_1\cr I&J_1&J}\right\}\
C(1\mu\ J_1M_1,JM)
\end{eqnarray}

\end{document}

%% file: logoFM.tex
$$
\vbox{\hsize=0.3\hsize 
\beginpicture
\setcoordinatesystem units <1 true cm,1 true cm> point at 0 0
\setplotarea x from -3. to 3., y from -4 to 4
 \linethickness 5 pt
 \setplotsymbol ({\Large .})
\setlinear
\plot -2 -3 -2 3 /
\plot  0 -3  0 3 /
\plot  .5 -1  .5 1 /
\plot  0 -1 .5 -1 /
\plot  0  1 .5  1 /
\plot  2 -3  2 3 /
\setdashes 
\plot -2 -1 0 -1 /
\plot  .5  1  2  1 /
\put {$\Delta$} at .25 0
\put {$\vec {Q'}$} at  1. .3
\put {$\vec Q$} at -1. -1.7
\put {$\pi$} at  1. 1.7
\put {$\pi$} at -1. -.3
\put {$<$} at  1. 1
\put {$>$} at -1. -1
\put {$\sigma (2)$} [l] at  2.25 1
\put {$\sigma (3)$} [r] at -2.25 -1
\put {$\bullet$} at -2 -1
\put {$\bullet$} at  2 1
\put {$3$} [lb] at -1.8 -3  
\put {$1$} [lb] at   .2 -3  
\put {$2$} [lb] at  2.2 -3  
\endpicture
}
$$

%% file: logo3NF.tex
$$
\vbox{\hsize=0.3\hsize 
\beginpicture
\setcoordinatesystem units <1 true cm,1 true cm> point at 0 0
\setplotarea x from -3. to 3., y from -4 to 4
 \linethickness 5 pt
 \setplotsymbol ({\Large .})
\setlinear
\plot -2 -3 -2 3 /
\plot  0 -3  0 -1 /
\plot  0  1  0 3 /
\plot  2 -3  2 3 /
\setdashes 
\plot -2 0 -.5 0 /
\plot  .5  0  2  0 /
\put {$\vec {Q'}$} at  1. -.7
\put {$\vec Q$} at -1. -.7
\put {$\pi$} at  1. .7
\put {$\pi$} at -1. .7
\put {$<$} at  1. 0
\put {$>$} at -1. 0
\put {$\sigma (2)$} [l] at  2.25 0
\put {$\sigma (3)$} [r] at -2.25 0
\put {$\bullet$} at -2 0
\put {$\bullet$} at  2 0
\setsolid
\ellipticalarc axes ratio .5:1 360 degrees from 0 -1 center at 0 0
\plot      0.000    -1.000     0.400    -0.600 /
\plot     -0.152    -0.952     0.472    -0.328 /
\plot     -0.257    -0.857     0.497    -0.103 /
\plot     -0.338    -0.738     0.498     0.098 /
\plot     -0.400    -0.600     0.480     0.280 /
\plot     -0.447    -0.447     0.447     0.447 /
\plot     -0.480    -0.280     0.400     0.600 /
\plot     -0.498    -0.098     0.338     0.738 /
\plot     -0.497     0.103     0.257     0.857 /
\plot     -0.472     0.328     0.152     0.952 /
\plot     -0.400     0.600     0.000     1.000 /
\put {$3$} [lb] at -1.8 -3  
\put {$1$} [lb] at   .2 -3  
\put {$2$} [lb] at  2.2 -3  
\endpicture
}
$$

%% file: figabcd.tex
\beginpicture
\setcoordinatesystem units < 1 true cm, 1 true cm >
\setplotarea x from 0 to 15, y from -7 to 0

\put { 
 \beginpicture
\setcoordinatesystem units < 0.03333333 true cm,    22.8710 true cm>
\setplotarea x from      0.0000 to    180.0000 , y from      0.0848 to      0.3468
\put {$ C_{xx}                 $} [lt] <10pt,-10pt> at      0.0000      0.3468
 \linethickness .8 pt
 \setplotsymbol ({.})
\axis bottom label {$\vartheta$ [deg]} ticks short numbered from      0.0000 to    180.0000 by         45. /
\axis top ticks short unlabeled from      0.0000 to    180.0000 by         45. /
\axis left ticks short numbered from        0.10 to        0.30 by        0.10 /
\axis right ticks short unlabeled from        0.10 to        0.30 by        0.10 /
 \setsolid                                                   
 \plot
     0.0000     0.23620000
     2.5000     0.23600000
     5.0000     0.23559999
     7.5000     0.23480000
    10.0000     0.23370001
    12.5000     0.23240000
    15.0000     0.23070000
    17.5000     0.22870000
    20.0000     0.22640000
    22.5000     0.22390001
    25.0000     0.22100000
    27.5000     0.21789999
    30.0000     0.21450000
    32.5000     0.21080001
    35.0000     0.20680000
    37.5000     0.20260000
    40.0000     0.19820000
    42.5000     0.19340000
    45.0000     0.18850000
    47.5000     0.18330000
    50.0000     0.17780000
    52.5000     0.17219999
    55.0000     0.16640000
    57.5000     0.16040000
    60.0000     0.15430000
    62.5000     0.14810000
    65.0000     0.14180000
    67.5000     0.13550000
    70.0000     0.12920000
    72.5000     0.12310000
    75.0000     0.11730000
    77.5000     0.11180000
    80.0000     0.10680000
    82.5000     0.10260000
    85.0000     0.09937000
    87.5000     0.09730000
    90.0000     0.09668000
    92.5000     0.09778000
    95.0000     0.10080000
    97.5000     0.10610000
   100.0000     0.11360000
   102.5000     0.12340000
   105.0000     0.13540000
   107.5000     0.14929999
   110.0000     0.16460000
   112.5000     0.18089999
   115.0000     0.19760001
   117.5000     0.21410000
   120.0000     0.22990000
   122.5000     0.24470000
   125.0000     0.25819999
   127.5000     0.27020001
   130.0000     0.28080001
   132.5000     0.28990000
   135.0000     0.29769999
   137.5000     0.30419999
   140.0000     0.30960000
   142.5000     0.31410000
   145.0000     0.31770000
   147.5000     0.32060000
   150.0000     0.32300001
   152.5000     0.32480001
   155.0000     0.32630000
   157.5000     0.32740000
   160.0000     0.32830000
   162.5000     0.32890001
   165.0000     0.32940000
   167.5000     0.32980001
   170.0000     0.33010000
   172.5000     0.33019999
   175.0000     0.33039999
   177.5000     0.33039999
   180.0000     0.33050001
 /
 \setdashpattern <8pt, 4pt>                                  
 \plot
     0.0000     0.23989999
     2.5000     0.23980001
     5.0000     0.23930000
     7.5000     0.23860000
    10.0000     0.23750000
    12.5000     0.23610000
    15.0000     0.23450001
    17.5000     0.23260000
    20.0000     0.23029999
    22.5000     0.22780000
    25.0000     0.22499999
    27.5000     0.22190000
    30.0000     0.21850000
    32.5000     0.21490000
    35.0000     0.21100000
    37.5000     0.20690000
    40.0000     0.20250000
    42.5000     0.19780000
    45.0000     0.19290000
    47.5000     0.18780001
    50.0000     0.18250000
    52.5000     0.17700000
    55.0000     0.17129999
    57.5000     0.16550000
    60.0000     0.15950000
    62.5000     0.15340000
    65.0000     0.14720000
    67.5000     0.14110000
    70.0000     0.13500001
    72.5000     0.12909999
    75.0000     0.12340000
    77.5000     0.11820000
    80.0000     0.11340000
    82.5000     0.10940000
    85.0000     0.10640000
    87.5000     0.10450000
    90.0000     0.10410000
    92.5000     0.10530000
    95.0000     0.10860000
    97.5000     0.11390000
   100.0000     0.12150000
   102.5000     0.13140000
   105.0000     0.14340000
   107.5000     0.15719999
   110.0000     0.17230000
   112.5000     0.18840000
   115.0000     0.20469999
   117.5000     0.22090000
   120.0000     0.23630001
   122.5000     0.25060001
   125.0000     0.26370001
   127.5000     0.27530000
   130.0000     0.28540000
   132.5000     0.29409999
   135.0000     0.30149999
   137.5000     0.30759999
   140.0000     0.31270000
   142.5000     0.31680000
   145.0000     0.32020000
   147.5000     0.32290000
   150.0000     0.32499999
   152.5000     0.32670000
   155.0000     0.32789999
   157.5000     0.32890001
   160.0000     0.32969999
   162.5000     0.33019999
   165.0000     0.33070001
   167.5000     0.33100000
   170.0000     0.33120000
   172.5000     0.33129999
   175.0000     0.33140001
   177.5000     0.33149999
   180.0000     0.33149999
 /
 \setdashpattern <4pt, 4pt>                                  
 \plot
     0.0000     0.25490001
     2.5000     0.25479999
     5.0000     0.25440001
     7.5000     0.25369999
    10.0000     0.25270000
    12.5000     0.25139999
    15.0000     0.24980000
    17.5000     0.24800000
    20.0000     0.24590001
    22.5000     0.24349999
    25.0000     0.24079999
    27.5000     0.23790000
    30.0000     0.23469999
    32.5000     0.23130000
    35.0000     0.22770000
    37.5000     0.22380000
    40.0000     0.21960001
    42.5000     0.21529999
    45.0000     0.21070001
    47.5000     0.20600000
    50.0000     0.20100001
    52.5000     0.19589999
    55.0000     0.19059999
    57.5000     0.18520001
    60.0000     0.17970000
    62.5000     0.17420000
    65.0000     0.16859999
    67.5000     0.16300000
    70.0000     0.15750000
    72.5000     0.15220000
    75.0000     0.14720000
    77.5000     0.14260000
    80.0000     0.13850001
    82.5000     0.13519999
    85.0000     0.13280000
    87.5000     0.13160001
    90.0000     0.13190000
    92.5000     0.13370000
    95.0000     0.13740000
    97.5000     0.14320000
   100.0000     0.15099999
   102.5000     0.16090000
   105.0000     0.17270000
   107.5000     0.18600000
   110.0000     0.20050000
   112.5000     0.21550000
   115.0000     0.23060000
   117.5000     0.24529999
   120.0000     0.25909999
   122.5000     0.27169999
   125.0000     0.28299999
   127.5000     0.29280001
   130.0000     0.30129999
   132.5000     0.30840001
   135.0000     0.31430000
   137.5000     0.31909999
   140.0000     0.32290000
   142.5000     0.32600001
   145.0000     0.32839999
   147.5000     0.33019999
   150.0000     0.33160001
   152.5000     0.33260000
   155.0000     0.33329999
   157.5000     0.33379999
   160.0000     0.33419999
   162.5000     0.33440000
   165.0000     0.33450001
   167.5000     0.33460000
   170.0000     0.33460000
   172.5000     0.33460000
   175.0000     0.33460000
   177.5000     0.33460000
   180.0000     0.33460000
 /
 \setdashpattern <1pt, 2pt>                                  
 \plot
     0.0000     0.25720000
     2.5000     0.25709999
     5.0000     0.25670001
     7.5000     0.25590000
    10.0000     0.25500000
    12.5000     0.25369999
    15.0000     0.25209999
    17.5000     0.25029999
    20.0000     0.24820000
    22.5000     0.24590001
    25.0000     0.24330001
    27.5000     0.24040000
    30.0000     0.23729999
    32.5000     0.23390000
    35.0000     0.23029999
    37.5000     0.22640000
    40.0000     0.22239999
    42.5000     0.21810000
    45.0000     0.21359999
    47.5000     0.20890000
    50.0000     0.20410000
    52.5000     0.19900000
    55.0000     0.19390000
    57.5000     0.18860000
    60.0000     0.18320000
    62.5000     0.17770000
    65.0000     0.17219999
    67.5000     0.16680001
    70.0000     0.16150001
    72.5000     0.15629999
    75.0000     0.15140000
    77.5000     0.14700000
    80.0000     0.14309999
    82.5000     0.14000000
    85.0000     0.13779999
    87.5000     0.13670000
    90.0000     0.13710000
    92.5000     0.13910000
    95.0000     0.14290000
    97.5000     0.14870000
   100.0000     0.15660000
   102.5000     0.16650000
   105.0000     0.17820001
   107.5000     0.19140001
   110.0000     0.20570000
   112.5000     0.22050001
   115.0000     0.23530000
   117.5000     0.24959999
   120.0000     0.26310000
   122.5000     0.27530000
   125.0000     0.28619999
   127.5000     0.29570001
   130.0000     0.30379999
   132.5000     0.31060001
   135.0000     0.31619999
   137.5000     0.32080001
   140.0000     0.32440001
   142.5000     0.32720000
   145.0000     0.32940000
   147.5000     0.33109999
   150.0000     0.33230001
   152.5000     0.33320001
   155.0000     0.33390000
   157.5000     0.33430001
   160.0000     0.33460000
   162.5000     0.33469999
   165.0000     0.33480000
   167.5000     0.33480000
   170.0000     0.33480000
   172.5000     0.33480000
   175.0000     0.33480000
   177.5000     0.33469999
   180.0000     0.33469999
 /
 \setdashpattern <8pt, 4pt, 4pt, 4pt>                        
 \plot
     0.0000     0.25690001
     2.5000     0.25680000
     5.0000     0.25630000
     7.5000     0.25560001
    10.0000     0.25459999
    12.5000     0.25340000
    15.0000     0.25180000
    17.5000     0.25000000
    20.0000     0.24789999
    22.5000     0.24550000
    25.0000     0.24290000
    27.5000     0.23999999
    30.0000     0.23690000
    32.5000     0.23350000
    35.0000     0.22990000
    37.5000     0.22600000
    40.0000     0.22190000
    42.5000     0.21760000
    45.0000     0.21310000
    47.5000     0.20840000
    50.0000     0.20350000
    52.5000     0.19840001
    55.0000     0.19320001
    57.5000     0.18790001
    60.0000     0.18240000
    62.5000     0.17690000
    65.0000     0.17140000
    67.5000     0.16590001
    70.0000     0.16050000
    72.5000     0.15530001
    75.0000     0.15040000
    77.5000     0.14590000
    80.0000     0.14190000
    82.5000     0.13869999
    85.0000     0.13640000
    87.5000     0.13530000
    90.0000     0.13560000
    92.5000     0.13760000
    95.0000     0.14139999
    97.5000     0.14720000
   100.0000     0.15500000
   102.5000     0.16490000
   105.0000     0.17659999
   107.5000     0.18990000
   110.0000     0.20420000
   112.5000     0.21910000
   115.0000     0.23400000
   117.5000     0.24840000
   120.0000     0.26190001
   122.5000     0.27430001
   125.0000     0.28529999
   127.5000     0.29490000
   130.0000     0.30320001
   132.5000     0.31000000
   135.0000     0.31569999
   137.5000     0.32040000
   140.0000     0.32409999
   142.5000     0.32699999
   145.0000     0.32920000
   147.5000     0.33090001
   150.0000     0.33219999
   152.5000     0.33320001
   155.0000     0.33379999
   157.5000     0.33430001
   160.0000     0.33460000
   162.5000     0.33469999
   165.0000     0.33480000
   167.5000     0.33489999
   170.0000     0.33489999
   172.5000     0.33489999
   175.0000     0.33480000
   177.5000     0.33480000
   180.0000     0.33480000
 /
 \endpicture
} [lt] at 0 0
\put {
 \beginpicture
\setcoordinatesystem units < 0.03333333 true cm,    23.5720 true cm>
\setplotarea x from      0.0000 to    180.0000 , y from      0.1616 to      0.4162
\put {$ K^{x'z'}_{y}(d)*10       $} [lt] <10pt,-10pt> at      0.0000      0.4162
 \linethickness .8 pt
 \setplotsymbol ({ .})
\axis bottom label {$\vartheta$ [deg]} ticks short numbered from      0.0000 to    180.0000 by         45. /
\axis top ticks short unlabeled from      0.0000 to    180.0000 by         45. /
\axis left ticks short numbered from        0.20 to        0.40 by        0.10 /
\axis right ticks short unlabeled from        0.20 to        0.40 by        0.10 /
 \setsolid                                                   
 \plot
     0.0000     0.21590000
     2.5000     0.21579999
     5.0000     0.21560000
     7.5000     0.21530001
    10.0000     0.21490000
    12.5000     0.21430001
    15.0000     0.21360001
    17.5000     0.21269999
    20.0000     0.21180001
    22.5000     0.21070001
    25.0000     0.20940000
    27.5000     0.20810001
    30.0000     0.20660000
    32.5000     0.20490000
    35.0000     0.20320000
    37.5000     0.20130000
    40.0000     0.19929999
    42.5000     0.19720000
    45.0000     0.19509999
    47.5000     0.19280000
    50.0000     0.19050001
    52.5000     0.18820001
    55.0000     0.18580002
    57.5000     0.18349999
    60.0000     0.18120000
    62.5000     0.17910001
    65.0000     0.17720000
    67.5000     0.17550001
    70.0000     0.17420000
    72.5000     0.17340001
    75.0000     0.17320000
    77.5000     0.17379999
    80.0000     0.17519999
    82.5000     0.17780000
    85.0000     0.18160000
    87.5000     0.18680000
    90.0000     0.19350000
    92.5000     0.20190001
    95.0000     0.21200000
    97.5000     0.22369999
   100.0000     0.23690000
   102.5000     0.25119999
   105.0000     0.26609999
   107.5000     0.28130001
   110.0000     0.29600000
   112.5000     0.30970001
   115.0000     0.32189998
   117.5000     0.33230001
   120.0000     0.34050000
   122.5000     0.34660000
   125.0000     0.35060000
   127.5000     0.35259998
   130.0000     0.35290000
   132.5000     0.35179999
   135.0000     0.34950000
   137.5000     0.34619999
   140.0000     0.34230000
   142.5000     0.33789998
   145.0000     0.33319998
   147.5000     0.32839999
   150.0000     0.32350001
   152.5000     0.31890002
   155.0000     0.31440002
   157.5000     0.31009999
   160.0000     0.30620000
   162.5000     0.30270001
   165.0000     0.29960001
   167.5000     0.29690000
   170.0000     0.29470000
   172.5000     0.29300001
   175.0000     0.29170001
   177.5000     0.29100001
   180.0000     0.29069999
 /
 \setdashpattern <8pt, 4pt>                                  
 \plot
     0.0000     0.22080001
     2.5000     0.22080001
     5.0000     0.22059999
     7.5000     0.22029999
    10.0000     0.21990000
    12.5000     0.21940000
    15.0000     0.21880001
    17.5000     0.21799999
    20.0000     0.21709999
    22.5000     0.21609999
    25.0000     0.21500000
    27.5000     0.21380000
    30.0000     0.21239999
    32.5000     0.21090001
    35.0000     0.20940000
    37.5000     0.20770000
    40.0000     0.20590000
    42.5000     0.20410000
    45.0000     0.20220000
    47.5000     0.20020001
    50.0000     0.19819999
    52.5000     0.19620000
    55.0000     0.19419999
    57.5000     0.19230001
    60.0000     0.19040000
    62.5000     0.18880001
    65.0000     0.18730000
    67.5000     0.18619999
    70.0000     0.18550000
    72.5000     0.18519999
    75.0000     0.18570000
    77.5000     0.18689999
    80.0000     0.18900000
    82.5000     0.19220001
    85.0000     0.19670001
    87.5000     0.20260000
    90.0000     0.20999999
    92.5000     0.21910000
    95.0000     0.22970000
    97.5000     0.24190000
   100.0000     0.25529999
   102.5000     0.26969999
   105.0000     0.28450000
   107.5000     0.29940000
   110.0000     0.31350002
   112.5000     0.32650000
   115.0000     0.33780003
   117.5000     0.34710002
   120.0000     0.35429999
   122.5000     0.35920000
   125.0000     0.36190000
   127.5000     0.36280000
   130.0000     0.36200002
   132.5000     0.35980001
   135.0000     0.35650000
   137.5000     0.35239998
   140.0000     0.34770000
   142.5000     0.34260002
   145.0000     0.33730000
   147.5000     0.33199999
   150.0000     0.32669997
   152.5000     0.32159999
   155.0000     0.31680000
   157.5000     0.31220001
   160.0000     0.30810001
   162.5000     0.30440000
   165.0000     0.30109999
   167.5000     0.29830000
   170.0000     0.29589999
   172.5000     0.29410002
   175.0000     0.29280001
   177.5000     0.29200000
   180.0000     0.29170001
 /
 \setdashpattern <4pt, 4pt>                                  
 \plot
     0.0000     0.23220001
     2.5000     0.23220001
     5.0000     0.23210001
     7.5000     0.23189999
    10.0000     0.23170000
    12.5000     0.23140000
    15.0000     0.23100001
    17.5000     0.23050001
    20.0000     0.23000000
    22.5000     0.22940001
    25.0000     0.22870001
    27.5000     0.22800000
    30.0000     0.22720000
    32.5000     0.22639999
    35.0000     0.22550000
    37.5000     0.22460000
    40.0000     0.22369999
    42.5000     0.22270000
    45.0000     0.22180000
    47.5000     0.22090000
    50.0000     0.22010000
    52.5000     0.21930000
    55.0000     0.21870001
    57.5000     0.21820000
    60.0000     0.21789999
    62.5000     0.21789999
    65.0000     0.21830000
    67.5000     0.21900000
    70.0000     0.22029999
    72.5000     0.22220001
    75.0000     0.22479999
    77.5000     0.22830001
    80.0000     0.23280001
    82.5000     0.23840001
    85.0000     0.24529999
    87.5000     0.25350001
    90.0000     0.26300001
    92.5000     0.27390000
    95.0000     0.28600001
    97.5000     0.29910001
   100.0000     0.31300002
   102.5000     0.32710001
   105.0000     0.34090000
   107.5000     0.35380000
   110.0000     0.36530000
   112.5000     0.37489998
   115.0000     0.38229999
   117.5000     0.38729998
   120.0000     0.38989997
   122.5000     0.39019999
   125.0000     0.38849998
   127.5000     0.38509998
   130.0000     0.38040000
   132.5000     0.37459999
   135.0000     0.36809999
   137.5000     0.36109999
   140.0000     0.35380000
   142.5000     0.34650001
   145.0000     0.33930001
   147.5000     0.33240002
   150.0000     0.32570001
   152.5000     0.31939998
   155.0000     0.31360000
   157.5000     0.30820000
   160.0000     0.30339998
   162.5000     0.29910001
   165.0000     0.29539999
   167.5000     0.29220000
   170.0000     0.28950000
   172.5000     0.28749999
   175.0000     0.28600001
   177.5000     0.28510001
   180.0000     0.28479999
 /
 \setdashpattern <1pt, 2pt>                                  
 \plot
     0.0000     0.24210000
     2.5000     0.24210000
     5.0000     0.24200000
     7.5000     0.24180001
    10.0000     0.24159999
    12.5000     0.24120000
    15.0000     0.24089999
    17.5000     0.24040000
    20.0000     0.23989999
    22.5000     0.23930000
    25.0000     0.23870000
    27.5000     0.23790000
    30.0000     0.23719999
    32.5000     0.23639999
    35.0000     0.23550001
    37.5000     0.23469999
    40.0000     0.23379999
    42.5000     0.23290001
    45.0000     0.23199999
    47.5000     0.23119999
    50.0000     0.23040000
    52.5000     0.22970000
    55.0000     0.22919999
    57.5000     0.22880000
    60.0000     0.22870001
    62.5000     0.22880000
    65.0000     0.22929999
    67.5000     0.23029999
    70.0000     0.23180000
    72.5000     0.23390001
    75.0000     0.23680000
    77.5000     0.24059999
    80.0000     0.24540000
    82.5000     0.25139999
    85.0000     0.25869998
    87.5000     0.26729998
    90.0000     0.27730000
    92.5000     0.28860000
    95.0000     0.30120000
    97.5000     0.31470001
   100.0000     0.32889998
   102.5000     0.34320000
   105.0000     0.35719997
   107.5000     0.37020001
   110.0000     0.38159999
   112.5000     0.39099997
   115.0000     0.39800000
   117.5000     0.40250000
   120.0000     0.40460002
   122.5000     0.40429997
   125.0000     0.40199998
   127.5000     0.39800000
   130.0000     0.39260000
   132.5000     0.38609999
   135.0000     0.37900001
   137.5000     0.37149999
   140.0000     0.36370003
   142.5000     0.35589999
   145.0000     0.34830001
   147.5000     0.34090000
   150.0000     0.33390000
   152.5000     0.32729998
   155.0000     0.32120001
   157.5000     0.31560001
   160.0000     0.31060001
   162.5000     0.30610001
   165.0000     0.30219999
   167.5000     0.29879999
   170.0000     0.29610002
   172.5000     0.29400000
   175.0000     0.29249999
   177.5000     0.29150000
   180.0000     0.29120001
 /
 \setdashpattern <8pt, 4pt, 4pt, 4pt>                        
 \plot
     0.0000     0.23840001
     2.5000     0.23840001
     5.0000     0.23830000
     7.5000     0.23809999
    10.0000     0.23790000
    12.5000     0.23760000
    15.0000     0.23730001
    17.5000     0.23690000
    20.0000     0.23639999
    22.5000     0.23590000
    25.0000     0.23530000
    27.5000     0.23469999
    30.0000     0.23400000
    32.5000     0.23330000
    35.0000     0.23250000
    37.5000     0.23180000
    40.0000     0.23100001
    42.5000     0.23019999
    45.0000     0.22950000
    47.5000     0.22880000
    50.0000     0.22819999
    52.5000     0.22770001
    55.0000     0.22730000
    57.5000     0.22709998
    60.0000     0.22709998
    62.5000     0.22740000
    65.0000     0.22810000
    67.5000     0.22929999
    70.0000     0.23100001
    72.5000     0.23330000
    75.0000     0.23639999
    77.5000     0.24040000
    80.0000     0.24540000
    82.5000     0.25149998
    85.0000     0.25889999
    87.5000     0.26760000
    90.0000     0.27760002
    92.5000     0.28900000
    95.0000     0.30149999
    97.5000     0.31500000
   100.0000     0.32900003
   102.5000     0.34320000
   105.0000     0.35689998
   107.5000     0.36960000
   110.0000     0.38069999
   112.5000     0.38980001
   115.0000     0.39649999
   117.5000     0.40070000
   120.0000     0.40250000
   122.5000     0.40189999
   125.0000     0.39940000
   127.5000     0.39519998
   130.0000     0.38959998
   132.5000     0.38309997
   135.0000     0.37590000
   137.5000     0.36830002
   140.0000     0.36049998
   142.5000     0.35270002
   145.0000     0.34500000
   147.5000     0.33759999
   150.0000     0.33059999
   152.5000     0.32409999
   155.0000     0.31800002
   157.5000     0.31239998
   160.0000     0.30739999
   162.5000     0.30290002
   165.0000     0.29899999
   167.5000     0.29570001
   170.0000     0.29300001
   172.5000     0.29089999
   175.0000     0.28929999
   177.5000     0.28839999
   180.0000     0.28810000
 /
 \endpicture
} [lt] at 7.5 0
\endpicture

%% file: figTM+.tex
\beginpicture
\setcoordinatesystem units < 1 true cm, 1 true cm >
\setplotarea x from 0 to 15, y from -7 to 0

\put { 
 \beginpicture
\setcoordinatesystem units < 0.03333333 true cm,    22.9135 true cm>
\setplotarea x from      0.0000 to    180.0000 , y from      0.0848 to      0.3466
\put {$ C_{xx}                 $} [lt] <10pt,-10pt> at      0.0000      0.3466
 \linethickness .8 pt
 \setplotsymbol ({ .})
\axis bottom label {$\vartheta$ [deg]} ticks short numbered from      0.0000 to    180.0000 by         45. /
\axis top ticks short unlabeled from      0.0000 to    180.0000 by         45. /
\axis left ticks short numbered from        0.10 to        0.30 by        0.10 /
\axis right ticks short unlabeled from        0.10 to        0.30 by        0.10 /
 \setsolid                                                   
 \plot
     0.0000     0.23620000
     2.5000     0.23600000
     5.0000     0.23559999
     7.5000     0.23480000
    10.0000     0.23370001
    12.5000     0.23240000
    15.0000     0.23070000
    17.5000     0.22870000
    20.0000     0.22640000
    22.5000     0.22390001
    25.0000     0.22100000
    27.5000     0.21789999
    30.0000     0.21450000
    32.5000     0.21080001
    35.0000     0.20680000
    37.5000     0.20260000
    40.0000     0.19820000
    42.5000     0.19340000
    45.0000     0.18850000
    47.5000     0.18330000
    50.0000     0.17780000
    52.5000     0.17219999
    55.0000     0.16640000
    57.5000     0.16040000
    60.0000     0.15430000
    62.5000     0.14810000
    65.0000     0.14180000
    67.5000     0.13550000
    70.0000     0.12920000
    72.5000     0.12310000
    75.0000     0.11730000
    77.5000     0.11180000
    80.0000     0.10680000
    82.5000     0.10260000
    85.0000     0.09937000
    87.5000     0.09730000
    90.0000     0.09668000
    92.5000     0.09778000
    95.0000     0.10080000
    97.5000     0.10610000
   100.0000     0.11360000
   102.5000     0.12340000
   105.0000     0.13540000
   107.5000     0.14929999
   110.0000     0.16460000
   112.5000     0.18089999
   115.0000     0.19760001
   117.5000     0.21410000
   120.0000     0.22990000
   122.5000     0.24470000
   125.0000     0.25819999
   127.5000     0.27020001
   130.0000     0.28080001
   132.5000     0.28990000
   135.0000     0.29769999
   137.5000     0.30419999
   140.0000     0.30960000
   142.5000     0.31410000
   145.0000     0.31770000
   147.5000     0.32060000
   150.0000     0.32300001
   152.5000     0.32480001
   155.0000     0.32630000
   157.5000     0.32740000
   160.0000     0.32830000
   162.5000     0.32890001
   165.0000     0.32940000
   167.5000     0.32980001
   170.0000     0.33010000
   172.5000     0.33019999
   175.0000     0.33039999
   177.5000     0.33039999
   180.0000     0.33050001
 /
 \setdashpattern <8pt, 4pt>                                  
 \plot
     0.0000     0.25369999
     2.5000     0.25360000
     5.0000     0.25319999
     7.5000     0.25250000
    10.0000     0.25150001
    12.5000     0.25020000
    15.0000     0.24860001
    17.5000     0.24680001
    20.0000     0.24470000
    22.5000     0.24230000
    25.0000     0.23970000
    27.5000     0.23680000
    30.0000     0.23360001
    32.5000     0.23029999
    35.0000     0.22660001
    37.5000     0.22280000
    40.0000     0.21870001
    42.5000     0.21439999
    45.0000     0.20990001
    47.5000     0.20520000
    50.0000     0.20029999
    52.5000     0.19530000
    55.0000     0.19010000
    57.5000     0.18480000
    60.0000     0.17940000
    62.5000     0.17389999
    65.0000     0.16850001
    67.5000     0.16300000
    70.0000     0.15770000
    72.5000     0.15250000
    75.0000     0.14760000
    77.5000     0.14309999
    80.0000     0.13920000
    82.5000     0.13609999
    85.0000     0.13380000
    87.5000     0.13280000
    90.0000     0.13310000
    92.5000     0.13500001
    95.0000     0.13880000
    97.5000     0.14460000
   100.0000     0.15250000
   102.5000     0.16240001
   105.0000     0.17410000
   107.5000     0.18740000
   110.0000     0.20170000
   112.5000     0.21670000
   115.0000     0.23170000
   117.5000     0.24620000
   120.0000     0.25990000
   122.5000     0.27239999
   125.0000     0.28360000
   127.5000     0.29339999
   130.0000     0.30170000
   132.5000     0.30880001
   135.0000     0.31459999
   137.5000     0.31940001
   140.0000     0.32319999
   142.5000     0.32620001
   145.0000     0.32859999
   147.5000     0.33039999
   150.0000     0.33170000
   152.5000     0.33270001
   155.0000     0.33340001
   157.5000     0.33390000
   160.0000     0.33430001
   162.5000     0.33450001
   165.0000     0.33460000
   167.5000     0.33460000
   170.0000     0.33469999
   172.5000     0.33469999
   175.0000     0.33460000
   177.5000     0.33460000
   180.0000     0.33460000
 /
 \setdashpattern <4pt, 4pt>                                  
 \plot
     0.0000     0.25569999
     2.5000     0.25549999
     5.0000     0.25510001
     7.5000     0.25440001
    10.0000     0.25350001
    12.5000     0.25220001
    15.0000     0.25070000
    17.5000     0.24890000
    20.0000     0.24680001
    22.5000     0.24450000
    25.0000     0.24200000
    27.5000     0.23920000
    30.0000     0.23610000
    32.5000     0.23280001
    35.0000     0.22930001
    37.5000     0.22560000
    40.0000     0.22160000
    42.5000     0.21750000
    45.0000     0.21310000
    47.5000     0.20860000
    50.0000     0.20389999
    52.5000     0.19900000
    55.0000     0.19410001
    57.5000     0.18900000
    60.0000     0.18380000
    62.5000     0.17860000
    65.0000     0.17340000
    67.5000     0.16820000
    70.0000     0.16320001
    72.5000     0.15830000
    75.0000     0.15380000
    77.5000     0.14960000
    80.0000     0.14610000
    82.5000     0.14320000
    85.0000     0.14129999
    87.5000     0.14060000
    90.0000     0.14120001
    92.5000     0.14330000
    95.0000     0.14730000
    97.5000     0.15320000
   100.0000     0.16110000
   102.5000     0.17090000
   105.0000     0.18240000
   107.5000     0.19540000
   110.0000     0.20930000
   112.5000     0.22380000
   115.0000     0.23819999
   117.5000     0.25209999
   120.0000     0.26510000
   122.5000     0.27689999
   125.0000     0.28740001
   127.5000     0.29660001
   130.0000     0.30430001
   132.5000     0.31090000
   135.0000     0.31619999
   137.5000     0.32049999
   140.0000     0.32400000
   142.5000     0.32670000
   145.0000     0.32879999
   147.5000     0.33030000
   150.0000     0.33149999
   152.5000     0.33230001
   155.0000     0.33289999
   157.5000     0.33320001
   160.0000     0.33350000
   162.5000     0.33360001
   165.0000     0.33360001
   167.5000     0.33370000
   170.0000     0.33360001
   172.5000     0.33360001
   175.0000     0.33360001
   177.5000     0.33350000
   180.0000     0.33350000
 /
 \setdashpattern <1pt, 2pt>                                  
 \plot
     0.0000     0.25479999
     2.5000     0.25459999
     5.0000     0.25420001
     7.5000     0.25350001
    10.0000     0.25250000
    12.5000     0.25119999
    15.0000     0.24969999
    17.5000     0.24789999
    20.0000     0.24580000
    22.5000     0.24339999
    25.0000     0.24079999
    27.5000     0.23790000
    30.0000     0.23480000
    32.5000     0.23140000
    35.0000     0.22780000
    37.5000     0.22400001
    40.0000     0.21990000
    42.5000     0.21560000
    45.0000     0.21110000
    47.5000     0.20649999
    50.0000     0.20160000
    52.5000     0.19650000
    55.0000     0.19140001
    57.5000     0.18600000
    60.0000     0.18060000
    62.5000     0.17510000
    65.0000     0.16960000
    67.5000     0.16410001
    70.0000     0.15870000
    72.5000     0.15340000
    75.0000     0.14839999
    77.5000     0.14380001
    80.0000     0.13980000
    82.5000     0.13650000
    85.0000     0.13410001
    87.5000     0.13290000
    90.0000     0.13310000
    92.5000     0.13490000
    95.0000     0.13860001
    97.5000     0.14430000
   100.0000     0.15210000
   102.5000     0.16190000
   105.0000     0.17370000
   107.5000     0.18700001
   110.0000     0.20140000
   112.5000     0.21640000
   115.0000     0.23140000
   117.5000     0.24600001
   120.0000     0.25970000
   122.5000     0.27219999
   125.0000     0.28340000
   127.5000     0.29310000
   130.0000     0.30149999
   132.5000     0.30849999
   135.0000     0.31430000
   137.5000     0.31900001
   140.0000     0.32280001
   142.5000     0.32580000
   145.0000     0.32810000
   147.5000     0.32990000
   150.0000     0.33129999
   152.5000     0.33230001
   155.0000     0.33300000
   157.5000     0.33350000
   160.0000     0.33379999
   162.5000     0.33399999
   165.0000     0.33410001
   167.5000     0.33419999
   170.0000     0.33419999
   172.5000     0.33419999
   175.0000     0.33419999
   177.5000     0.33419999
   180.0000     0.33419999
 /
 \endpicture
} [lt] at 0 0
\put {

 \beginpicture
\setcoordinatesystem units < 0.03333333 true cm,    11.6476 true cm>
\setplotarea x from      0.0000 to    180.0000 , y from     -0.0234 to      0.4917
\put {$ A_y          *10       $} [lt] <10pt,-10pt> at      0.0000      0.4917
 \linethickness .8 pt
 \setplotsymbol ({ .})
\axis bottom label {$\vartheta$ [deg]} ticks short numbered from      0.0000 to    180.0000 by         45. /
\axis top ticks short unlabeled from      0.0000 to    180.0000 by         45. /
\axis left ticks short numbered from        0.00 to        0.40 by        0.10 /
\axis right ticks short unlabeled from        0.00 to        0.40 by        0.10 /
 \setsolid                                                   
 \plot
     0.0000     0.00000000
     2.5000     0.00645600
     5.0000     0.01293000
     7.5000     0.01945000
    10.0000     0.02601000
    12.5000     0.03264000
    15.0000     0.03934000
    17.5000     0.04611000
    20.0000     0.05297000
    22.5000     0.05992000
    25.0000     0.06698000
    27.5000     0.07418000
    30.0000     0.08155000
    32.5000     0.08913000
    35.0000     0.09694000
    37.5000     0.10500000
    40.0000     0.11340000
    42.5000     0.12210000
    45.0000     0.13110000
    47.5000     0.14050001
    50.0000     0.15030000
    52.5000     0.16049999
    55.0000     0.17130001
    57.5000     0.18260001
    60.0000     0.19450000
    62.5000     0.20710000
    65.0000     0.22050001
    67.5000     0.23469999
    70.0000     0.24970001
    72.5000     0.26550001
    75.0000     0.28209999
    77.5000     0.29929999
    80.0000     0.31709999
    82.5000     0.33530000
    85.0000     0.35380000
    87.5000     0.37220001
    90.0000     0.39019999
    92.5000     0.40720001
    95.0000     0.42250001
    97.5000     0.43540001
   100.0000     0.44510001
   102.5000     0.45069999
   105.0000     0.45179999
   107.5000     0.44810000
   110.0000     0.43940002
   112.5000     0.42620000
   115.0000     0.40899998
   117.5000     0.38849998
   120.0000     0.36559999
   122.5000     0.34119999
   125.0000     0.31590000
   127.5000     0.29060000
   130.0000     0.26569998
   132.5000     0.24169999
   135.0000     0.21890000
   137.5000     0.19750001
   140.0000     0.17760000
   142.5000     0.15920000
   145.0000     0.14229999
   147.5000     0.12680000
   150.0000     0.11260000
   152.5000     0.09948000
   155.0000     0.08741000
   157.5000     0.07622000
   160.0000     0.06581000
   162.5000     0.05607000
   165.0000     0.04693000
   167.5000     0.03830000
   170.0000     0.03010000
   172.5000     0.02225000
   175.0000     0.01468000
   177.5000     0.00729300
   180.0000     0.00000000
 /
 \setdashpattern <8pt, 4pt>                                  
 \plot
     0.0000     0.00000000
     2.5000     0.00590600
     5.0000     0.01183000
     7.5000     0.01778000
    10.0000     0.02377000
    12.5000     0.02980000
    15.0000     0.03588000
    17.5000     0.04200000
    20.0000     0.04818000
    22.5000     0.05441000
    25.0000     0.06072000
    27.5000     0.06712000
    30.0000     0.07364000
    32.5000     0.08032000
    35.0000     0.08718000
    37.5000     0.09425000
    40.0000     0.10150000
    42.5000     0.10910000
    45.0000     0.11690000
    47.5000     0.12500000
    50.0000     0.13339999
    52.5000     0.14220001
    55.0000     0.15130000
    57.5000     0.16100001
    60.0000     0.17109999
    62.5000     0.18189999
    65.0000     0.19340000
    67.5000     0.20559999
    70.0000     0.21840000
    72.5000     0.23199999
    75.0000     0.24620000
    77.5000     0.26100001
    80.0000     0.27629998
    82.5000     0.29200000
    85.0000     0.30789998
    87.5000     0.32380000
    90.0000     0.33930001
    92.5000     0.35390002
    95.0000     0.36710000
    97.5000     0.37799999
   100.0000     0.38600001
   102.5000     0.39039999
   105.0000     0.39069998
   107.5000     0.38660002
   110.0000     0.37819999
   112.5000     0.36579999
   115.0000     0.34999999
   117.5000     0.33160001
   120.0000     0.31110001
   122.5000     0.28950000
   125.0000     0.26740000
   127.5000     0.24540000
   130.0000     0.22390001
   132.5000     0.20320000
   135.0000     0.18370001
   137.5000     0.16550002
   140.0000     0.14860001
   142.5000     0.13300000
   145.0000     0.11880000
   147.5000     0.10580000
   150.0000     0.09387001
   152.5000     0.08293000
   155.0000     0.07285000
   157.5000     0.06351000
   160.0000     0.05482000
   162.5000     0.04669000
   165.0000     0.03906000
   167.5000     0.03186000
   170.0000     0.02503000
   172.5000     0.01850000
   175.0000     0.01220000
   177.5000     0.00606000
   180.0000     0.00000000
 /
 \setdashpattern <4pt, 4pt>                                  
 \plot
     0.0000     0.00000000
     2.5000     0.00679100
     5.0000     0.01360000
     7.5000     0.02044000
    10.0000     0.02733000
    12.5000     0.03427000
    15.0000     0.04127000
    17.5000     0.04833000
    20.0000     0.05547000
    22.5000     0.06268001
    25.0000     0.06999000
    27.5000     0.07742000
    30.0000     0.08501000
    32.5000     0.09278999
    35.0000     0.10080001
    37.5000     0.10910000
    40.0000     0.11760000
    42.5000     0.12650000
    45.0000     0.13570000
    47.5000     0.14520000
    50.0000     0.15510000
    52.5000     0.16550002
    55.0000     0.17640001
    57.5000     0.18789999
    60.0000     0.20010000
    62.5000     0.21300000
    65.0000     0.22670001
    67.5000     0.24130000
    70.0000     0.25669998
    72.5000     0.27300000
    75.0000     0.29010001
    77.5000     0.30800000
    80.0000     0.32660002
    82.5000     0.34559998
    85.0000     0.36500001
    87.5000     0.38440001
    90.0000     0.40340000
    92.5000     0.42129999
    95.0000     0.43740001
    97.5000     0.45090002
   100.0000     0.46090001
   102.5000     0.46649998
   105.0000     0.46720001
   107.5000     0.46270001
   110.0000     0.45289999
   112.5000     0.43830001
   115.0000     0.41960001
   117.5000     0.39749998
   120.0000     0.37310001
   122.5000     0.34729999
   125.0000     0.32079998
   127.5000     0.29440001
   130.0000     0.26860002
   132.5000     0.24390000
   135.0000     0.22050001
   137.5000     0.19859999
   140.0000     0.17839999
   142.5000     0.15979999
   145.0000     0.14270000
   147.5000     0.12700000
   150.0000     0.11270000
   152.5000     0.09958000
   155.0000     0.08746000
   157.5000     0.07625000
   160.0000     0.06582000
   162.5000     0.05608000
   165.0000     0.04693000
   167.5000     0.03829000
   170.0000     0.03009000
   172.5000     0.02225000
   175.0000     0.01468000
   177.5000     0.00729100
   180.0000     0.00000000
 /
 \setdashpattern <1pt, 2pt>                                  
 \plot
     0.0000     0.00000000
     2.5000     0.00658600
     5.0000     0.01319000
     7.5000     0.01983000
    10.0000     0.02651000
    12.5000     0.03325000
    15.0000     0.04005000
    17.5000     0.04691000
    20.0000     0.05384000
    22.5000     0.06086000
    25.0000     0.06797000
    27.5000     0.07521001
    30.0000     0.08261000
    32.5000     0.09019001
    35.0000     0.09801000
    37.5000     0.10610000
    40.0000     0.11450000
    42.5000     0.12310000
    45.0000     0.13220000
    47.5000     0.14150000
    50.0000     0.15130000
    52.5000     0.16159999
    55.0000     0.17240001
    57.5000     0.18370001
    60.0000     0.19579999
    62.5000     0.20860000
    65.0000     0.22229999
    67.5000     0.23680000
    70.0000     0.25229999
    72.5000     0.26860002
    75.0000     0.28590000
    77.5000     0.30390000
    80.0000     0.32269999
    82.5000     0.34200001
    85.0000     0.36180001
    87.5000     0.38159999
    90.0000     0.40100002
    92.5000     0.41949999
    95.0000     0.43620002
    97.5000     0.45030001
   100.0000     0.46090001
   102.5000     0.46709999
   105.0000     0.46829998
   107.5000     0.46420002
   110.0000     0.45469999
   112.5000     0.44039997
   115.0000     0.42170000
   117.5000     0.39969999
   120.0000     0.37530002
   122.5000     0.34940001
   125.0000     0.32280001
   127.5000     0.29629999
   130.0000     0.27039999
   132.5000     0.24560001
   135.0000     0.22210000
   137.5000     0.20010000
   140.0000     0.17980000
   142.5000     0.16100001
   145.0000     0.14390001
   147.5000     0.12810001
   150.0000     0.11370000
   152.5000     0.10050000
   155.0000     0.08827000
   157.5000     0.07697000
   160.0000     0.06646000
   162.5000     0.05663000
   165.0000     0.04740000
   167.5000     0.03869000
   170.0000     0.03041000
   172.5000     0.02249000
   175.0000     0.01484000
   177.5000     0.00737300
   180.0000     0.00000000
 /
 \endpicture
} [lt] at 7.5 0
\endpicture

%% file: logoTLA.tex
$$
\vbox{\hsize=0.3\hsize 
\beginpicture
\setcoordinatesystem units <1 true cm,1 true cm> point at 0 0
\setplotarea x from -3. to 3., y from -4 to 4
 \linethickness 5 pt
 \setplotsymbol ({\Large .})
\setlinear
\plot -2 -3 -1.5 0 /
\plot -1.5 0 -2 3 /
\plot  0 -3  0 -1 /
\plot  0  1  0 3 /
\plot  2 -3  2 3 /
\setdashes 
\plot -1.5 0 -.5 0 /
\plot  .5  0  2  0 /
\put {$\vec {Q'}$} at  1.25 -.7
\put {$\vec Q$} at -1. -.7
\put {$\pi$} at  1.25 .7
\put {$sr$} at -1. .7
\put {$<$} at  1.25 0
\put {$>$} at -1. 0
\put {$\sigma (2)$} [l] at  2.25 0
\put {$\sigma (3)$} [r] at -2. 0
\put {$\bullet$} at -1.5 0
\put {$\bullet$} at  2 0
\setsolid
\ellipticalarc axes ratio .5:1 360 degrees from 0 -1 center at 0 0
\plot      0.000    -1.000     0.400    -0.600 /
\plot     -0.152    -0.952     0.472    -0.328 /
\plot     -0.257    -0.857     0.497    -0.103 /
\plot     -0.338    -0.738     0.498     0.098 /
\plot     -0.400    -0.600     0.480     0.280 /
\plot     -0.447    -0.447     0.447     0.447 /
\plot     -0.480    -0.280     0.400     0.600 /
\plot     -0.498    -0.098     0.338     0.738 /
\plot     -0.497     0.103     0.257     0.857 /
\plot     -0.472     0.328     0.152     0.952 /
\plot     -0.400     0.600     0.000     1.000 /
\put {$3$} [lb] at -1.8 -3  
\put {$1$} [lb] at   .2 -3  
\put {$2$} [lb] at  2.2 -3  
\endpicture
}
$$

%% file: logoTLA2.tex
$$
\vbox{\hsize=0.3\hsize 
\beginpicture
\setcoordinatesystem units <1 true cm,1 true cm> point at 0 0
\setplotarea x from -3. to 3., y from -4 to 4
 \linethickness 5 pt
 \setplotsymbol ({\Large .})
\setlinear
\plot -2 -3 -2 3 /
\plot  0 -3  0 -1 /
\plot  0  1  0 3 /
\plot  2 -3  1.5 0 /
\plot 1.5 0 2 3 /
\setdashes 
\plot -2  0 -.5 0 /
\plot  .5  0  1.5 0 /
\put {$\vec {Q'}$} at  1. -.7
\put {$\vec Q$} at -1.25 -.7
\put {$\pi$} at -1.25 .7
\put {$sr$} at  1. .7
\put {$<$} at  1. 0
\put {$>$} at -1.25 0
\put {$\sigma (2)$} [l] at  2. 0
\put {$\sigma (3)$} [r] at -2.25 0
\put {$\bullet$} at -2 0
\put {$\bullet$} at  1.5 0
\setsolid
\ellipticalarc axes ratio .5:1 360 degrees from 0 -1 center at 0 0
\plot      0.000    -1.000     0.400    -0.600 /
\plot     -0.152    -0.952     0.472    -0.328 /
\plot     -0.257    -0.857     0.497    -0.103 /
\plot     -0.338    -0.738     0.498     0.098 /
\plot     -0.400    -0.600     0.480     0.280 /
\plot     -0.447    -0.447     0.447     0.447 /
\plot     -0.480    -0.280     0.400     0.600 /
\plot     -0.498    -0.098     0.338     0.738 /
\plot     -0.497     0.103     0.257     0.857 /
\plot     -0.472     0.328     0.152     0.952 /
\plot     -0.400     0.600     0.000     1.000 /
\put {$3$} [lb] at -1.8 -3  
\put {$1$} [lb] at   .2 -3  
\put {$2$} [lb] at  2.2 -3  
\endpicture
}
$$

%% file: figay3.tex
\beginpicture
\setcoordinatesystem units < 1 true cm, 1 true cm >
\setplotarea x from 0 to 15, y from -7 to 0

\put { 
 \beginpicture
\setcoordinatesystem units <  .03333333 true cm,     9.6015 true cm>
\setplotarea x from       .0000 to    180.0000 , y from      -.0239 to       .6010
\put {$ A_y          *10       $} [lt] <10pt,-10pt> at       .0000       .6010
 \linethickness .8 pt
 \setplotsymbol ({.})
\axis bottom label {$\vartheta$ [deg]} ticks short numbered from       .0000 to    180.0000 by         45. /
\axis top ticks short unlabeled from       .0000 to    180.0000 by         45. /
\axis left ticks short numbered from         .00 to         .60 by         .20 /
\axis right ticks short unlabeled from         .00 to         .60 by         .20 /
 \setsolid                                                   
 \plot
      .0000      .00000000
     2.5000      .00646900
     5.0000      .01296000
     7.5000      .01949000
    10.0000      .02607000
    12.5000      .03271000
    15.0000      .03942000
    17.5000      .04621000
    20.0000      .05308000
    22.5000      .06005000
    25.0000      .06712000
    27.5000      .07434000
    30.0000      .08173000
    32.5000      .08933000
    35.0000      .09717000
    37.5000      .10529999
    40.0000      .11370000
    42.5000      .12240000
    45.0000      .13150001
    47.5000      .14090000
    50.0000      .15070000
    52.5000      .16100001
    55.0000      .17179999
    57.5000      .18309999
    60.0000      .19509999
    62.5000      .20780000
    65.0000      .22129999
    67.5000      .23560001
    70.0000      .25070000
    72.5000      .26660001
    75.0000      .28330001
    77.5000      .30059999
    80.0000      .31860000
    82.5000      .33690003
    85.0000      .35560000
    87.5000      .37409997
    90.0000      .39220002
    92.5000      .40930000
    95.0000      .42470002
    97.5000      .43760002
   100.0000      .44720000
   102.5000      .45269999
   105.0000      .45359999
   107.5000      .44950002
   110.0000      .44050002
   112.5000      .42679998
   115.0000      .40920001
   117.5000      .38839999
   120.0000      .36509997
   122.5000      .34040001
   125.0000      .31500000
   127.5000      .28950000
   130.0000      .26450002
   132.5000      .24040000
   135.0000      .21760000
   137.5000      .19620000
   140.0000      .17629999
   142.5000      .15799999
   145.0000      .14120001
   147.5000      .12580000
   150.0000      .11160000
   152.5000      .09861000
   155.0000      .08662999
   157.5000      .07553000
   160.0000      .06520000
   162.5000      .05555000
   165.0000      .04649000
   167.5000      .03793000
   170.0000      .02981000
   172.5000      .02204000
   175.0000      .01454000
   177.5000      .00722200
   180.0000      .00000000
 /
 \setdashpattern <8pt, 4pt>                                  
 \plot
      .0000      .00000000
     2.5000      .00652100
     5.0000      .01306000
     7.5000      .01965000
    10.0000      .02628000
    12.5000      .03299000
    15.0000      .03976000
    17.5000      .04662000
    20.0000      .05356000
    22.5000      .06061000
    25.0000      .06778000
    27.5000      .07510000
    30.0000      .08261000
    32.5000      .09033000
    35.0000      .09832001
    37.5000      .10660000
    40.0000      .11520000
    42.5000      .12410000
    45.0000      .13339999
    47.5000      .14310001
    50.0000      .15330000
    52.5000      .16389999
    55.0000      .17510000
    57.5000      .18689999
    60.0000      .19940001
    62.5000      .21269999
    65.0000      .22679999
    67.5000      .24190000
    70.0000      .25790000
    72.5000      .27480000
    75.0000      .29249999
    77.5000      .31110001
    80.0000      .33039999
    82.5000      .35020000
    85.0000      .37039998
    87.5000      .39060000
    90.0000      .41030002
    92.5000      .42899999
    95.0000      .44589999
    97.5000      .46010000
   100.0000      .47060001
   102.5000      .47670001
   105.0000      .47780001
   107.5000      .47340000
   110.0000      .46360001
   112.5000      .44880000
   115.0000      .42970002
   117.5000      .40720001
   120.0000      .38229999
   122.5000      .35580003
   125.0000      .32869998
   127.5000      .30170000
   130.0000      .27530000
   132.5000      .24990000
   135.0000      .22600001
   137.5000      .20360000
   140.0000      .18290000
   142.5000      .16380000
   145.0000      .14630000
   147.5000      .13020000
   150.0000      .11560000
   152.5000      .10210000
   155.0000      .08966000
   157.5000      .07816000
   160.0000      .06747000
   162.5000      .05749000
   165.0000      .04811000
   167.5000      .03926000
   170.0000      .03086000
   172.5000      .02282000
   175.0000      .01505000
   177.5000      .00748000
   180.0000      .00000000
 /
 \setdashpattern <4pt, 4pt>                                  
 \plot
      .0000      .00000000
     2.5000      .00652200
     5.0000      .01307000
     7.5000      .01965000
    10.0000      .02628000
    12.5000      .03298000
    15.0000      .03974000
    17.5000      .04658000
    20.0000      .05350000
    22.5000      .06052000
    25.0000      .06765000
    27.5000      .07491000
    30.0000      .08235000
    32.5000      .08999000
    35.0000      .09788000
    37.5000      .10600000
    40.0000      .11450000
    42.5000      .12320000
    45.0000      .13229999
    47.5000      .14180000
    50.0000      .15160000
    52.5000      .16190000
    55.0000      .17270000
    57.5000      .18400000
    60.0000      .19600001
    62.5000      .20870000
    65.0000      .22220001
    67.5000      .23639999
    70.0000      .25139999
    72.5000      .26719999
    75.0000      .28380001
    77.5000      .30100000
    80.0000      .31870002
    82.5000      .33690003
    85.0000      .35530001
    87.5000      .37369999
    90.0000      .39159998
    92.5000      .40840000
    95.0000      .42360002
    97.5000      .43630001
   100.0000      .44569999
   102.5000      .45120001
   105.0000      .45210001
   107.5000      .44810000
   110.0000      .43919998
   112.5000      .42580003
   115.0000      .40840000
   117.5000      .38789999
   120.0000      .36489999
   122.5000      .34040001
   125.0000      .31520003
   127.5000      .28979999
   130.0000      .26490000
   132.5000      .24100000
   135.0000      .21820000
   137.5000      .19680001
   140.0000      .17690000
   142.5000      .15860000
   145.0000      .14170000
   147.5000      .12629999
   150.0000      .11210001
   152.5000      .09904000
   155.0000      .08701000
   157.5000      .07587000
   160.0000      .06550000
   162.5000      .05581000
   165.0000      .04670000
   167.5000      .03811000
   170.0000      .02995000
   172.5000      .02214000
   175.0000      .01461000
   177.5000      .00725600
   180.0000      .00000000
 /
  \setsolid
  \setlinear
 \put {$\circ          $} at     44.5300      .1515000
  \plot
    44.5300      .1582000
    44.5300      .1448000
  /
  \setsolid
  \setlinear
 \put {$\circ          $} at     65.7800      .2762000
  \plot
    65.7800      .2846000
    65.7800      .2678000
  /
  \setsolid
  \setlinear
 \put {$\circ          $} at     85.7400      .4670000
  \plot
    85.7400      .4780000
    85.7400      .4560000
  /
  \setsolid
  \setlinear
 \put {$\circ          $} at    103.9700      .5860000
  \plot
   103.9700      .5990000
   103.9700      .5730000
  /
  \setsolid
  \setlinear
 \put {$\circ          $} at    120.0900      .4660000
  \plot
   120.0900      .4790000
   120.0900      .4530000
  /
  \setsolid
  \setlinear
 \put {$\circ          $} at    145.7300      .1570000
  \plot
   145.7300      .1690000
   145.7300      .1450000
  /
 \endpicture
} [lt] at 0 0
\put {

 \beginpicture
\setcoordinatesystem units <  .03333333 true cm,     9.5770 true cm>
\setplotarea x from       .0000 to    180.0000 , y from      -.0255 to       .6010
\put {$ A_y          *10       $} [lt] <10pt,-10pt> at       .0000       .6010
 \linethickness .8 pt
 \setplotsymbol ({.})
\axis bottom label {$\vartheta$ [deg]} ticks short numbered from       .0000 to    180.0000 by         45. /
\axis top ticks short unlabeled from       .0000 to    180.0000 by         45. /
\axis left ticks short numbered from         .00 to         .60 by         .20 /
\axis right ticks short unlabeled from         .00 to         .60 by         .20 /
 \setsolid                                                   
 \plot
      .0000      .00000000
     2.5000      .00646900
     5.0000      .01296000
     7.5000      .01949000
    10.0000      .02607000
    12.5000      .03271000
    15.0000      .03942000
    17.5000      .04621000
    20.0000      .05308000
    22.5000      .06005000
    25.0000      .06712000
    27.5000      .07434000
    30.0000      .08173000
    32.5000      .08933000
    35.0000      .09717000
    37.5000      .10529999
    40.0000      .11370000
    42.5000      .12240000
    45.0000      .13150001
    47.5000      .14090000
    50.0000      .15070000
    52.5000      .16100001
    55.0000      .17179999
    57.5000      .18309999
    60.0000      .19509999
    62.5000      .20780000
    65.0000      .22129999
    67.5000      .23560001
    70.0000      .25070000
    72.5000      .26660001
    75.0000      .28330001
    77.5000      .30059999
    80.0000      .31860000
    82.5000      .33690003
    85.0000      .35560000
    87.5000      .37409997
    90.0000      .39220002
    92.5000      .40930000
    95.0000      .42470002
    97.5000      .43760002
   100.0000      .44720000
   102.5000      .45269999
   105.0000      .45359999
   107.5000      .44950002
   110.0000      .44050002
   112.5000      .42679998
   115.0000      .40920001
   117.5000      .38839999
   120.0000      .36509997
   122.5000      .34040001
   125.0000      .31500000
   127.5000      .28950000
   130.0000      .26450002
   132.5000      .24040000
   135.0000      .21760000
   137.5000      .19620000
   140.0000      .17629999
   142.5000      .15799999
   145.0000      .14120001
   147.5000      .12580000
   150.0000      .11160000
   152.5000      .09861000
   155.0000      .08662999
   157.5000      .07553000
   160.0000      .06520000
   162.5000      .05555000
   165.0000      .04649000
   167.5000      .03793000
   170.0000      .02981000
   172.5000      .02204000
   175.0000      .01454000
   177.5000      .00722200
   180.0000      .00000000
 /
 \setdashpattern <8pt, 4pt>                                  
 \plot
      .0000      .00000000
     2.5000      .00664400
     5.0000      .01331000
     7.5000      .02001000
    10.0000      .02677000
    12.5000      .03360000
    15.0000      .04050000
    17.5000      .04748000
    20.0000      .05455000
    22.5000      .06172000
    25.0000      .06902000
    27.5000      .07647000
    30.0000      .08410999
    32.5000      .09197000
    35.0000      .10010000
    37.5000      .10850000
    40.0000      .11720000
    42.5000      .12629999
    45.0000      .13580000
    47.5000      .14560001
    50.0000      .15590000
    52.5000      .16670001
    55.0000      .17810000
    57.5000      .19010000
    60.0000      .20280001
    62.5000      .21630001
    65.0000      .23060000
    67.5000      .24590001
    70.0000      .26220000
    72.5000      .27930000
    75.0000      .29729998
    77.5000      .31610000
    80.0000      .33569998
    82.5000      .35580003
    85.0000      .37620002
    87.5000      .39660001
    90.0000      .41659999
    92.5000      .43540001
    95.0000      .45220000
    97.5000      .46629998
   100.0000      .47649997
   102.5000      .48220003
   105.0000      .48250002
   107.5000      .47729999
   110.0000      .46670002
   112.5000      .45090002
   115.0000      .43090001
   117.5000      .40750000
   120.0000      .38180000
   122.5000      .35470003
   125.0000      .32699999
   127.5000      .29960001
   130.0000      .27289999
   132.5000      .24739999
   135.0000      .22340000
   137.5000      .20099999
   140.0000      .18030001
   142.5000      .16140001
   145.0000      .14399999
   147.5000      .12810001
   150.0000      .11360000
   152.5000      .10030000
   155.0000      .08804000
   157.5000      .07672000
   160.0000      .06620000
   162.5000      .05639000
   165.0000      .04718000
   167.5000      .03849000
   170.0000      .03025000
   172.5000      .02236000
   175.0000      .01475000
   177.5000      .00732900
   180.0000      .00000000
 /
 \setdashpattern <4pt, 4pt>                                  
 \plot
      .0000      .00000000
     2.5000      .00680700
     5.0000      .01364000
     7.5000      .02051000
    10.0000      .02743000
    12.5000      .03443000
    15.0000      .04150000
    17.5000      .04866000
    20.0000      .05592000
    22.5000      .06329000
    25.0000      .07080000
    27.5000      .07845999
    30.0000      .08633000
    32.5000      .09443000
    35.0000      .10280000
    37.5000      .11149999
    40.0000      .12050000
    42.5000      .13000000
    45.0000      .13980000
    47.5000      .14999999
    50.0000      .16070001
    52.5000      .17200001
    55.0000      .18390000
    57.5000      .19640000
    60.0000      .20980000
    62.5000      .22399999
    65.0000      .23920000
    67.5000      .25529999
    70.0000      .27249998
    72.5000      .29069999
    75.0000      .30990002
    77.5000      .33010000
    80.0000      .35100001
    82.5000      .37269998
    85.0000      .39469999
    87.5000      .41680002
    90.0000      .43839997
    92.5000      .45880002
    95.0000      .47720000
    97.5000      .49250001
   100.0000      .50379997
   102.5000      .50999999
   105.0000      .51059997
   107.5000      .50510001
   110.0000      .49369997
   112.5000      .47680002
   115.0000      .45540002
   117.5000      .43040001
   120.0000      .40290001
   122.5000      .37400001
   125.0000      .34460002
   127.5000      .31550002
   130.0000      .28720000
   132.5000      .26030001
   135.0000      .23499998
   137.5000      .21140000
   140.0000      .18959999
   142.5000      .16960001
   145.0000      .15140000
   147.5000      .13470000
   150.0000      .11939999
   152.5000      .10540000
   155.0000      .09257000
   157.5000      .08067000
   160.0000      .06962000
   162.5000      .05931000
   165.0000      .04963000
   167.5000      .04050000
   170.0000      .03183000
   172.5000      .02354000
   175.0000      .01553000
   177.5000      .00771600
   180.0000      .00000000
 /
  \setsolid
  \setlinear
 \put {$\circ          $} at     44.5300      .1515000
  \plot
    44.5300      .1582000
    44.5300      .1448000
  /
  \setsolid
  \setlinear
 \put {$\circ          $} at     65.7800      .2762000
  \plot
    65.7800      .2846000
    65.7800      .2678000
  /
  \setsolid
  \setlinear
 \put {$\circ          $} at     85.7400      .4670000
  \plot
    85.7400      .4780000
    85.7400      .4560000
  /
  \setsolid
  \setlinear
 \put {$\circ          $} at    103.9700      .5860000
  \plot
   103.9700      .5990000
   103.9700      .5730000
  /
  \setsolid
  \setlinear
 \put {$\circ          $} at    120.0900      .4660000
  \plot
   120.0900      .4790000
   120.0900      .4530000
  /
  \setsolid
  \setlinear
 \put {$\circ          $} at    145.7300      .1570000
  \plot
   145.7300      .1690000
   145.7300      .1450000
  /
 \endpicture
} [lt] at 7.5 0
\endpicture

%% file: figay10.tex
\beginpicture
\setcoordinatesystem units < 1 true cm, 1 true cm >
\setplotarea x from 0 to 15, y from -7 to 0

\put { 
 \beginpicture
\setcoordinatesystem units <  .03333333 true cm,    33.0033 true cm>
\setplotarea x from       .0000 to    180.0000 , y from      -.0068 to       .1750
\put {$ A_y                    $} [lt] <10pt,-10pt> at       .0000       .1750
 \linethickness .8 pt
 \setplotsymbol ({.})
\axis bottom label {$\vartheta$ [deg]} ticks short numbered from       .0000 to    180.0000 by         45. /
\axis top ticks short unlabeled from       .0000 to    180.0000 by         45. /
\axis left ticks short numbered from         .00 to         .17 by         .05 /
\axis right ticks short unlabeled from         .00 to         .17 by         .05 /
 \setsolid                                                   
 \plot
      .0000      .00000000
     2.5000      .00151300
     5.0000      .00302600
     7.5000      .00453600
    10.0000      .00604300
    12.5000      .00754400
    15.0000      .00903800
    17.5000      .01052000
    20.0000      .01199000
    22.5000      .01345000
    25.0000      .01490000
    27.5000      .01632000
    30.0000      .01773000
    32.5000      .01911000
    35.0000      .02047000
    37.5000      .02181000
    40.0000      .02310000
    42.5000      .02437000
    45.0000      .02559000
    47.5000      .02677000
    50.0000      .02791000
    52.5000      .02901000
    55.0000      .03008000
    57.5000      .03111000
    60.0000      .03212000
    62.5000      .03311000
    65.0000      .03409000
    67.5000      .03507000
    70.0000      .03607000
    72.5000      .03710000
    75.0000      .03818000
    77.5000      .03937000
    80.0000      .04068000
    82.5000      .04218000
    85.0000      .04391000
    87.5000      .04595000
    90.0000      .04837000
    92.5000      .05125000
    95.0000      .05470000
    97.5000      .05883000
   100.0000      .06376000
   102.5000      .06960000
   105.0000      .07644000
   107.5000      .08430000
   110.0000      .09309000
   112.5000      .10250000
   115.0000      .11190000
   117.5000      .12030000
   120.0000      .12650000
   122.5000      .12940000
   125.0000      .12819999
   127.5000      .12290000
   130.0000      .11420000
   132.5000      .10340000
   135.0000      .09172000
   137.5000      .08015000
   140.0000      .06932000
   142.5000      .05954000
   145.0000      .05092000
   147.5000      .04342000
   150.0000      .03693000
   152.5000      .03134000
   155.0000      .02652000
   157.5000      .02234000
   160.0000      .01870000
   162.5000      .01552000
   165.0000      .01269000
   167.5000      .01016000
   170.0000      .00786900
   172.5000      .00575200
   175.0000      .00376500
   177.5000      .00186200
   180.0000      .00000000
 /
 \setdashpattern <8pt, 4pt>                                  
 \plot
      .0000      .00000000
     2.5000      .00152800
     5.0000      .00305500
     7.5000      .00458000
    10.0000      .00610200
    12.5000      .00761700
    15.0000      .00912500
    17.5000      .01062000
    20.0000      .01211000
    22.5000      .01358000
    25.0000      .01504000
    27.5000      .01648000
    30.0000      .01789000
    32.5000      .01929000
    35.0000      .02066000
    37.5000      .02200000
    40.0000      .02331000
    42.5000      .02458000
    45.0000      .02582000
    47.5000      .02701000
    50.0000      .02816000
    52.5000      .02927000
    55.0000      .03034000
    57.5000      .03139000
    60.0000      .03240000
    62.5000      .03340000
    65.0000      .03439000
    67.5000      .03539000
    70.0000      .03639000
    72.5000      .03744000
    75.0000      .03855000
    77.5000      .03976000
    80.0000      .04111000
    82.5000      .04265000
    85.0000      .04445000
    87.5000      .04657000
    90.0000      .04910000
    92.5000      .05213000
    95.0000      .05577000
    97.5000      .06015000
   100.0000      .06541000
   102.5000      .07166000
   105.0000      .07903000
   107.5000      .08755000
   110.0000      .09712000
   112.5000      .10740000
   115.0000      .11770000
   117.5000      .12700000
   120.0000      .13390000
   122.5000      .13699999
   125.0000      .13560000
   127.5000      .12970001
   130.0000      .12020000
   132.5000      .10860000
   135.0000      .09609000
   137.5000      .08380000
   140.0000      .07236000
   142.5000      .06208000
   145.0000      .05305000
   147.5000      .04521000
   150.0000      .03845000
   152.5000      .03263000
   155.0000      .02761000
   157.5000      .02326000
   160.0000      .01948000
   162.5000      .01616000
   165.0000      .01322000
   167.5000      .01059000
   170.0000      .00820100
   172.5000      .00599600
   175.0000      .00392500
   177.5000      .00194100
   180.0000      .00000000
 /
 \setdashpattern <4pt, 4pt>                                  
 \plot
      .0000      .00000000
     2.5000      .00148900
     5.0000      .00297700
     7.5000      .00446300
    10.0000      .00594500
    12.5000      .00742000
    15.0000      .00888800
    17.5000      .01035000
    20.0000      .01179000
    22.5000      .01322000
    25.0000      .01464000
    27.5000      .01603000
    30.0000      .01741000
    32.5000      .01876000
    35.0000      .02009000
    37.5000      .02139000
    40.0000      .02265000
    42.5000      .02388000
    45.0000      .02506000
    47.5000      .02621000
    50.0000      .02731000
    52.5000      .02837000
    55.0000      .02940000
    57.5000      .03039000
    60.0000      .03136000
    62.5000      .03231000
    65.0000      .03324000
    67.5000      .03418000
    70.0000      .03513000
    72.5000      .03612000
    75.0000      .03716000
    77.5000      .03830000
    80.0000      .03957000
    82.5000      .04103000
    85.0000      .04272000
    87.5000      .04471000
    90.0000      .04708000
    92.5000      .04992000
    95.0000      .05332000
    97.5000      .05739000
   100.0000      .06225000
   102.5000      .06801000
   105.0000      .07476000
   107.5000      .08252000
   110.0000      .09120000
   112.5000      .10050000
   115.0000      .10980000
   117.5000      .11820000
   120.0000      .12460000
   122.5000      .12770000
   125.0000      .12680000
   127.5000      .12180000
   130.0000      .11350000
   132.5000      .10310000
   135.0000      .09169000
   137.5000      .08030000
   140.0000      .06957000
   142.5000      .05984000
   145.0000      .05124000
   147.5000      .04372000
   150.0000      .03722000
   152.5000      .03160000
   155.0000      .02675000
   157.5000      .02254000
   160.0000      .01887000
   162.5000      .01566000
   165.0000      .01281000
   167.5000      .01026000
   170.0000      .00794200
   172.5000      .00580600
   175.0000      .00380000
   177.5000      .00187900
   180.0000      .00000000
 /
  \setsolid
  \setlinear
 \put {$\circ       $} at     30.4300      .01850000
  \plot
    30.4300      .02290000
    30.4300      .01410000
  /
  \setsolid
  \setlinear
 \put {$\circ       $} at     37.3400      .02230000
  \plot
    37.3400      .02560000
    37.3400      .01900000
  /
  \setsolid
  \setlinear
 \put {$\circ       $} at     44.5500      .02390000
  \plot
    44.5500      .02690000
    44.5500      .02090000
  /
  \setsolid
  \setlinear
 \put {$\circ       $} at     58.8700      .03180000
  \plot
    58.8700      .03490000
    58.8700      .02870000
  /
  \setsolid
  \setlinear
 \put {$\circ       $} at     72.6500      .03780000
  \plot
    72.6500      .04100000
    72.6500      .03460000
  /
  \setsolid
  \setlinear
 \put {$\circ       $} at     79.2100      .04000000
  \plot
    79.2100      .04280000
    79.2100      .03720000
  /
  \setsolid
  \setlinear
 \put {$\circ       $} at     85.6000      .04470000
  \plot
    85.6000      .04780000
    85.6000      .04160000
  /
  \setsolid
  \setlinear
 \put {$\circ       $} at     91.9000      .05140000
  \plot
    91.9000      .05460000
    91.9000      .04820000
  /
  \setsolid
  \setlinear
 \put {$\circ       $} at     97.8800      .06310000
  \plot
    97.8800      .06690000
    97.8800      .05930000
  /
  \setsolid
  \setlinear
 \put {$\circ       $} at    103.7000      .08440000
  \plot
   103.7000      .08760000
   103.7000      .08120000
  /
  \setsolid
  \setlinear
 \put {$\circ       $} at    109.3000      .11530000
  \plot
   109.3000      .11950000
   109.3000      .11110000
  /
  \setsolid
  \setlinear
 \put {$\circ       $} at    114.7600      .13249999
  \plot
   114.7600      .13649999
   114.7600      .12850000
  /
  \setsolid
  \setlinear
 \put {$\circ       $} at    119.9600      .16480000
  \plot
   119.9600      .16980000
   119.9600      .15980001
  /
  \setsolid
  \setlinear
 \put {$\circ       $} at    124.9200      .17020001
  \plot
   124.9200      .17500001
   124.9200      .16540000
  /
  \setsolid
  \setlinear
 \put {$\circ       $} at    129.6200      .14830001
  \plot
   129.6200      .15310001
   129.6200      .14350000
  /
  \setsolid
  \setlinear
 \put {$\circ       $} at    134.0800      .12230000
  \plot
   134.0800      .12690000
   134.0800      .11770000
  /
  \setsolid
  \setlinear
 \put {$\circ       $} at    138.1500      .11190000
  \plot
   138.1500      .11640000
   138.1500      .10740000
  /
  \setsolid
  \setlinear
 \put {$\circ       $} at    145.7400      .06450000
  \plot
   145.7400      .06910000
   145.7400      .05990000
  /
  \setsolid
  \setlinear
 \endpicture
} [lt] at 0 0
\put {
 \beginpicture
\setcoordinatesystem units <  .03333333 true cm,    32.8767 true cm>
\setplotarea x from       .0000 to    180.0000 , y from      -.0075 to       .1750
\put {$ A_y                    $} [lt] <10pt,-10pt> at       .0000       .1750
 \linethickness .8 pt
 \setplotsymbol ({.})
\axis bottom label {$\vartheta$ [deg]} ticks short numbered from       .0000 to    180.0000 by         45. /
\axis top ticks short unlabeled from       .0000 to    180.0000 by         45. /
\axis left ticks short numbered from         .00 to         .17 by         .05 /
\axis right ticks short unlabeled from         .00 to         .17 by         .05 /
 \setsolid                                                   
 \plot
      .0000      .00000000
     2.5000      .00151300
     5.0000      .00302600
     7.5000      .00453600
    10.0000      .00604300
    12.5000      .00754400
    15.0000      .00903800
    17.5000      .01052000
    20.0000      .01199000
    22.5000      .01345000
    25.0000      .01490000
    27.5000      .01632000
    30.0000      .01773000
    32.5000      .01911000
    35.0000      .02047000
    37.5000      .02181000
    40.0000      .02310000
    42.5000      .02437000
    45.0000      .02559000
    47.5000      .02677000
    50.0000      .02791000
    52.5000      .02901000
    55.0000      .03008000
    57.5000      .03111000
    60.0000      .03212000
    62.5000      .03311000
    65.0000      .03409000
    67.5000      .03507000
    70.0000      .03607000
    72.5000      .03710000
    75.0000      .03818000
    77.5000      .03937000
    80.0000      .04068000
    82.5000      .04218000
    85.0000      .04391000
    87.5000      .04595000
    90.0000      .04837000
    92.5000      .05125000
    95.0000      .05470000
    97.5000      .05883000
   100.0000      .06376000
   102.5000      .06960000
   105.0000      .07644000
   107.5000      .08430000
   110.0000      .09309000
   112.5000      .10250000
   115.0000      .11190000
   117.5000      .12030000
   120.0000      .12650000
   122.5000      .12940000
   125.0000      .12819999
   127.5000      .12290000
   130.0000      .11420000
   132.5000      .10340000
   135.0000      .09172000
   137.5000      .08015000
   140.0000      .06932000
   142.5000      .05954000
   145.0000      .05092000
   147.5000      .04342000
   150.0000      .03693000
   152.5000      .03134000
   155.0000      .02652000
   157.5000      .02234000
   160.0000      .01870000
   162.5000      .01552000
   165.0000      .01269000
   167.5000      .01016000
   170.0000      .00786900
   172.5000      .00575200
   175.0000      .00376500
   177.5000      .00186200
   180.0000      .00000000
 /
 \setdashpattern <8pt, 4pt>                                  
 \plot
      .0000      .00000000
     2.5000      .00161200
     5.0000      .00322300
     7.5000      .00483100
    10.0000      .00643700
    12.5000      .00803700
    15.0000      .00963000
    17.5000      .01121000
    20.0000      .01279000
    22.5000      .01435000
    25.0000      .01589000
    27.5000      .01742000
    30.0000      .01893000
    32.5000      .02042000
    35.0000      .02188000
    37.5000      .02331000
    40.0000      .02472000
    42.5000      .02609000
    45.0000      .02741000
    47.5000      .02870000
    50.0000      .02995000
    52.5000      .03116000
    55.0000      .03234000
    57.5000      .03348000
    60.0000      .03460000
    62.5000      .03570000
    65.0000      .03680000
    67.5000      .03790000
    70.0000      .03901000
    72.5000      .04017000
    75.0000      .04139000
    77.5000      .04272000
    80.0000      .04419000
    82.5000      .04586000
    85.0000      .04779000
    87.5000      .05005000
    90.0000      .05272000
    92.5000      .05591000
    95.0000      .05972000
    97.5000      .06428000
   100.0000      .06973000
   102.5000      .07619000
   105.0000      .08378000
   107.5000      .09252000
   110.0000      .10230000
   112.5000      .11280000
   115.0000      .12320000
   117.5000      .13249999
   120.0000      .13920000
   122.5000      .14200000
   125.0000      .14010000
   127.5000      .13349999
   130.0000      .12330000
   132.5000      .11090000
   135.0000      .09784000
   137.5000      .08505000
   140.0000      .07323000
   142.5000      .06267000
   145.0000      .05343000
   147.5000      .04544000
   150.0000      .03858000
   152.5000      .03269000
   155.0000      .02763000
   157.5000      .02325000
   160.0000      .01945000
   162.5000      .01613000
   165.0000      .01319000
   167.5000      .01056000
   170.0000      .00817200
   172.5000      .00597200
   175.0000      .00390900
   177.5000      .00193300
   180.0000      .00000000
 /
 \setdashpattern <4pt, 4pt>                                  
 \plot
      .0000      .00000000
     2.5000      .00159000
     5.0000      .00317900
     7.5000      .00476600
    10.0000      .00634800
    12.5000      .00792500
    15.0000      .00949400
    17.5000      .01105000
    20.0000      .01260000
    22.5000      .01413000
    25.0000      .01565000
    27.5000      .01714000
    30.0000      .01862000
    32.5000      .02007000
    35.0000      .02150000
    37.5000      .02290000
    40.0000      .02426000
    42.5000      .02558000
    45.0000      .02687000
    47.5000      .02811000
    50.0000      .02931000
    52.5000      .03048000
    55.0000      .03160000
    57.5000      .03270000
    60.0000      .03377000
    62.5000      .03482000
    65.0000      .03587000
    67.5000      .03692000
    70.0000      .03800000
    72.5000      .03911000
    75.0000      .04031000
    77.5000      .04161000
    80.0000      .04308000
    82.5000      .04475000
    85.0000      .04671000
    87.5000      .04902000
    90.0000      .05178000
    92.5000      .05510000
    95.0000      .05910000
    97.5000      .06392000
   100.0000      .06971000
   102.5000      .07663000
   105.0000      .08481000
   107.5000      .09430000
   110.0000      .10500000
   112.5000      .11650000
   115.0000      .12809999
   117.5000      .13860001
   120.0000      .14620000
   122.5000      .14970000
   125.0000      .14800000
   127.5000      .14129999
   130.0000      .13060001
   132.5000      .11760000
   135.0000      .10370000
   137.5000      .09021000
   140.0000      .07771000
   142.5000      .06655000
   145.0000      .05678000
   147.5000      .04833000
   150.0000      .04106000
   152.5000      .03481000
   155.0000      .02944000
   157.5000      .02479000
   160.0000      .02075000
   162.5000      .01721000
   165.0000      .01408000
   167.5000      .01128000
   170.0000      .00873100
   172.5000      .00638300
   175.0000      .00417800
   177.5000      .00206600
   180.0000      .00000000
 /
  \setsolid
  \setlinear
 \put {$\circ       $} at     30.4300      .01850000
  \plot
    30.4300      .02290000
    30.4300      .01410000
  /
  \setsolid
  \setlinear
 \put {$\circ       $} at     37.3400      .02230000
  \plot
    37.3400      .02560000
    37.3400      .01900000
  /
  \setsolid
  \setlinear
 \put {$\circ       $} at     44.5500      .02390000
  \plot
    44.5500      .02690000
    44.5500      .02090000
  /
  \setsolid
  \setlinear
 \put {$\circ       $} at     58.8700      .03180000
  \plot
    58.8700      .03490000
    58.8700      .02870000
  /
  \setsolid
  \setlinear
 \put {$\circ       $} at     72.6500      .03780000
  \plot
    72.6500      .04100000
    72.6500      .03460000
  /
  \setsolid
  \setlinear
 \put {$\circ       $} at     79.2100      .04000000
  \plot
    79.2100      .04280000
    79.2100      .03720000
  /
  \setsolid
  \setlinear
 \put {$\circ       $} at     85.6000      .04470000
  \plot
    85.6000      .04780000
    85.6000      .04160000
  /
  \setsolid
  \setlinear
 \put {$\circ       $} at     91.9000      .05140000
  \plot
    91.9000      .05460000
    91.9000      .04820000
  /
  \setsolid
  \setlinear
 \put {$\circ       $} at     97.8800      .06310000
  \plot
    97.8800      .06690000
    97.8800      .05930000
  /
  \setsolid
  \setlinear
 \put {$\circ       $} at    103.7000      .08440000
  \plot
   103.7000      .08760000
   103.7000      .08120000
  /
  \setsolid
  \setlinear
 \put {$\circ       $} at    109.3000      .11530000
  \plot
   109.3000      .11950000
   109.3000      .11110000
  /
  \setsolid
  \setlinear
 \put {$\circ       $} at    114.7600      .13249999
  \plot
   114.7600      .13649999
   114.7600      .12850000
  /
  \setsolid
  \setlinear
 \put {$\circ       $} at    119.9600      .16480000
  \plot
   119.9600      .16980000
   119.9600      .15980001
  /
  \setsolid
  \setlinear
 \put {$\circ       $} at    124.9200      .17020001
  \plot
   124.9200      .17500001
   124.9200      .16540000
  /
  \setsolid
  \setlinear
 \put {$\circ       $} at    129.6200      .14830001
  \plot
   129.6200      .15310001
   129.6200      .14350000
  /
  \setsolid
  \setlinear
 \put {$\circ       $} at    134.0800      .12230000
  \plot
   134.0800      .12690000
   134.0800      .11770000
  /
  \setsolid
  \setlinear
 \put {$\circ       $} at    138.1500      .11190000
  \plot
   138.1500      .11640000
   138.1500      .10740000
  /
  \setsolid
  \setlinear
 \put {$\circ       $} at    145.7400      .06450000
  \plot
   145.7400      .06910000
   145.7400      .05990000
  /
  \setsolid
  \setlinear
 \endpicture

} [lt] at 7.5 0
\endpicture

%% file: figay50.tex
\beginpicture
\setcoordinatesystem units < 1 true cm, 1 true cm >
\setplotarea x from 0 to 15, y from -7 to 0

\put { 
 \beginpicture
\setcoordinatesystem units <  .03333333 true cm,     6.9671 true cm>
\setplotarea x from       .0000 to    180.0000 , y from      -.5997 to       .2614
\put {$ A_y                    $} [lt] <10pt,-10pt> at       .0000       .2614
 \linethickness .8 pt
 \setplotsymbol ({.})
\axis bottom label {$\vartheta$ [deg]} ticks short numbered from       .0000 to    180.0000 by         45. /
\axis top ticks short unlabeled from       .0000 to    180.0000 by         45. /
\axis left ticks short numbered from        -.50 to         .20 by         .20 /
\axis right ticks short unlabeled from        -.50 to         .20 by         .20 /
 \setsolid                                                   
 \plot
      .0000      .00000000
     2.5000      .00962400
     5.0000      .01926000
     7.5000      .02893000
    10.0000      .03864000
    12.5000      .04840000
    15.0000      .05819000
    17.5000      .06800000
    20.0000      .07780000
    22.5000      .08753000
    25.0000      .09712000
    27.5000      .10650000
    30.0000      .11550000
    32.5000      .12400000
    35.0000      .13180000
    37.5000      .13880000
    40.0000      .14480001
    42.5000      .14940000
    45.0000      .15260001
    47.5000      .15390000
    50.0000      .15330000
    52.5000      .15040000
    55.0000      .14500000
    57.5000      .13689999
    60.0000      .12590000
    62.5000      .11180000
    65.0000      .09431000
    67.5000      .07347000
    70.0000      .04916000
    72.5000      .02141000
    75.0000     -.00970400
    77.5000     -.04397000
    80.0000     -.08107000
    82.5000     -.12050000
    85.0000     -.16180000
    87.5000     -.20410000
    90.0000     -.24680001
    92.5000     -.28880000
    95.0000     -.32929999
    97.5000     -.36739999
   100.0000     -.40230000
   102.5000     -.43309999
   105.0000     -.45910001
   107.5000     -.47970000
   110.0000     -.49410000
   112.5000     -.50160003
   115.0000     -.50129998
   117.5000     -.49190000
   120.0000     -.47170001
   122.5000     -.43830001
   125.0000     -.38929999
   127.5000     -.32229999
   130.0000     -.23710001
   132.5000     -.13789999
   135.0000     -.03476000
   137.5000      .05794000
   140.0000      .12740000
   142.5000      .16840000
   145.0000      .18359999
   147.5000      .18040000
   150.0000      .16630000
   152.5000      .14720000
   155.0000      .12680000
   157.5000      .10710000
   160.0000      .08913000
   162.5000      .07308000
   165.0000      .05895000
   167.5000      .04653000
   170.0000      .03554000
   172.5000      .02569000
   175.0000      .01667000
   177.5000      .00820100
   180.0000      .00000000
 /
 \setdashpattern <8pt, 4pt>                                  
 \plot
      .0000      .00000000
     2.5000      .00900800
     5.0000      .01803000
     7.5000      .02706000
    10.0000      .03612000
    12.5000      .04519000
    15.0000      .05428000
    17.5000      .06335000
    20.0000      .07237000
    22.5000      .08129000
    25.0000      .09003000
    27.5000      .09851000
    30.0000      .10660000
    32.5000      .11420000
    35.0000      .12110000
    37.5000      .12720001
    40.0000      .13220000
    42.5000      .13590001
    45.0000      .13820000
    47.5000      .13880000
    50.0000      .13750000
    52.5000      .13400000
    55.0000      .12809999
    57.5000      .11950000
    60.0000      .10810000
    62.5000      .09371000
    65.0000      .07605000
    67.5000      .05502000
    70.0000      .03051000
    72.5000      .00248000
    75.0000     -.02903000
    77.5000     -.06388000
    80.0000     -.10180000
    82.5000     -.14250000
    85.0000     -.18529999
    87.5000     -.22960000
    90.0000     -.27469999
    92.5000     -.31959999
    95.0000     -.36340001
    97.5000     -.40520000
   100.0000     -.44380000
   102.5000     -.47850001
   105.0000     -.50819999
   107.5000     -.53219998
   110.0000     -.54949999
   112.5000     -.55930001
   115.0000     -.56059998
   117.5000     -.55210000
   120.0000     -.53230000
   122.5000     -.49919999
   125.0000     -.45089999
   127.5000     -.38609999
   130.0000     -.30509999
   132.5000     -.21170001
   135.0000     -.11380000
   137.5000     -.02243000
   140.0000      .05214000
   142.5000      .10400000
   145.0000      .13310000
   147.5000      .14370000
   150.0000      .14129999
   152.5000      .13120000
   155.0000      .11720000
   157.5000      .10190000
   160.0000      .08670000
   162.5000      .07237000
   165.0000      .05920000
   167.5000      .04725000
   170.0000      .03641000
   172.5000      .02648000
   175.0000      .01726000
   177.5000      .00851300
   180.0000      .00000000
 /
 \setdashpattern <4pt, 4pt>                                  
 \plot
      .0000      .00000000
     2.5000      .00972400
     5.0000      .01947000
     7.5000      .02924000
    10.0000      .03907000
    12.5000      .04894000
    15.0000      .05886000
    17.5000      .06882000
    20.0000      .07877000
    22.5000      .08868000
    25.0000      .09847000
    27.5000      .10800000
    30.0000      .11730000
    32.5000      .12600000
    35.0000      .13420001
    37.5000      .14150000
    40.0000      .14770000
    42.5000      .15270001
    45.0000      .15610000
    47.5000      .15780000
    50.0000      .15750000
    52.5000      .15490000
    55.0000      .14980000
    57.5000      .14190000
    60.0000      .13100000
    62.5000      .11690000
    65.0000      .09950000
    67.5000      .07856000
    70.0000      .05406000
    72.5000      .02599000
    75.0000     -.00555300
    77.5000     -.04036000
    80.0000     -.07811000
    82.5000     -.11830000
    85.0000     -.16040000
    87.5000     -.20360000
    90.0000     -.24720000
    92.5000     -.29020000
    95.0000     -.33170000
    97.5000     -.37090001
   100.0000     -.40680000
   102.5000     -.43889999
   105.0000     -.46630001
   107.5000     -.48840001
   110.0000     -.50459999
   112.5000     -.51400000
   115.0000     -.51569998
   117.5000     -.50830001
   120.0000     -.48980001
   122.5000     -.45719999
   125.0000     -.40720001
   127.5000     -.33620000
   130.0000     -.24290000
   132.5000     -.13120000
   135.0000     -.01317000
   137.5000      .09251000
   140.0000      .16930000
   142.5000      .21100000
   145.0000      .22229999
   147.5000      .21300000
   150.0000      .19280000
   152.5000      .16820000
   155.0000      .14330000
   157.5000      .12010000
   160.0000      .09923000
   162.5000      .08096000
   165.0000      .06506000
   167.5000      .05121000
   170.0000      .03904000
   172.5000      .02817000
   175.0000      .01826000
   177.5000      .00897800
   180.0000      .00000000
 /
  \setsolid
  \setlinear
 \put {$\circ       $} at    149.8000      .20000000
  \plot
   149.8000      .23000000
   149.8000      .17000000
  /
  \setsolid
  \setlinear
 \put {$\circ       $} at    144.8000      .16000000
  \plot
   144.8000      .19999999
   144.8000      .12000000
  /
  \setsolid
  \setlinear
 \put {$\circ       $} at    139.8000      .14000000
  \plot
   139.8000      .18000001
   139.8000      .10000000
  /
  \setsolid
  \setlinear
 \put {$\circ       $} at    134.7000      .00000000
  \plot
   134.7000      .08000000
   134.7000     -.08000000
  /
  \setsolid
  \setlinear
 \put {$\circ       $} at    129.7000     -.23999999
  \plot
   129.7000     -.19999999
   129.7000     -.28000000
  /
  \setsolid
  \setlinear
 \put {$\circ       $} at    124.7000     -.31000000
  \plot
   124.7000     -.27000001
   124.7000     -.34999999
  /
  \setsolid
  \setlinear
 \put {$\circ       $} at    119.7000     -.41999999
  \plot
   119.7000     -.38999999
   119.7000     -.44999999
  /
  \setsolid
  \setlinear
 \put {$\circ       $} at    114.7000     -.44999999
  \plot
   114.7000     -.41000000
   114.7000     -.48999998
  /
  \setsolid
  \setlinear
 \put {$\circ       $} at    109.7000     -.52999997
  \plot
   109.7000     -.50999999
   109.7000     -.54999995
  /
  \setsolid
  \setlinear
 \put {$\circ       $} at    104.7000     -.44999999
  \plot
   104.7000     -.38999999
   104.7000     -.50999999
  /
  \setsolid
  \setlinear
 \put {$\circ       $} at     99.7000     -.43000001
  \plot
    99.7000     -.41000000
    99.7000     -.45000002
  /
  \setsolid
  \setlinear
 \put {$\circ       $} at     94.7000     -.37000000
  \plot
    94.7000     -.33000001
    94.7000     -.41000000
  /
  \setsolid
  \setlinear
 \put {$\circ       $} at     89.7000     -.25999999
  \plot
    89.7000     -.23999999
    89.7000     -.28000000
  /
  \setsolid
  \setlinear
 \put {$\circ       $} at     37.6000      .14000000
  \plot
    37.6000      .17000000
    37.6000      .11000000
  /
  \setsolid
  \setlinear
 \put {$\circ       $} at     45.0000      .16000000
  \plot
    45.0000      .19999999
    45.0000      .12000000
  /
  \setsolid
  \setlinear
 \put {$\circ       $} at     52.1000      .17000000
  \plot
    52.1000      .20000000
    52.1000      .14000000
  /
  \setsolid
  \setlinear
 \put {$\circ       $} at     59.3000      .12000000
  \plot
    59.3000      .16000000
    59.3000      .08000000
  /
  \setsolid
  \setlinear
 \put {$\circ       $} at     73.2000      .02000000
  \plot
    73.2000      .06000000
    73.2000     -.02000000
  /
  \setsolid
  \setlinear
 \put {$\circ       $} at     79.8000     -.14000000
  \plot
    79.8000     -.08000000
    79.8000     -.20000000
  /
  \setsolid
  \setlinear
 \put {$\circ       $} at     86.3000     -.27000001
  \plot
    86.3000     -.22000001
    86.3000     -.32000002
  /
  \setsolid
  \setlinear
 \put {$\circ       $} at     92.5000     -.33000001
  \plot
    92.5000     -.25000000
    92.5000     -.41000003
  /
 \endpicture
} [lt] at 0 0
\put {
 \beginpicture
\setcoordinatesystem units <  .03333333 true cm,     7.5726 true cm>
\setplotarea x from       .0000 to    180.0000 , y from      -.5727 to       .2196
\put {$ A_y                    $} [lt] <10pt,-10pt> at       .0000       .2196
 \linethickness .8 pt
 \setplotsymbol ({.})
\axis bottom label {$\vartheta$ [deg]} ticks short numbered from       .0000 to    180.0000 by         45. /
\axis top ticks short unlabeled from       .0000 to    180.0000 by         45. /
\axis left ticks short numbered from        -.50 to         .20 by         .20 /
\axis right ticks short unlabeled from        -.50 to         .20 by         .20 /
 \setsolid                                                   
 \plot
      .0000      .00000000
     2.5000      .00962400
     5.0000      .01926000
     7.5000      .02893000
    10.0000      .03864000
    12.5000      .04840000
    15.0000      .05819000
    17.5000      .06800000
    20.0000      .07780000
    22.5000      .08753000
    25.0000      .09712000
    27.5000      .10650000
    30.0000      .11550000
    32.5000      .12400000
    35.0000      .13180000
    37.5000      .13880000
    40.0000      .14480001
    42.5000      .14940000
    45.0000      .15260001
    47.5000      .15390000
    50.0000      .15330000
    52.5000      .15040000
    55.0000      .14500000
    57.5000      .13689999
    60.0000      .12590000
    62.5000      .11180000
    65.0000      .09431000
    67.5000      .07347000
    70.0000      .04916000
    72.5000      .02141000
    75.0000     -.00970400
    77.5000     -.04397000
    80.0000     -.08107000
    82.5000     -.12050000
    85.0000     -.16180000
    87.5000     -.20410000
    90.0000     -.24680001
    92.5000     -.28880000
    95.0000     -.32929999
    97.5000     -.36739999
   100.0000     -.40230000
   102.5000     -.43309999
   105.0000     -.45910001
   107.5000     -.47970000
   110.0000     -.49410000
   112.5000     -.50160003
   115.0000     -.50129998
   117.5000     -.49190000
   120.0000     -.47170001
   122.5000     -.43830001
   125.0000     -.38929999
   127.5000     -.32229999
   130.0000     -.23710001
   132.5000     -.13789999
   135.0000     -.03476000
   137.5000      .05794000
   140.0000      .12740000
   142.5000      .16840000
   145.0000      .18359999
   147.5000      .18040000
   150.0000      .16630000
   152.5000      .14720000
   155.0000      .12680000
   157.5000      .10710000
   160.0000      .08913000
   162.5000      .07308000
   165.0000      .05895000
   167.5000      .04653000
   170.0000      .03554000
   172.5000      .02569000
   175.0000      .01667000
   177.5000      .00820100
   180.0000      .00000000
 /
 \setdashpattern <8pt, 4pt>                                  
 \plot
      .0000      .00000000
     2.5000      .00941700
     5.0000      .01885000
     7.5000      .02830000
    10.0000      .03778000
    12.5000      .04728000
    15.0000      .05680000
    17.5000      .06632000
    20.0000      .07579000
    22.5000      .08517000
    25.0000      .09438000
    27.5000      .10330000
    30.0000      .11190000
    32.5000      .11990000
    35.0000      .12729999
    37.5000      .13380000
    40.0000      .13930000
    42.5000      .14350000
    45.0000      .14620000
    47.5000      .14720000
    50.0000      .14630000
    52.5000      .14330000
    55.0000      .13779999
    57.5000      .12980001
    60.0000      .11900000
    62.5000      .10530000
    65.0000      .08841000
    67.5000      .06828000
    70.0000      .04481000
    72.5000      .01797000
    75.0000     -.01220000
    77.5000     -.04555000
    80.0000     -.08184000
    82.5000     -.12070000
    85.0000     -.16159999
    87.5000     -.20400000
    90.0000     -.24699999
    92.5000     -.28979999
    95.0000     -.33140001
    97.5000     -.37079999
   100.0000     -.40709999
   102.5000     -.43910000
   105.0000     -.46590000
   107.5000     -.48670000
   110.0000     -.50059998
   112.5000     -.50660002
   115.0000     -.50379997
   117.5000     -.49110001
   120.0000     -.46730000
   122.5000     -.43110001
   125.0000     -.38119999
   127.5000     -.31750000
   130.0000     -.24140000
   132.5000     -.15710001
   135.0000     -.07221000
   137.5000      .00434200
   140.0000      .06480000
   142.5000      .10540000
   145.0000      .12690000
   147.5000      .13310000
   150.0000      .12890001
   152.5000      .11850000
   155.0000      .10520000
   157.5000      .09107000
   160.0000      .07720000
   162.5000      .06425000
   165.0000      .05243000
   167.5000      .04177000
   170.0000      .03213000
   172.5000      .02334000
   175.0000      .01520000
   177.5000      .00749200
   180.0000      .00000000
 /
 \setdashpattern <4pt, 4pt>                                  
 \plot
      .0000      .00000000
     2.5000      .00393100
     5.0000      .00785400
     7.5000      .01176000
    10.0000      .01563000
    12.5000      .01946000
    15.0000      .02323000
    17.5000      .02692000
    20.0000      .03051000
    22.5000      .03395000
    25.0000      .03719000
    27.5000      .04018000
    30.0000      .04284000
    32.5000      .04507000
    35.0000      .04677000
    37.5000      .04781000
    40.0000      .04804000
    42.5000      .04729000
    45.0000      .04541000
    47.5000      .04219000
    50.0000      .03744000
    52.5000      .03099000
    55.0000      .02264000
    57.5000      .01224000
    60.0000     -.00035870
    62.5000     -.01523000
    65.0000     -.03244000
    67.5000     -.05196000
    70.0000     -.07373000
    72.5000     -.09762000
    75.0000     -.12340000
    77.5000     -.15099999
    80.0000     -.17990001
    82.5000     -.20990001
    85.0000     -.24079999
    87.5000     -.27210000
    90.0000     -.30350000
    92.5000     -.33480000
    95.0000     -.36550000
    97.5000     -.39530000
   100.0000     -.42390001
   102.5000     -.45080000
   105.0000     -.47549999
   107.5000     -.49730000
   110.0000     -.51560003
   112.5000     -.52920002
   115.0000     -.53670001
   117.5000     -.53630000
   120.0000     -.52560002
   122.5000     -.50199997
   125.0000     -.46250001
   127.5000     -.40540001
   130.0000     -.33100000
   132.5000     -.24400000
   135.0000     -.15310000
   137.5000     -.06922000
   140.0000     -.00116600
   142.5000      .04704000
   145.0000      .07624000
   147.5000      .09020000
   150.0000      .09344000
   152.5000      .08991000
   155.0000      .08256000
   157.5000      .07338000
   160.0000      .06360000
   162.5000      .05392000
   165.0000      .04470000
   167.5000      .03606000
   170.0000      .02803000
   172.5000      .02052000
   175.0000      .01344000
   177.5000      .00664700
   180.0000      .00000000
 /
  \setsolid
  \setlinear
 \put {$\circ       $} at    149.8000      .20000000
  \plot
   149.8000      .23000000
   149.8000      .17000000
  /
  \setsolid
  \setlinear
 \put {$\circ       $} at    144.8000      .16000000
  \plot
   144.8000      .19999999
   144.8000      .12000000
  /
  \setsolid
  \setlinear
 \put {$\circ       $} at    139.8000      .14000000
  \plot
   139.8000      .18000001
   139.8000      .10000000
  /
  \setsolid
  \setlinear
 \put {$\circ       $} at    134.7000      .00000000
  \plot
   134.7000      .08000000
   134.7000     -.08000000
  /
  \setsolid
  \setlinear
 \put {$\circ       $} at    129.7000     -.23999999
  \plot
   129.7000     -.19999999
   129.7000     -.28000000
  /
  \setsolid
  \setlinear
 \put {$\circ       $} at    124.7000     -.31000000
  \plot
   124.7000     -.27000001
   124.7000     -.34999999
  /
  \setsolid
  \setlinear
 \put {$\circ       $} at    119.7000     -.41999999
  \plot
   119.7000     -.38999999
   119.7000     -.44999999
  /
  \setsolid
  \setlinear
 \put {$\circ       $} at    114.7000     -.44999999
  \plot
   114.7000     -.41000000
   114.7000     -.48999998
  /
  \setsolid
  \setlinear
 \put {$\circ       $} at    109.7000     -.52999997
  \plot
   109.7000     -.50999999
   109.7000     -.54999995
  /
  \setsolid
  \setlinear
 \put {$\circ       $} at    104.7000     -.44999999
  \plot
   104.7000     -.38999999
   104.7000     -.50999999
  /
  \setsolid
  \setlinear
 \put {$\circ       $} at     99.7000     -.43000001
  \plot
    99.7000     -.41000000
    99.7000     -.45000002
  /
  \setsolid
  \setlinear
 \put {$\circ       $} at     94.7000     -.37000000
  \plot
    94.7000     -.33000001
    94.7000     -.41000000
  /
  \setsolid
  \setlinear
 \put {$\circ       $} at     89.7000     -.25999999
  \plot
    89.7000     -.23999999
    89.7000     -.28000000
  /
  \setsolid
  \setlinear
 \put {$\circ       $} at     37.6000      .14000000
  \plot
    37.6000      .17000000
    37.6000      .11000000
  /
  \setsolid
  \setlinear
 \put {$\circ       $} at     45.0000      .16000000
  \plot
    45.0000      .19999999
    45.0000      .12000000
  /
  \setsolid
  \setlinear
 \put {$\circ       $} at     52.1000      .17000000
  \plot
    52.1000      .20000000
    52.1000      .14000000
  /
  \setsolid
  \setlinear
 \put {$\circ       $} at     59.3000      .12000000
  \plot
    59.3000      .16000000
    59.3000      .08000000
  /
  \setsolid
  \setlinear
 \put {$\circ       $} at     73.2000      .02000000
  \plot
    73.2000      .06000000
    73.2000     -.02000000
  /
  \setsolid
  \setlinear
 \put {$\circ       $} at     79.8000     -.14000000
  \plot
    79.8000     -.08000000
    79.8000     -.20000000
  /
  \setsolid
  \setlinear
 \put {$\circ       $} at     86.3000     -.27000001
  \plot
    86.3000     -.22000001
    86.3000     -.32000002
  /
  \setsolid
  \setlinear
 \put {$\circ       $} at     92.5000     -.33000001
  \plot
    92.5000     -.25000000
    92.5000     -.41000003
  /
 \endpicture

} [lt] at 7.5 0
\endpicture

%% file: texla_ar.bbl
\begin{thebibliography}{99}
\bibitem{total}
H.\ Wita\l a, H.\ Kamada, A.\ Nogga, W.\ Gl\"ockle, Ch.\ Elster, and D.\
H\"uber,
{\it Phys.\ Rev.\ } {\bf C59}, 3035 (1999);\\
W.\ P.\ Abfalterer {\it et al.}, {\it Phys.\ Rev.\ Lett.\ } {\bf 81}, 57
(1998).
\bibitem{diff}
H.\ Wita\l a, W.\ Gl\"ockle, D.\ H\"uber, J.\ Golak, and H.\ Kamada, {\it
Phys.\
Rev.\ Lett.\ } {\bf 81}, 1183 (1998);\\
H.\ Rohdjess {\it et al.}, {\it Phys.\ Rev.\ } {\bf C57}, 2111 (1998).
\bibitem{report}
W.\ Gl\"ockle, H.\ Wita\l a, D.\ H\"uber, H.\ Kamada, and J.\ Golak, {\it
Phys.\
Rep.\ }{\bf 274}, 107 (1996).
\bibitem{Ay}
D.\ H\"uber and J.\ L.\ Friar, {\it Phys.\ Rev.\ }{\bf C58}, 674 (1998).
\bibitem{FM}
J.-I. Fujita and H. Miyazawa, {\it Prog.\ Theor.\ Phys.\ }{\bf 17}, 360
(1957).
\bibitem{TM}
S.\ A.\ Coon, M.\ D.\ Scadron, P.\ C.\ McNamee, B.\ R.\ Barrett,
D.\ W.\ E.\ Blatt, and B.\ H.\ J.\ McKellar, {\it Nucl.\ Phys.\ }{\bf
A317},
242 (1979);\\
S.\ A.\ Coon and W.\ Gl\"ockle, {\it Phys.\ Rev.\ }{\bf C23}, 1790 (1981).
\bibitem{Br}
H.\ T.\ Coelho, T.\ K.\ Das, and M.\ R.\ Robilotta, {\it Phys.\
Rev.\ C} {\bf 28}, 1812 (1983); M.\ R.\ Robilotta and H.\ T.\ Coelho,
{\it Nucl.\ Phys.\ }{\bf A460}, 645 (1986).
\bibitem{Ur}
J.\ Carlson, V.R.\ Pandharipande, and R.\ B.\ Wiringa,
{\it Nucl.\ Phys.\ }{\bf A401}, 59 (1983).
\bibitem{Texas}
U.\ van Kolck, Thesis, University of Texas, (1993);\\
C.\ Ord\'o\~nez and U.\ van Kolck, {\it Phys.\ Lett.\ }{\bf B 291},
459 (1992); \\
U.\ van Kolck, {\it Phys.\ Rev.\ C} {\bf 49}, 2932 (1994).
\bibitem{Rp}
J.\ A.\ Eden and M.\ F.\ Gari, {\it Phys.\ Rev.\ C} {\bf 53},
1510 (1996); \\
J.\ A.\ Eden, (private communication).
\bibitem{cs3nf}
J.\ L.\ Friar, D.\ H\"uber, and U.\ van Kolck, {\it Phys.\ Rev.\ }{\bf
C59},
53 (1999).
\bibitem{triton}
J.\ L.\ Friar, B.\ F.\ Gibson, G.\ L.\ Payne, and S.\ A.\ Coon, {\it
Few-Body
Systems }{\bf 5}, 13 (1988).
\bibitem{trifit}
A.\ Nogga, D.\ H\"uber, H.\ Kamada, and W.\ Gl\"ockle, {\it Phys.\ Lett.\ }
{\bf B409}, 19 (1997).
\bibitem{ich}
D.\ H\"uber, private communication.
\bibitem{AV18}
R.\ B.\ Wiringa, V.\ G.\ J.\ Stoks, and R.\ Schiavilla, {\it Phys.\ Rev.\ }
{\bf C51}, 38 (1995).
\bibitem{rhos}
R.\ G.\ Ellis, S.\ A.\ Coon, and B.\ H.\ J.\ McKellar, {\it Nucl.\ Phys.\ }
{\bf A438}, 631 (1985).
\bibitem{TMrhos}
B.\ H.\ J.\ McKellar, {\it Lecture Notes in Physics }{\bf 260}, 7 (1986);\\
S.\ A.\ Coon, {\it Lecture Notes in Physics }{\bf 260}, 92 (1986);\\
S.\ A.\ Coon, {\it Few-Body Systems, Suppl.\ }{\bf 1}, 41 (1986);\\
S.\ A.\ Coon and M.\ T.\ Pe\~na, {\it Phys.\ Rev.\ }{\bf C48}, 2559 (1993).
\bibitem{3Nrhos}
H.\ Wita\l a, D.\ H\"uber, W.\ Gl\"ockle, J.\ Golak, A.\ Stadler, and J.\
Adam Jr. , {\it Phys.\ Rev.\ }{\bf C52}, 1254 (1995).
\bibitem{d1meson}
S.\ A.\ Coon, M.\ T.\ Pe\~na, and D.\ O.\ Riska, {\it Phys.\ Rev.\ }{\bf
C52},
2925 (1995).
\bibitem{first}
D.\ H\"uber, H.\ Wita\l a, and W.\ Gl\"ockle, {\it Few-Body Systems} {\bf
14},
171 (1993).
\bibitem{algorithm}
D.\ H\"uber, H.\ Kamada, H.\ Wita\l a, and W.\ Gl\"ockle, {\it Acta Phys.\
Pol.\ } {\bf B28}, 1677 (1997).
\bibitem{book} W.\ Gl\"{o}ckle, {\it The Quantum Mechanical Few-Body
Problem}
(Springer-Verlag, Heidelberg, New York, 1983), pp 132-137. This
comprehensive
examination of the three-nucleon problem has an extensive discussion of
permutations.
\bibitem{pwd}
D.\ H\"uber, H.\ Wita\l a, A.\ Nogga, W.\ Gl\"ockle, and H.\ Kamada, {\it
Few-Body Systems }{\bf 22}, 107 (1997).
\bibitem{PhD}
D.\ H\"uber, PhD Thesis, Bochum 1993, unpublished.
\bibitem{ay3data}
J.\ E.\ McAninch, W.\ Haeberli, H.\ Wita\l a, W.\ Gl\"ockle, and J.\ Golak,
{\it Phys.\ Lett.\ }{\bf B307}, 13 (1993);\\
J.\ E.\ McAninch, L.\ O.\ Lamm, and W.\ Haeberli, {\it Phys.\ Rev.\ }{\bf
C50},
589 (1994).
\bibitem{ay10data}
C.\ R.\ Howell {\it et al.}, {\it Few-Body Systems  }{\bf 2}, 19 (1987).
\bibitem{ay50data}
J.\ L.\ Romero {\it et al.}, {\it Phys.\ Rev.\ }{\bf C25}, 2214 (1982);\\
J.\ W.\ Watson {\it et al.}, {\it Phys.\ Rev.\ }{\bf C25}, 2219 (1982).
\bibitem{cf}  S.\ A.\ Coon and J.\ L.\ Friar, {\it Phys.\ Rev.\ C} {\bf
34},
1060 (1986). Eq.~(33b) should equal Eq.~(27b), but contains a typographical
error: the indices $i$ and $j$ should be interchanged in Eq.~(33b) [J.
Adam,
Private Communication].

\end{thebibliography}
